\documentclass[
amsmath,
prd,nofootinbib,floatfix,12pt,
]{revtex4}

\newcommand\beq{\begin{eqnarray}}
\newcommand\eeq{\end{eqnarray}}

\newcommand\missET{E_T^{\rm miss}}

\def\lsim{\mathrel{\rlap{\lower4pt\hbox{$\sim$}}
    \raise1pt\hbox{$<$}}}                
\def\gsim{\mathrel{\rlap{\lower4pt\hbox{$\sim$}}
    \raise1pt\hbox{$>$}}}            

\allowdisplaybreaks
\usepackage{graphicx}
\usepackage{setspace}
\usepackage{dcolumn}
\usepackage{bm}
\usepackage{verbatim}

\usepackage[dvips]{color}
\definecolor{Red}{cmyk}{0,1,1,0}
\definecolor{BrickRed}{cmyk}{0,0.89,0.94,0.28}
\definecolor{Blue}{cmyk}{1,1,0,0}
\definecolor{Green}{cmyk}{1,0,1,0}

\renewcommand{\theequation}{\arabic{section}.\arabic{equation}}

\begin{document}

\title{\large \baselineskip=20pt 
Prospects for vectorlike leptons at future proton-proton colliders}

\author{Prudhvi N.~Bhattiprolu and Stephen P.~Martin}
\affiliation{\it 
Department of Physics, Northern Illinois University, DeKalb IL 60115}

\begin{abstract}\normalsize \baselineskip=15.5pt
Vectorlike leptons are an intriguing possibility for physics beyond the Standard Model. We study the reach for discovering (at 5$\sigma$ significance) or excluding (at 95\% confidence) models of charged vectorlike leptons that mix predominantly with the tau, using multi-lepton signatures at various future proton-proton collider options: a high-luminosity LHC with $\sqrt{s} =$ 14 TeV, a high-energy LHC with $\sqrt{s} =$ 27 TeV, and possible new longer-tunnel colliders with $\sqrt{s} =$ 70 or 100 TeV. For weak isodoublet vectorlike leptons, we estimate that a 27 TeV high-energy LHC with 15 ab$^{-1}$ could exclude masses up to about 2300 GeV, or discover them if the mass is less than about 1700 GeV, while a 100 TeV collider with 30 ab$^{-1}$ could exclude masses up to about 5750 GeV, or make a discovery if the mass is less than about 4000 GeV. However, we find that weak isosinglet vectorlike leptons present a much more difficult challenge, with some reach for exclusion, but not for discovery at any of the collider options considered.
\end{abstract}

\maketitle

\vspace{-1.2cm}

\baselineskip=13.5pt

\tableofcontents
\baselineskip=15.4pt
\newpage

\section{Introduction \label{sec:intro}}
\setcounter{equation}{0}
\setcounter{figure}{0}
\setcounter{table}{0}
\setcounter{footnote}{1}

The Large Hadron Collider (LHC) has conducted searches for new physics beyond 
the Standard Model, at proton-proton center of mass energies up to 
$\sqrt{s} = 13$ TeV. Following the discovery \cite{LHCdiscovery}
of the Higgs scalar boson with mass near $M_h = 125$ GeV, 
the present evidence from the LHC is consistent with the minimal version
of the Standard Model, with direct and indirect limits on new particles 
extending well into the TeV range. However, there are many candidate new 
physics models that obey decoupling as the mass scale of new physics is raised.
These include supersymmetric theories, where agreement with the Standard Model
gets better as the scale of supersymmetry breaking is raised. An example of a 
non-decoupling theory is a new chiral 4th family of fermions. 
Because of the necessity of generating the large chiral fermion 
masses through Yukawa couplings and the Higgs mechanism, decoupling is violated,
which causes significant contributions to precision electroweak observables
as well as Higgs boson production and decay rates, in conflict 
\cite{Lenz:2013iha} with the observations.

New fermions are still allowed, because they will obey decoupling, if they obtain 
their mass entirely or mostly as bare electroweak-singlet terms in the Lagrangian, 
rather than from the Yukawa couplings to the Higgs field. 
This implies that the new fermions are vectorlike (the antonym of chiral).
It is a notable common feature of many 
new physics theories that are motivated for other reasons (such as the need for 
a cosmological and astrophysical dark matter candidate, or to address 
the hierarchy problem) that they often contain vectorlike fermions. For example,
in supersymmetry \cite{Primer}, the Higgsinos have the same 
gauge quantum numbers as would a vectorlike
pair of weak isodoublet leptons. 
It is plausible that additional vectorlike fermion masses are at the
TeV scale, because whatever mechanism is responsible for the Higgsino masses
can also be applied to the vectorlike fermion masses, 
which have the same structure. 
Vectorlike fermions can therefore be added on to the minimal version of 
supersymmetry, with one possible benefit that they can 
contribute to the lightest Higgs boson mass
\cite{Moroi:1991mg,Moroi:1992zk,Babu:2004xg,Babu:2008ge,Martin:2009bg,
Graham:2009gy,Martin:2010dc,Endo:2011mc,Evans:2011uq,Li:2011ab,
Moroi:2011aa,Endo:2011xq,Endo:2012rd,Nakayama:2012zc,Martin:2012dg}, 
which allows a lower supersymmetry breaking scale consistent
with $M_h = 125$ GeV. For example, adding vectorlike quarks to  
the simplest models of gauge-mediated supersymmetry breaking allows
\cite{Endo:2011xq,Endo:2012rd,Nakayama:2012zc,Martin:2012dg}
them to be still discoverable at the LHC.
However, even ignoring such motivations,
it is also worthwhile to consider vectorlike fermions merely on the 
merits of being a simple and consistent candidate extension 
of the Standard Model.

For vectorlike quarks, the LHC pair production
cross-section is determined from QCD, and is therefore 
large and essentially model-independent.\footnote{Single 
production of vectorlike quarks is also subject to LHC limits, but in a much more 
model-dependent way.}
The searches for vectorlike quarks are consequently quite advanced, with 
current LHC limits (depending on the specific decay modes) on 
pair-produced vectorlike quarks
that can be found in refs.~\cite{Aaboud:2018pii,Sirunyan:2018omb}. 

In contrast, vectorlike leptons (VLLs) 
\cite{Dimopoulos:1990kc,
Sher:1995tc,
Thomas:1998wy,
Frampton:1999xi,
delAguila:2008pw,
delAguila:2010vg,
Ishiwata:2011hr,
Dermisek:2012as,
Joglekar:2012vc,
Kearney:2012zi,
Arina:2012aj,
Batell:2012zw,
Dermisek:2012ke,
Feng:2013mea,
Joglekar:2013zya,
Schwaller:2013hqa,
Dermisek:2013gta,
Ishiwata:2013gma,
Fairbairn:2013xaa,
Altmannshofer:2013zba,
Fichet:2013gsa,
Falkowski:2013jya,
Halverson:2014nwa,
Dobrescu:2014fca,
Ellis:2014dza,
Falkowski:2014ffa,
Holdom:2014rsa,
Dermisek:2014cia,
Dobrescu:2014esa,
Dermisek:2014qca,
Holdom:2014boa,
Aad:2015dha,
Ishiwata:2015cga,
Bizot:2015zaa, 
Dermisek:2015oja,
Bhattacharya:2015qpa,
Kumar:2015tna, 
Dermisek:2015hue,
Abdullah:2016avr,
Dermisek:2016via,
Feng:2016ysn,
Angelescu:2016mhl,
Bahrami:2016has,
Choudhury:2017fuu,
Poh:2017tfo,
Lu:2017uur,
Araz:2018uyi,
Bell:2019mbn,
Zheng:2019kqu,
CMS:2018cgi,
Sirunyan:2019ofn}
are pair-produced by $s$-channel
electroweak vector boson diagrams, leading to cross-sections that 
are much smaller and
dependent on the choices of weak isospin and weak hypercharge 
of the new states.\footnote{Here we concentrate on models with new 
charged vectorlike leptons. Signatures from models with neutrino masses
from electroweak singlet states at the TeV scale are reviewed in 
\cite{Cai:2017jrq,Pascoli:2018heg}, for example.} 
For this reason, until recently the LHC detector collaborations had not put
limits on VLLs beyond those that could be inferred from the CERN LEP 
$e^-e^+$ collider, which excluded \cite{Achard:2001qw} masses up to 
$101.2$ GeV. 
There had been several phenomenological
studies that looked into the LHC discovery and exclusion
capabilities. In ref.~\cite{Dermisek:2014qca}, search strategies and capabilities
were discussed for the optimistic
case that the VLLs decayed mostly to muons, and an ATLAS search
was conducted for that case in \cite{Aad:2015dha}. 
In ref.~\cite{Kumar:2015tna}, 
the more pessimistic case of decays to tau-rich final states were considered,
and it was argued that even with the (then current, but now old) 
data at $\sqrt{s}=8$ TeV it should
be possible to exclude such 
$SU(2)_L$-doublet VLLs up to about $M_{\tau'} = 275$ GeV, 
but not $SU(2)_L$-singlet charged VLLs, which have a much smaller production
cross-section and unfavorable branching ratios. 
Since then, the CMS collaboration
has published \cite{CMS:2018cgi,Sirunyan:2019ofn}
the results of dedicated searches for doublet VLLs,
based on 41.4 fb$^{-1}$ and 77.4 fb$^{-1}$ data samples at $\sqrt{s}=13$ TeV. 
Aided in part by a deficit of events in the signal regions
compared to the background expectation, 
CMS was able to obtain an exclusion for 
120 GeV $< M_{\tau'} <$ 790 GeV
for a pure $SU(2)_L$ doublet VLL pair that mixes with and decays to tau leptons.

In this paper, we will consider the prospects for exclusion or discovery
of VLLs at planned future proton-proton colliders. While there are a wide
variety of possible VLL models, we will consider as benchmark models
two simple minimal cases as defined in ref.~\cite{Kumar:2015tna},
a ``Singlet VLL" and a ``Doublet VLL" model, with the mixing to the Standard Model leptons assumed to be entirely with the tau, and small. 
One advantage of these models as benchmarks
is that the pair production cross-sections are uniquely
determined as a function of the mass, as 
discussed in ref.~\cite{Kumar:2015tna} and reviewed in the next section.
As was the case in ref.~\cite{Kumar:2015tna}, our analysis below shows that the minimal 
Singlet VLL model is very difficult to probe at proton-proton colliders. 
We will therefore also consider a class of non-minimal models
which are similar to the minimal Singlet VLL model and have the same production
cross-section, but have different branching ratios for the lightest
charged new lepton.

We will consider the following proton-proton
collider options: 
\begin{itemize}
\item a High-Luminosity LHC (HL-LHC),
  defined as 3 ab$^{-1}$ of proton-proton collisions at $\sqrt{s} = 14$ TeV,
\item a High-Energy LHC (HE-LHC),
  defined as 15 ab$^{-1}$ at $\sqrt{s} = 27$ TeV, 
\item a very high energy $pp$ collider,
defined as up to 30 ab$^{-1}$ 
at either $\sqrt{s} = 70$ or 100 TeV, which could be realized as e.g.~a CERN
hadron-hadron Future Circular Collider (FCC-hh) \cite{FCChh} 
or a Super proton proton Collider (SppC) \cite{SppC} in China.
\end{itemize}
The physics potential for HL-LHC and HE-LHC has been studied for various
other beyond-the-Standard-Model scenarios in 
ref.~\cite{CidVidal:2018eel} and references therein.

For reach estimates, we will use a 
simple cut-based counting experiment criteria,
where the significance
\beq
Z = \sqrt{2}\, {\rm erfc}^{-1}(2p).
\eeq 
of an experimental result is related to the probability $p$ of observing, 
in an ensemble of many repeated experiments, a result
of equal or greater incompatibility with the null hypothesis.
Let the number of background events be Poisson distributed with mean 
$b$, with the variance in $b$ (corresponding intuitively to an uncertainty 
in the expected number of background events) given by $\Delta_b$, 
and suppose the number of 
signal events is Poisson distributed with predicted mean $s$. 
Then the median expected
significance for discovery (where the null hypothesis is background only,
and the signal is assumed to be present in the data) 
is approximately \cite{LiMa,Cousins:2008zz,Cowan:2010js,Cowan}:
\beq
Z_{\rm disc} &=& \biggl [2 \biggl ((s+b) \ln \biggl [
\frac{(s+b) (b + \Delta_b^2)}{b^2 + (s+b) \Delta_b^2} \biggr ]
- \frac {b^2}{\Delta_b^2} \ln \biggl [
1+ \frac{s \Delta_b^2}{b (b+\Delta_b^2) }
\biggr ] \biggr ) \biggr ]^{1/2} .
\label{eq:Zdisc}
\eeq
The median expected significance for exclusion (where the null hypothesis is
now the background + signal model, 
but the signal is assumed to be actually absent in the data) 
can be approximated by \cite{Kumar:2015tna}:
\beq
Z_{\rm exc} =
\biggl \{
2 \biggl [ s-b \ln \biggl (\frac{b+s+x}{2b} \biggr )
- \frac{b^2}{\Delta_b^2} \ln \biggl (\frac{b-s+x}{2b} \biggr ) \biggr ] -
(b + s - x) (1 + b/\Delta_b^2) 
\biggr \}^{1/2},
\phantom{xx}
\label{eq:Zexc}
\eeq
where
\beq
x = \sqrt{(s+b)^2 - 4 s \Delta_b^2/(1 + \Delta_b^2/b)} .
\eeq
In the idealized limit of a perfectly known background prediction, $\Delta_b=0$,
these would reduce to 
\beq
Z_{\rm disc} &=& \sqrt{2 [(s+b) \ln(1+s/b) - s]},
\label{eq:ZdiscnoDeltab}
\\
Z_{\rm exc} &=& \sqrt{2 [s - b \ln(1 + s/b)]}.
\label{eq:ZexcnoDeltab}
\eeq
In the further limit of large $b$, these both approach 
$Z_{\rm disc} = Z_{\rm exc} = s/\sqrt{b}$.
In the following, we will use as a criteria for expected 5-sigma 
discovery that $s$ is at least 10 and 
eq.~(\ref{eq:Zdisc}) should exceed $Z_{\rm disc} > 5$,
and for an expected 95\% exclusion 
that eq~(\ref{eq:Zexc}) should exceed $Z_{\rm exc} > 1.645$ corresponding
to $p= 0.05$. We will use the somewhat arbitrary (since the capabilities of the detectors
are unknown at present)
choices $\Delta_b/b =$ 0.1, 0.2, and 0.5, 
corresponding to a 10\%, 20\%, and 50\% uncertainty in the background. 
We have also assumed that $b$ is always at least 1 event, to be conservative.
 
\section{Minimal models for production and decay of vectorlike leptons\label{sec:productiondecay}}
\setcounter{equation}{0}
\setcounter{figure}{0}
\setcounter{table}{0}
\setcounter{footnote}{1}

Following the terminology and definitions in ref.~\cite{Kumar:2015tna},
the ``Singlet VLL Model" 
contains the Standard Model fields and interactions plus
an $SU(2)_L$-singlet charged VLL $\tau^{\prime -}$ and its
antiparticle $\tau^{\prime +}$, which consist of 2-component left-handed
fermions transforming
under $SU(3)_c \times SU(2)_L \times U(1)_Y$ as
\beq
\tau^{\prime} + \overline \tau^{\prime} &=&
({\bf 1}, {\bf 1}, -1) + ({\bf 1}, {\bf 1}, +1).
\eeq
The ``Doublet VLL Model" contains a new charged 
lepton and its neutral heavy Dirac neutrino partner,
realized as 2-component left-handed fermions transforming as
\beq
L + \overline L 
\,=\, 
\begin{pmatrix} \nu'\cr \tau' \end{pmatrix} + 
\begin{pmatrix} 
\overline \tau'
\cr 
\overline \nu' 
\end{pmatrix}
&=& ({\bf 1}, {\bf 2}, -\frac{1}{2}) + ({\bf 1}, {\bf 2}, +\frac{1}{2}).
\eeq
In both models, a single weak-isosinglet bare fermion mass parameter $M$ 
is mostly responsible for the new fermion masses, 
with a small Yukawa couplings $\epsilon$ to the Higgs field
providing the mixing mass with
the Standard Model $\tau$ lepton, which also 
has its own Yukawa coupling $y_\tau$
to the Higgs field. 
The charged fermion mass matrix for the 
$\tau, \tau'$ system 
in each case then can be written in the form
\beq
{\cal M} &=& \begin{pmatrix} y_\tau v & 0 \\
\epsilon v & M 
\end{pmatrix},
\eeq
where $v \approx 174$ GeV is the Standard Model Higgs vacuum expectation value.
The tree-level mass eigenvalues, obtained from the square roots of
the eigenvalues of ${\cal M}^\dagger {\cal M}$ after expanding for 
$y_\tau v, \epsilon v \ll M$, are:
\beq
M_\tau &=& y_\tau v (1 - \epsilon^2 v^2/2 M^2 + \ldots)
,
\\
M_{\tau'} &=& M (1 + \epsilon^2 v^2/2 M^2 + \ldots)
,
\eeq
where the ellipses in both cases represent terms suppressed by 
$\epsilon^4 v^4/M^4$ or 
$\epsilon^2 y_\tau^2 v^4/M^4$,
while the tree-level neutral VLL mass in the Doublet model 
is simply $M_{\nu'} = M$.
There is also a 1-loop radiative correction to the 
physical mass splitting $M_{\nu'} - M_{\tau'}$, but it is also small, only
of order a few hundred MeV \cite{Thomas:1998wy}.

In the special case of no mixing with the tau lepton, $\epsilon = 0$, the 
lightest VLL would be absolutely stable
due to a conserved global symmetry. 
Here, we assume instead that $\epsilon$ 
is small enough to be treated as a tiny perturbation 
in the mass matrix, but that it exceeds
about $2 \times 10^{-7}$, which is large enough to allow \cite{Kumar:2015tna} 
the VLLs to decay promptly\footnote{If the $\tau'$ in the Singlet VLL model is quasi-stable,
then search strategies based on time-of-flight and ionization rate should apply, as in the
ATLAS search in ref.~\cite{Aaboud:2019trc}. Assuming that the $\tau'$ interacts similarly to a chargino,
we infer from the cross-section limit in Figure 10 of \cite{Aaboud:2019trc}
that its mass should exceed about 750 GeV. This does not apply to the Doublet VLL model, where
the $\tau'$ can decay to $\nu'$ inside the detector.}
on collider
detector length scales 
to Standard Model states,
with widths that dominate over 
the competing mode $\tau^{-\prime} \rightarrow \pi^- \nu'$ \cite{Thomas:1998wy}
in the Doublet VLL case.
The fermion mass eigenstates then consist of, besides
the usual $\tau^+, \tau^-, \nu_\tau$ and the rest of the Standard Model states, a charged 
Dirac pair $\tau^{\prime -}, \tau^{\prime +}$ in both models, and a neutral Dirac pair
$\nu', \overline \nu'$ in the Doublet VLL model only. Due to the 
small size of $\epsilon$ and the 1-loop radiative splitting,
we can take $M_{\nu'} \approx M_{\tau'}$ for the purposes of collider simulations.

The collider pair production modes are
\beq
pp &\rightarrow& \gamma^*, Z^* \>\rightarrow\> \tau^{\prime -} \tau^{\prime +} \label{tau'tau'}
,
\\
pp &\rightarrow& Z^* \>\rightarrow\> \nu'\, \overline \nu' \label{nu'nu'}
,
\\
pp &\rightarrow& W^{-*} \>\rightarrow\> \tau^{\prime -}\overline \nu' \label{tau'mnu'}
,
\\
pp &\rightarrow& W^{+*} \>\rightarrow\> \tau^{\prime +} \nu' \label{tau'pnu'}
,
\eeq
involving couplings to the $\gamma$, $Z$, and $W^\pm$ vector bosons, 
which are listed in
ref.~\cite{Kumar:2015tna}. Of course, only the first of these processes 
occurs for the Singlet VLL model.

\begin{figure}[!tb]
  \begin{minipage}[]{0.495\linewidth}
    \includegraphics[width=8.0cm,angle=0]{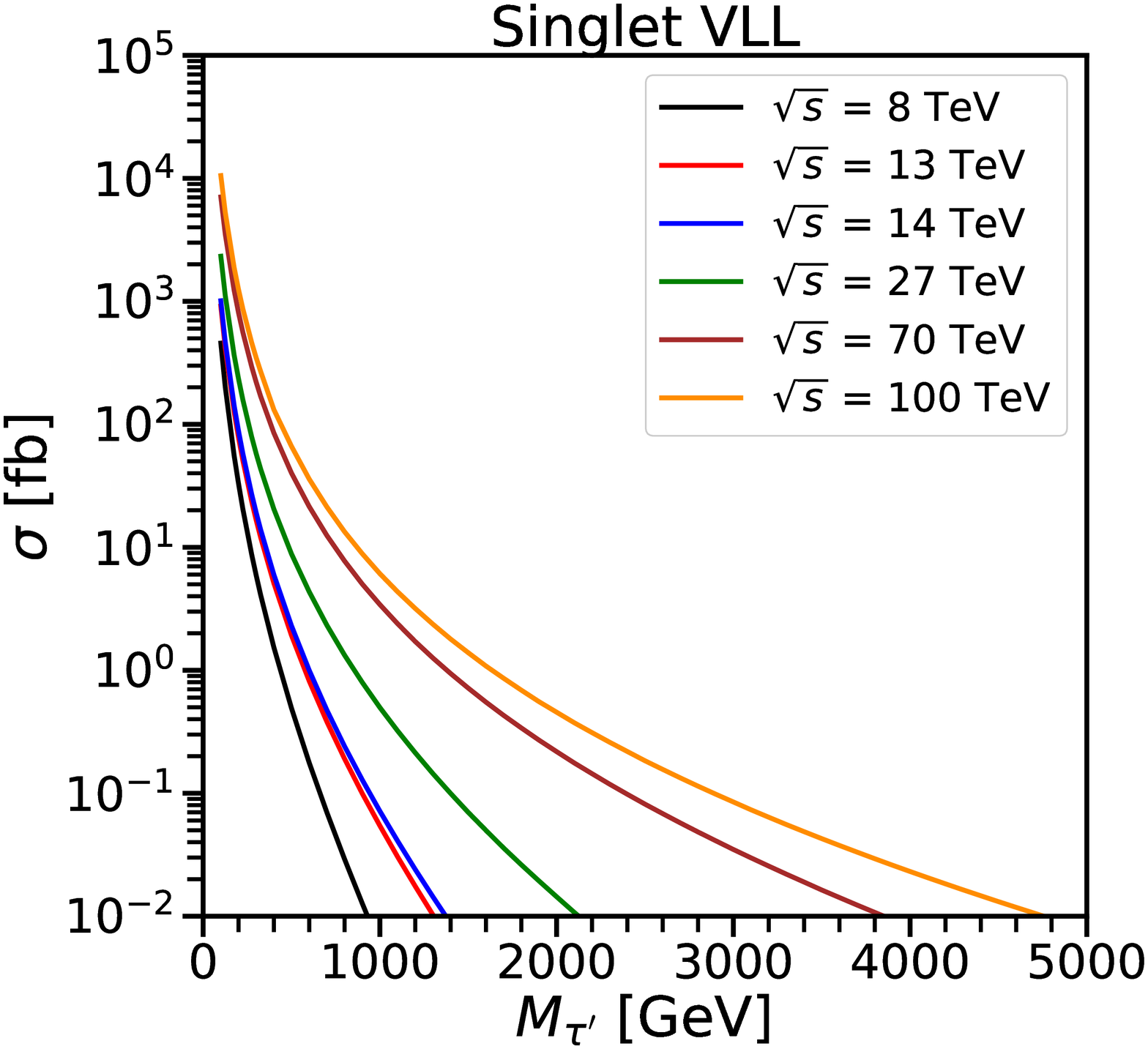}
  \end{minipage}
    \begin{minipage}[]{0.495\linewidth}
    \includegraphics[width=8.0cm,angle=0]{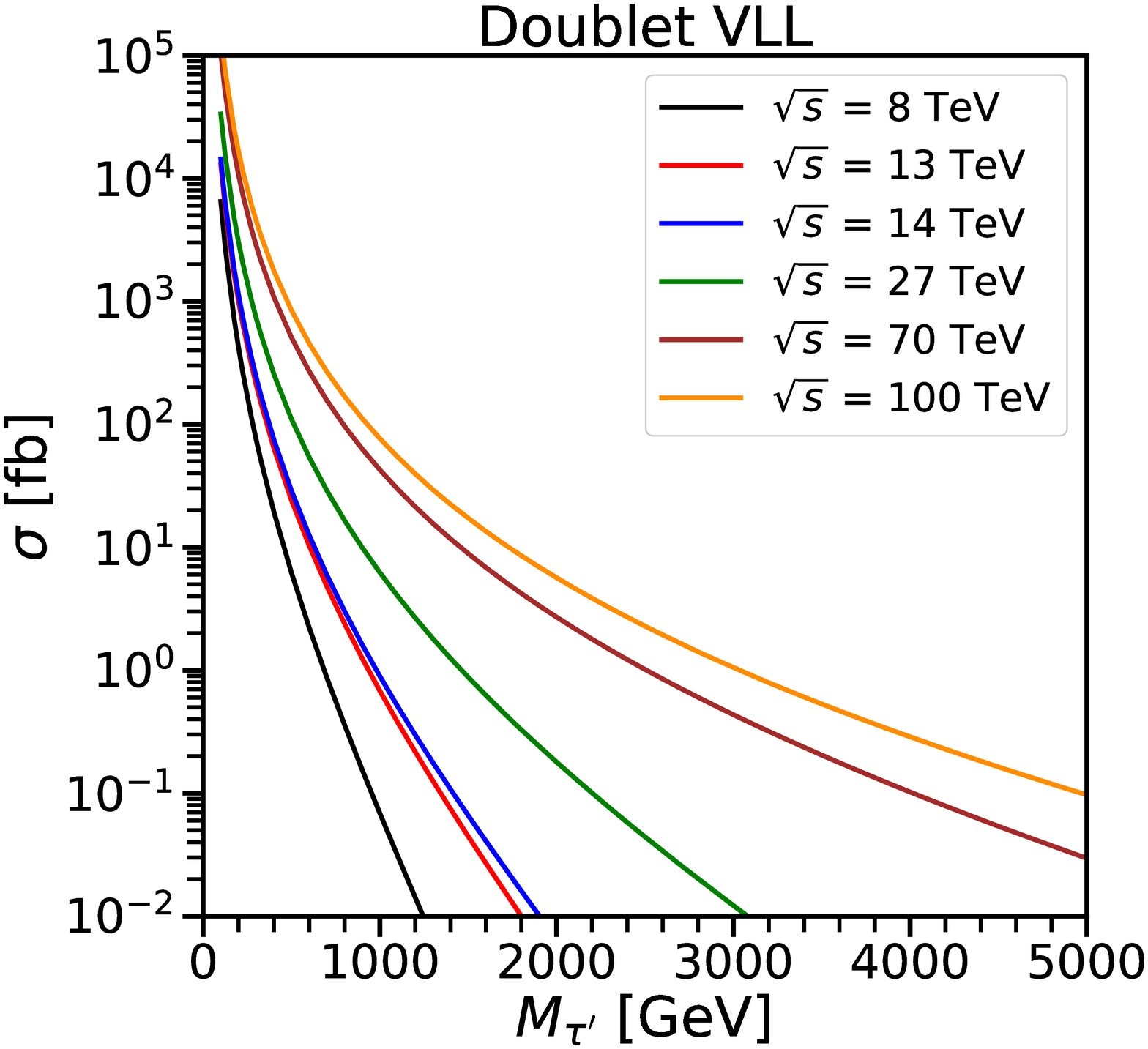}
  \end{minipage}
\begin{center}\begin{minipage}[]{0.95\linewidth}
\caption{\label{fig:totalsigmas} The total pair production cross-sections 
for $\tau^{\prime -} \tau^{\prime +}$ in the Singlet VLL Model (left panel)
and for the sum of $\tau^{\prime -} \tau^{\prime +}$ and 
$\nu' \overline\nu'$ and 
$\tau^{\prime-} \overline \nu'$ and $\tau^{\prime +} \nu'$ 
in the Doublet VLL Model (right panel),
for proton-proton collisions at $\sqrt{s} = 8, 13, 14, 27, 70, 100$ TeV (bottom to top).}
\end{minipage}\end{center}
\end{figure}

\begin{figure}[!tb]
  \begin{minipage}[]{0.505\linewidth}
  \begin{flushleft}
    \includegraphics[width=8.0cm,angle=0]{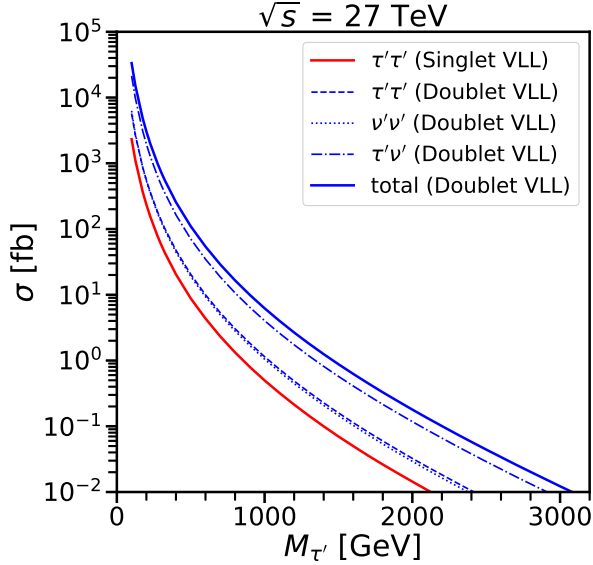}
  \end{flushleft}
  \end{minipage}
\begin{minipage}[]{0.445\linewidth}
\caption{\label{fig:sigmas27TeV} The individual pair production cross-sections 
in the Singlet VLL Model for $\tau^{\prime -} \tau^{\prime +}$ (solid red line) 
and in the Doublet VLL Model
for $\tau^{\prime -} \tau^{\prime +}$ (dashed blue line) and 
$\nu' \overline\nu'$ (dotted blue line) and 
$\tau^{\prime-} \overline \nu'$ plus $\tau^{\prime +} \nu'$ (dot-dashed blue line)
and total (solid blue line),
for proton-proton collisions at $\sqrt{s} = 27$ TeV.}   
  \end{minipage}
\end{figure}

In Figure \ref{fig:totalsigmas}, 
we show the total pair production cross-sections at a 
proton-proton collider as a function of $M_{\tau'} = M_{\nu'}$, for 
the choices $\sqrt{s} = 8, 13, 14, 27, 70, 100$ TeV.
Note that the Doublet VLL model has a considerably 
larger total cross-section, which
is partly responsible for the much better search prospects as found in ref.~\cite{Kumar:2015tna} and below in the present paper. For example, at 
$M_{\tau'} = 1000$ GeV, the total cross-section is 
about 12.5 times larger 
in the Doublet VLL model than for the Singlet VLL model,
nearly independent of the proton-proton collision energy over the range from 14 to 100 TeV.
In the Doublet VLL model,
the total contribution is dominated by the $W$-boson-mediated 
$\tau^{\prime-} \overline \nu'$ and $\tau^{\prime +} \nu'$ final states. However,
even if one restricts attention
to only the $\tau^{\prime -} \tau^{\prime +}$ final state
common to both models, the Doublet VLL model has a significantly 
larger cross-section. This is illustrated in 
Figure \ref{fig:sigmas27TeV}, which compares the individual cross-sections for
each of the final states at $\sqrt{s} = 27$ TeV. (Results are similar
at other values of $\sqrt{s}$.)

\begin{figure}[!tb]
  \begin{minipage}[]{0.495\linewidth}
    \includegraphics[width=8.0cm,angle=0]{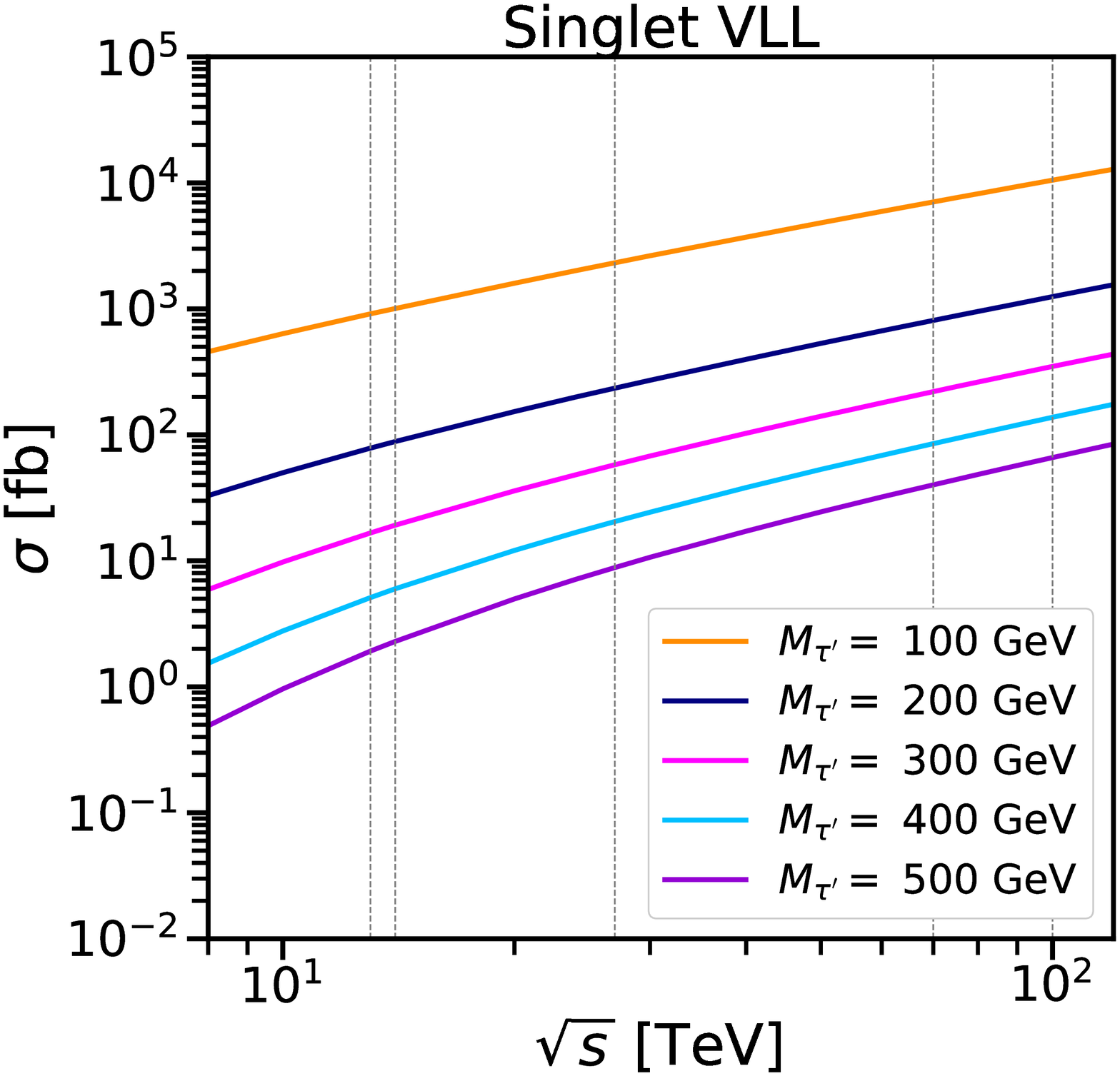}
  \end{minipage}
    \begin{minipage}[]{0.495\linewidth}
    \includegraphics[width=8.0cm,angle=0]{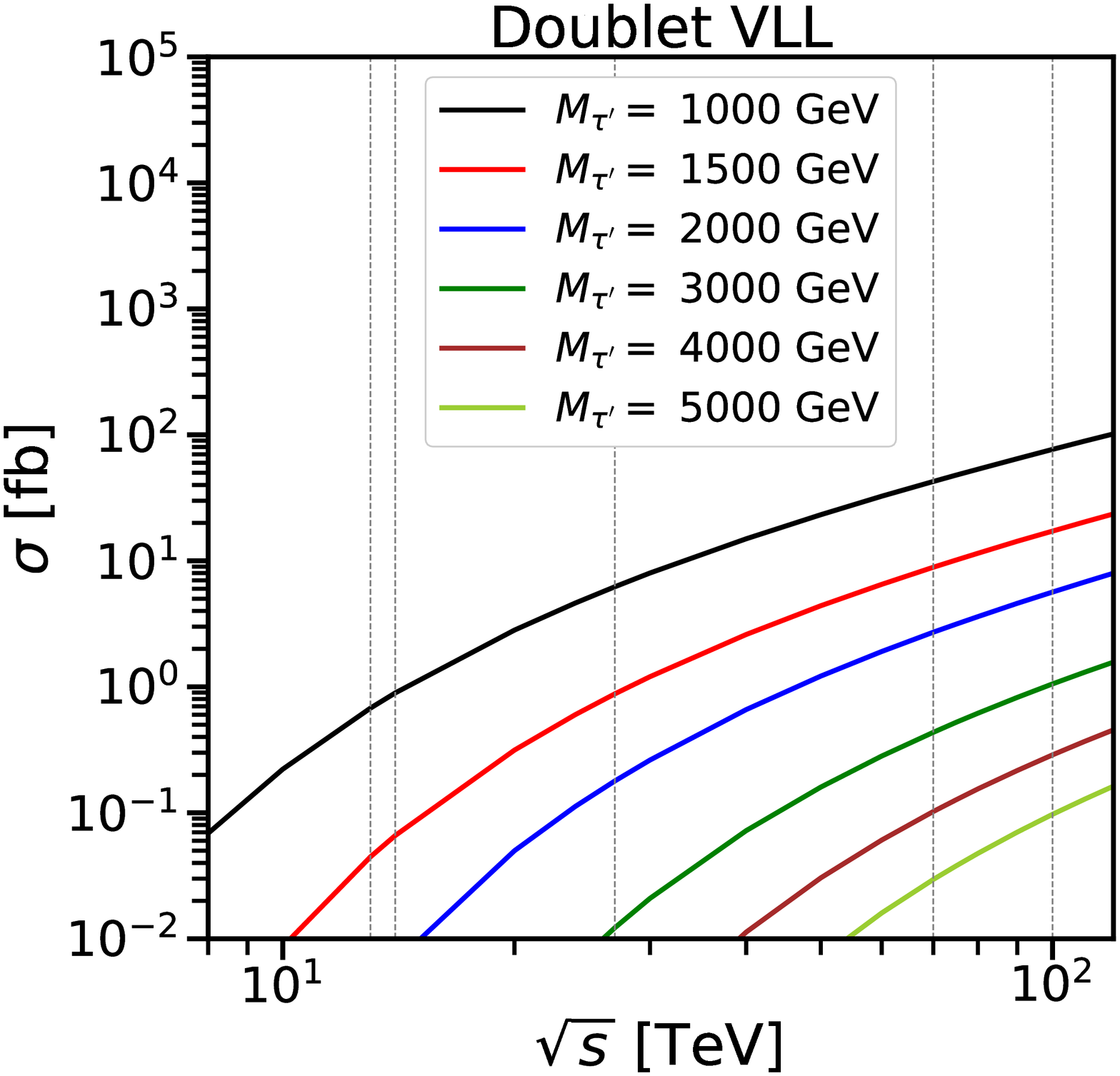}
  \end{minipage}
\begin{center}\begin{minipage}[]{0.95\linewidth}
\caption{\label{fig:sigmavssqrts} The total pair production cross-sections 
for $\tau^{\prime -} \tau^{\prime +}$ in the Singlet VLL Model (left panel)
and for the sum of $\tau^{\prime -} \tau^{\prime +}$ and 
$\nu' \overline\nu'$ and 
$\tau^{\prime-} \overline \nu'$ and $\tau^{\prime +} \nu'$ 
in the Doublet VLL Model (right panel),
as a function of $\sqrt{s}$, for various values of $M_{\tau'} = M_{\nu'}$
as labeled.
}
\end{minipage}\end{center}
\end{figure}

The total pair-production cross-sections for these models are also shown in 
Figure \ref{fig:sigmavssqrts} as a function of $\sqrt{s}$, for 
various choices of $M_{\tau'} = M_{\nu'}$. 
Clearly, the cross-sections are getting larger at higher collider energies. 
In the Doublet VLL model,
one obtains roughly the same cross-section for 
$M_{\tau'} = 4000$ GeV at $\sqrt{s}= 100$ TeV
as one would have for 
$M_{\tau'} = 1870$ GeV at $\sqrt{s}= 27$ TeV, or for 
$M_{\tau'} = 1220$ GeV at $\sqrt{s} = 14$ TeV.

As for the decays, the lepton-flavor conserving charged current
process $\tau' \rightarrow \nu' \pi^-$
is kinematically allowed in the Doublet VLL model, but has a width smaller than 
the direct lepton-flavor-violating decays to Standard Model 
fermions as long as $\epsilon \gsim 2 \times 10^{-7}$ \cite{Kumar:2015tna},
and is therefore neglected. 
In the Doublet VLL model, the neutral
VLLs decay 100\% of the time to a $W$ boson and an ordinary  tau lepton:
\beq
{\rm BR}(\nu' \rightarrow W^+ \tau^-) = 
{\rm BR}(\overline \nu' \rightarrow W^- \tau^+) = 1. 
\eeq 
This reflects our assumptions that there is no $\nu'$ mixing with the left-handed neutrinos
of the Standard Model and that the mixing Yukawa coupling $\epsilon$ involves the 
tau lepton, rather than muon or the electron; otherwise the discovery and 
exclusion strategies would be much easier, due to the higher 
detection efficiency and lower fake rates for $e,\mu$ compared to $\tau$. 

\begin{figure}[!tb]
  \begin{minipage}[]{0.495\linewidth}
    \includegraphics[width=8.0cm,angle=0]{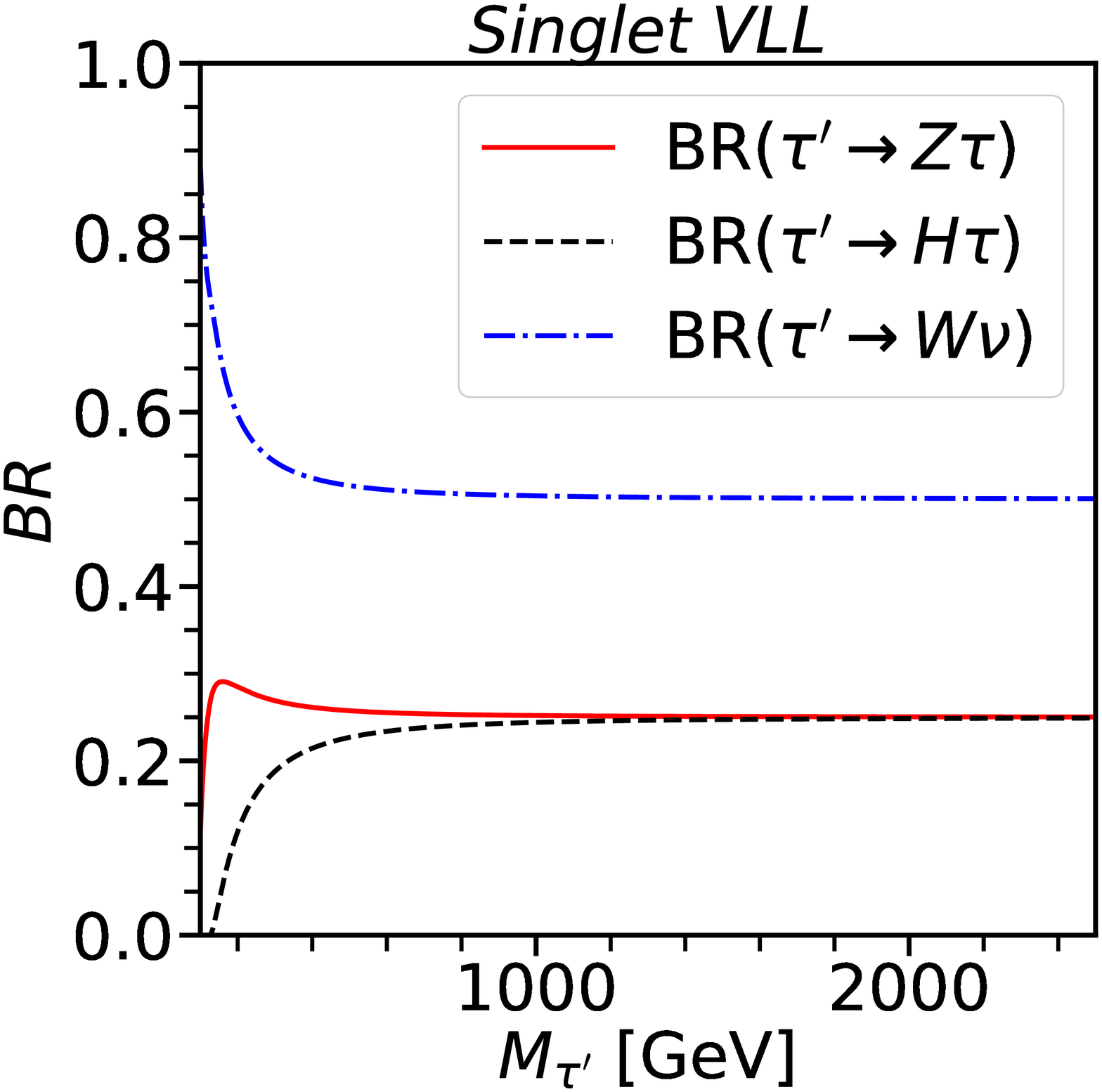}
  \end{minipage}
    \begin{minipage}[]{0.495\linewidth}
    \includegraphics[width=8.0cm,angle=0]{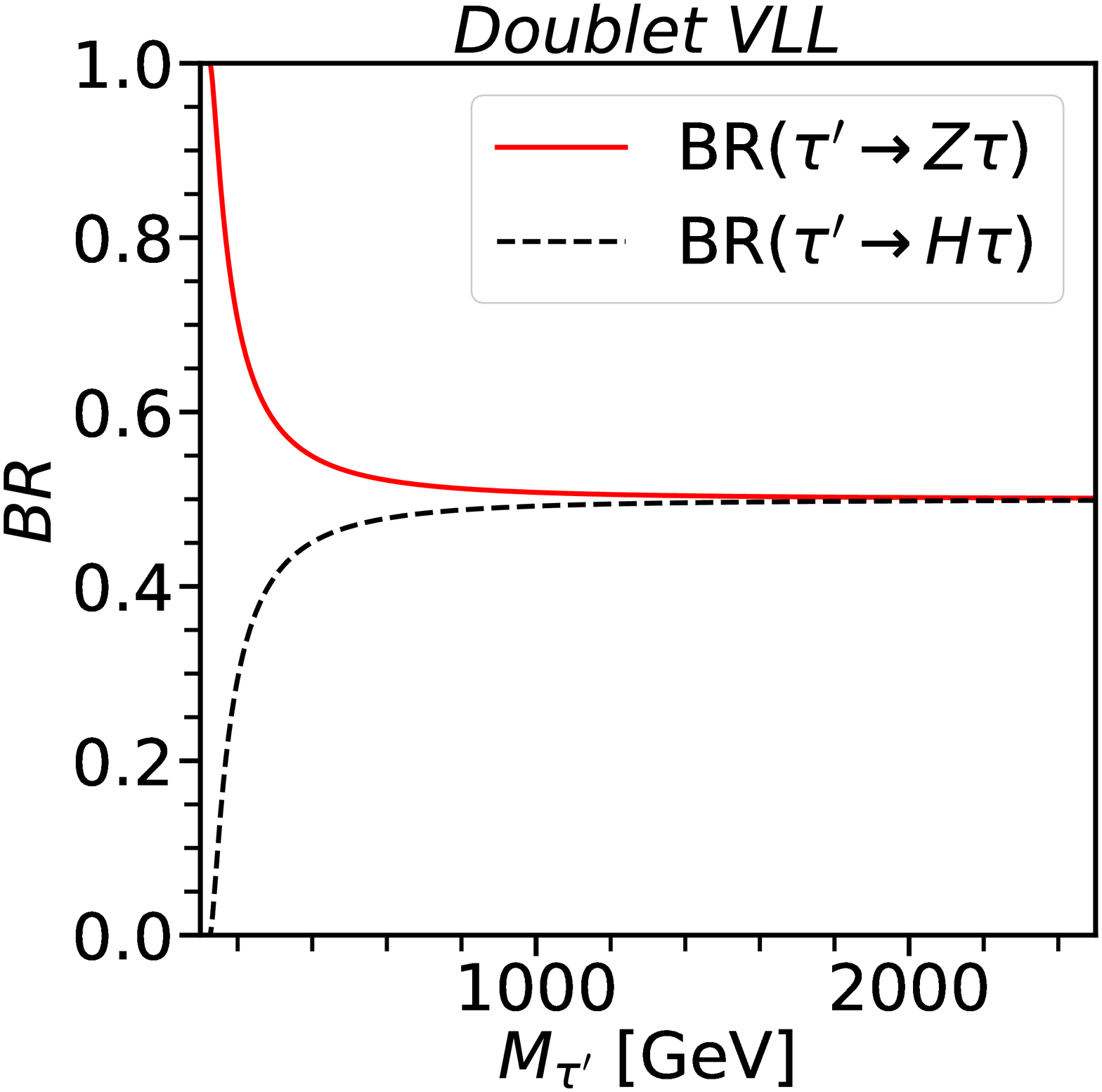}
  \end{minipage}
\begin{center}\begin{minipage}[]{0.95\linewidth}  
\caption{\label{fig:BRs} The branching ratios for $\tau' \rightarrow W\nu$ and
$Z\tau$ and $h\tau$, as a function of $M_{\tau'}$,
for the Singlet VLL model (left panel) and
the Doublet VLL model (right panel), showing the rapid approach to the Goldstone boson 
equivalence limit for larger masses.}
\end{minipage}\end{center}
\end{figure}

In both models, the charged VLLs can decay into the final states
$Z \tau$, $h \tau$, and $W \nu_\tau$. The decay widths 
(neglecting the tau lepton mass) are \cite{Kumar:2015tna}:
\beq
\Gamma (\tau' \rightarrow h \tau) &=& 
\frac{\epsilon^2}{64 \pi} M_{\tau'} (1 - r_h)^2,
\\
\Gamma (\tau' \rightarrow Z \tau) &=& 
\frac{\epsilon^2}{64 \pi} M_{\tau'} (1 + 2 r_Z) (1 - r_Z)^2,
\eeq
in both the Singlet and Doublet VLL models, while
\beq
\Gamma (\tau' \rightarrow W \nu) &=& 
\left \{ \begin{array}{ll} \displaystyle
\frac{\epsilon^2}{32 \pi} M_{\tau'} (1 + 2 r_W)  (1 - r_W)^2, \qquad\quad &\mbox{(Singlet VLL model)},
\\[4pt]
\phantom{xxxxx} 0, \qquad &\mbox{(Doublet VLL model),}
\end{array} \right.
\eeq
where $r_X \equiv M_X^2/M_{\tau'}^2$ for each of $X = h,Z,W$. 
In the decays to $Z$ and $W$, the
factors $(1 + 2 r_Z)$ and $(1 + 2 r_W)$ can be understood as coming from the 
longitudinal (1) and transverse $(2 r_X)$ 
components of the weak vector bosons. The longitudinal
components can in turn be understood as essentially the 
Goldstone modes that are eaten by the
vector bosons to obtain their masses. This illustrates the usual
Goldstone equivalence principle, which implies that 
for the limit of large $M_{\tau'}$ the branching ratios should approach:
\beq
\left [\mbox{BR}(\tau' \rightarrow h\tau),\> \mbox{BR}(\tau' \rightarrow Z\tau),\>
 \mbox{BR}(\tau' \rightarrow W \nu)\right ]
&=&
\biggl \{ \begin{array}{ll}
{}[ 0.25,\, 0.25,\, 0.5 ]\quad\!& \mbox{(Singlet VLL model)},\\
{}[ 0.5,\> 0.5,\> 0 ]\quad\!& \mbox{(Doublet VLL model)} .
\end{array} \biggr.
\nonumber \\ &&
\eeq
In Figure \ref{fig:BRs}, we plot the branching ratios for the $\tau'$ decays
in the two models, as a function of the mass $M_{\tau'}$, showing the asymptotic
approach to the Goldstone equivalence limit.

The Doublet VLL model therefore has the following final states:
\beq
ZZ\tau^{-} \tau^{+}, \quad Zh\tau^{-} \tau^{+},\quad hh\tau^{-} \tau^{+} \label{doublet_tau'tau'}
\eeq
from $\tau^{\prime}$ pair production eq. (\ref{tau'tau'}),
\beq
W^{+}W^{-}\tau^{-} \tau^{+}
\eeq
from $\nu^{\prime}$ pair production eq. (\ref{nu'nu'}), and
\beq
ZW^{\pm}\tau^{-} \tau^{+}, \quad hW^{\pm}\tau^{-} \tau^{+}
\eeq
from $\tau^{\prime}\nu^{\prime}$ production eq. (\ref{tau'mnu'}) and eq. (\ref{tau'pnu'}). 
Besides the three signals in eq. (\ref{doublet_tau'tau'}), the Singlet VLL model also has final states 
\beq
ZW^{\pm}\tau^{\mp}+\missET, \quad hW^{\pm}\tau^{\mp}+\missET, \quad W^{+}W^{-} + \missET 
,
\eeq
with the last being the largest in the minimal Singlet VLL model.

With the above branching ratios for the minimal Singlet VLL model, 
the analysis below does not find any 
reach for any of the signal regions at any of the collider options considered 
below, due to the low cross-section and the unfortunate large branching ratio for
$\tau' \rightarrow W \nu$. Therefore, we instead consider a class of non-minimal
models which have the feature 
that the lightest new fermion is still a charged, 
mostly isosinglet $\tau'$, but with other new 
fermions much heavier and therefore not contributing to the production. 
Thus, the new particle content first accessible to colliders is 
the same as in the minimal Singlet VLL model, but the mixing with 
the much heavier vectorlike fermions allows the branching ratios of $\tau'$ 
to be essentially arbitrary among the three final states listed above. 
We therefore present results for 
modified Singlet VLL models, with the same $\tau^{\prime +} \tau^{\prime -}$ 
production cross-section as the minimal Singlet VLL model, but with 
branching ratios set to the following three sub-cases:
\beq
{\rm BR }(\tau' \rightarrow h \tau) &=& 1,\!\!\!\qquad\qquad\qquad\qquad\qquad\mbox{``Higgs-philic Singlet VLL"},
\\
{\rm BR }(\tau' \rightarrow Z \tau) &=& 1,\!\!\!\qquad\qquad\qquad\qquad\qquad\mbox{``$Z$-philic Singlet VLL"},
\\
{\rm BR }(\tau' \rightarrow h \tau) &=& {\rm BR }(\tau' \rightarrow Z \tau) \>=\> 0.5,
\quad\mbox{``$W$-phobic Singlet VLL"}.
\phantom{x}
\eeq
Realizations of these simplified models, from appropriate limits of 
mixing a vectorlike isosinglet fermion with a much heavier 
vectorlike isodoublet fermion, are discussed in an Appendix.

For signal simulation, we have input a Lagrangian which governs the production and decay 
of the Doublet VLL and the Singlet VLL models 
(as discussed above) into {\sc FeynRules} v2.3 \cite{Alloul:2013bka}, 
a {\sc Mathematica} package, to obtain the Feynman rules as a
\textit{Universal FeynRules Output (UFO)} 
file, which is then imported into {\sc MadGraph5{\_}aMC@NLO} v2.6.5 \cite{Alwall:2011uj} 
\footnote{Electronic input files relevant for VLL models are available from the authors on request.}
We considered Standard Model production of the final states
$WZ$, $ZZ$, $t\overline t Z$, $t\overline t W$, $hW$, $hZ$, $t\overline t h$, 
$WWW$, $WWZ$, $ZZW$, and $ZZZ$ 
as the main backgrounds contributing to multi-lepton final states. We did not include 
reducible backgrounds such as $W+jets$, $Z+jets$, and $WZ+jets$, where one or more jets 
fake a tau. While they could be important, we do not have a way of reliably 
estimating them for unknown future detectors. Therefore it must be kept in mind 
that if they are large, our projections may be too optimistic.

Both signal and background events were generated using {\sc MadGraph} at leading order (LO). 
The decay couplings in the (VLL) Lagrangian are left as free parameters in the model file 
of {\sc FeynRules}, so that one has the flexibility to adjust them 
during the run time of {\sc MadGraph}. The 
numerical values of the decay couplings that are used
are actually not important 
(as long as they are not too small or too large), because in each separate Madgraph run
the decays of VLL were forced into one of the individual channels mentioned above, and later the event
samples were normalized using the branching ratios shown in Figure \ref{fig:BRs}. 
We calculated the cross-sections for signal at LO and normalized the background 
cross-sections as calculated at next-to-leading order (NLO), using {\sc MadGraph}.  
While using {\sc MadGraph}, we used their default set of parton distribution functions (PDFs) based on NNPDF2.3 set \cite{nnpdf23qed_lo}.
We used {\sc Pythia} 8.2 \cite{Sjostrand:2006za,Sjostrand:2015za} for showering and hadronization. 
Then, {\sc Delphes} 3.4 \cite{deFavereau:2013fsa} was used for detector simulation. 
We have used the default Delphes tagging efficiencies and misidentification rates 
for taus and b-jets, where the tau tagging efficiency is 0.6, 
and the tau misidentification rate for QCD jets is 0.01.
To increase the yield for background events to pass cuts described below, 
we forced every particle, except for the SM Higgs boson ($h$), 
to decay into leptons (including taus) and/or jets, such that they contribute to 
multi-lepton final states with at least 3 leptons. 
For generation of both signal and background events, 
we only considered the decay of $h$ into $W^{+}W^{-}$, $ZZ$, and $\tau^{+} \tau^{-}$, 
ignoring all other decays, and have normalized using the 
branching ratios of $h$ from HDECAY \cite{Djouadi:1997yw}. 
Thus each signal/background containing one or more $h$ is split into 3 signal/background components 
per $h$. We generated at least $10^5$ events for each signal component at each mass point, 
and at least $10^6$ events for each background component, at each of $\sqrt{s} = (14, 27, 70, 100)$ TeV. These numbers of generated events are usually sufficient, with a few exceptions
noted below. In the real world, better background determinations may come from data.

In our analysis, we first impose object cuts on leptons $\ell = e,\mu,\tau$, 
where, from now on, $\tau$ refers to a tau lepton that undergoes a hadronic decay. 
These include minimum
$p_T$ cuts that depend on the collider option, 
and are therefore listed in the subsections below. In all cases,
the lepton candidates are required to pass the 
following further cuts on pseudo-rapidity $\eta$
and isolation from other lepton candidates or jets:
\beq
 |\eta|  & < &  2.5 \label{eta}\\
 \Delta R_{l,l'} & > & 0.1 \quad \textrm{ (for each } \ell,\ell'= e,\mu,\tau) \\
 \Delta R_{l,j} & > & 0.3 \quad \textrm{ (for each jet and } \ell= e,\mu,\tau) .
\eeq
Here, $\Delta R = \sqrt{(\Delta\phi)^2 + (\Delta \eta)^2}$ as is usual.

Events are then selected with at least 3 three leptons, of which at least 2 must be $e$ or $\mu$.
The leading $e$ or $\mu$ lepton is also required to satisfy a minimum $p_T$ trigger requirement that
depends on the collider option, and is therefore listed separately in each of the subsections below. 
We also veto $b$-jets to reduce large backgrounds from $t\overline t$ production processes. Thus our event pre-selection common to all signal regions is:
\beq
N(e,\mu,\tau) & \geq & 3\\
N(e,\mu) & \geq & 2 \\
p_T^{e_1}\>\,\mbox{or}\>\, p_T^{\mu_1} &>& p_T^{\rm trigger}\\
N(\mbox{$b$-jets}) & = & 0 \label{nbjets}
\eeq
Events with no $e^+ e^-$ or $\mu^+\mu^-$ pair with 
invariant mass within 15 GeV of $M_Z$ are referred to below as ``no-$Z$", and events with exactly
two $e/\mu$ that have same-sign charges will be labeled as ``SS".
We then considered the following 6 distinct types of signal regions:
\beq
&&
\mbox{2 SS $e/\mu\> +\!\geq 1 \tau$}
\label{eq:signal1}
\\
&&
\mbox{2 SS $e/\mu\> +\!\geq 1 \tau$ with $E_T^{\rm miss} > 150$ GeV}
\\
&&
\mbox{$\geq 3e/\mu + 1 \tau$}
\\
&&
\mbox{$\geq 3e/\mu + 1 \tau$, no-$Z$}
\\
&&
\mbox{$\geq 2e/\mu + 2 \tau$}
\\
&&
\mbox{$\geq 2e/\mu + 2 \tau$, no-$Z$}
\label{eq:signal6}
\eeq
Finally, in each of these signal regions, we imposed a minimum lower bound cut on $L_T$, 
which is defined to be the sum of the transverse momentum of all leptons:
\beq
L_T = \sum_{\ell = e,\mu,\tau} p_T (\ell).  
\label{lt}
\eeq
We varied the choice of this cut to obtain good 
exclusion and discovery reach simultaneously for each signal region
and collider option and assumed fractional
background uncertainty $\Delta_b/b = (0.1,\> 0.2, \> 0.5)$. 
However, we have only done a very rough optimization 
for the $L_T$ cut, in part because the optimization is different for exclusion and for discovery,
and also because
the Monte Carlo simulations are only an approximation to the actual experimental capabilities, 
which will rely on detector designs yet to be determined. 
In general, the results found below reflect that the choice of cut on $L_T$ 
increases with the mass $M_{\tau'}$ at the edge of the reach. 
Also, for convenience, we always chose the same $L_T$ cut for both exclusion and discovery, 
even though an optimized cut would likely be somewhat different for the two cases. 

Note that the signal regions considered in eqs.~(\ref{eq:signal1})-(\ref{eq:signal6}) 
are far from exclusive of each other. Therefore,
to be conservative we have not attempted to combine them, 
although doing so could lead to some extension of the reach prospects.

\section{Results for the HL-LHC collider\label{sec:pp14TeV}}
\setcounter{equation}{0}
\setcounter{figure}{0}
\setcounter{table}{0}
\setcounter{footnote}{1}

In this section, we discuss the possibility of exclusion or discovery 
of both Doublet and Singlet VLL (including the minimal and non-minimal versions) 
at $\sqrt{s} = 14$ TeV with 3 ab$^{-1}$ of $pp$ collisions. 
In addition to the pseudo-rapidity, isolation and other requirements of 
eqs.~(\ref{eta})-(\ref{nbjets}), we require all leptons including hadronic tau candidates to satisfy 
\beq
p_T^\ell > \mbox{15 GeV}.
\eeq 
Additionally, the leading $e$ or $\mu$ 
in each event is required to satisfy a trigger requirement:
\beq
p_T^{e_1}\>\,\mbox{or}\>\, p_T^{\mu_1} > 30 \textrm{ GeV}.
\eeq
We then considered the six signal regions mentioned in 
eqs.~(\ref{eq:signal1})-(\ref{eq:signal6}). 

\subsection{Doublet VLL model}

\begin{figure}[!tb]
  \begin{minipage}[]{0.495\linewidth}
    \includegraphics[width=8.0cm,angle=0]{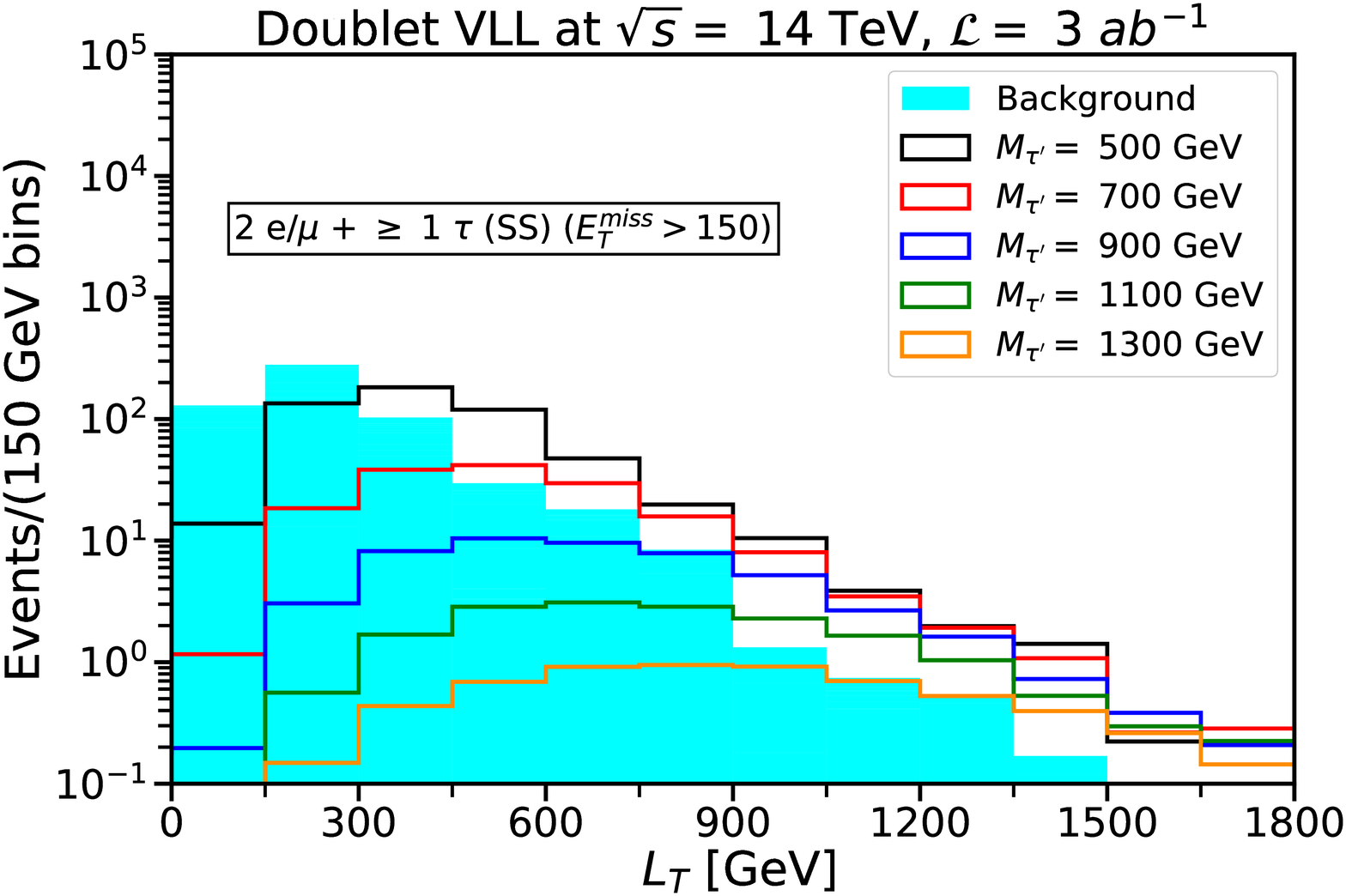}
  \end{minipage}
    \begin{minipage}[]{0.495\linewidth}
    \includegraphics[width=8.0cm,angle=0]{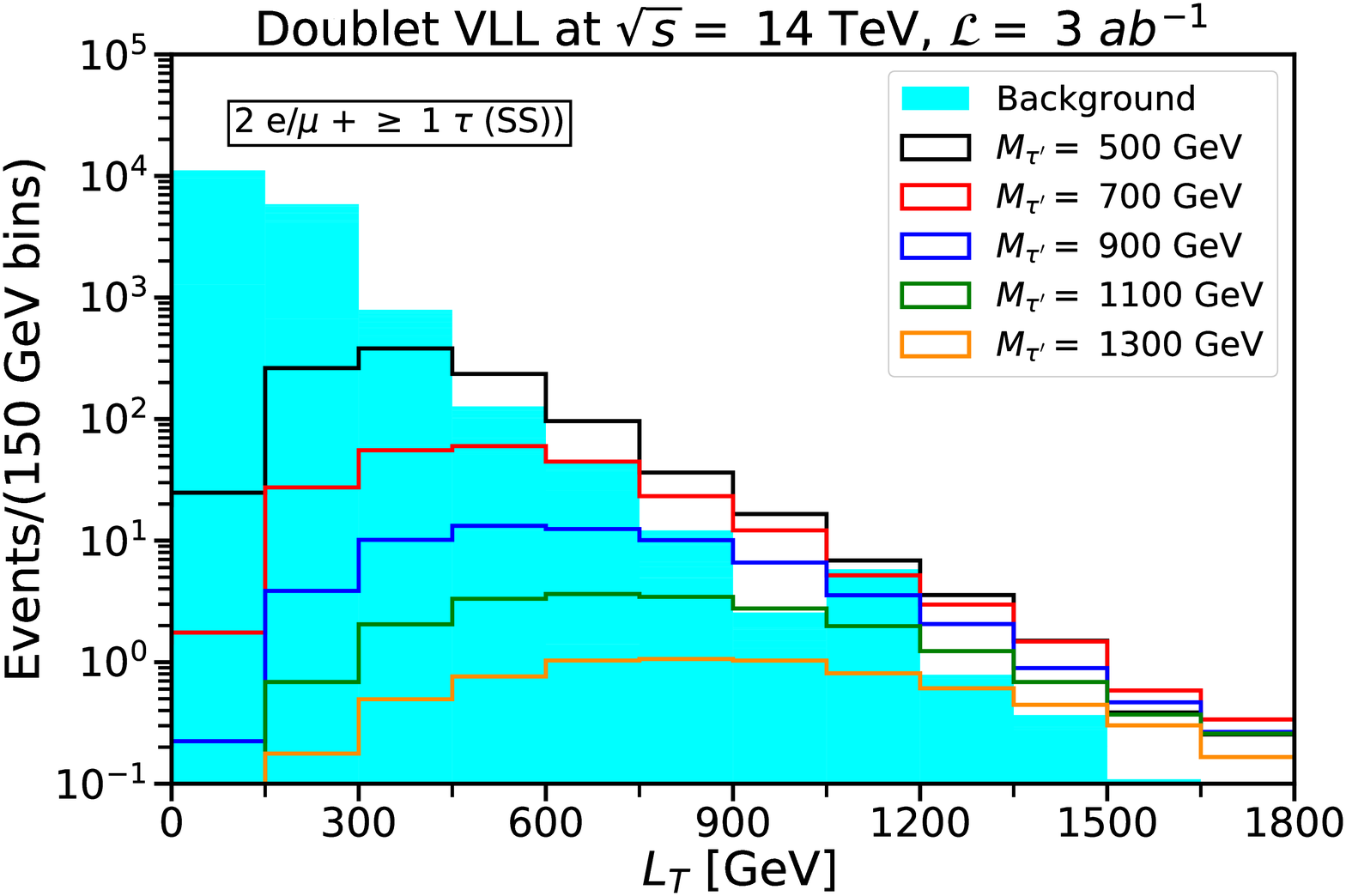}
  \end{minipage}
  \begin{minipage}[]{0.495\linewidth}
    \includegraphics[width=8.0cm,angle=0]{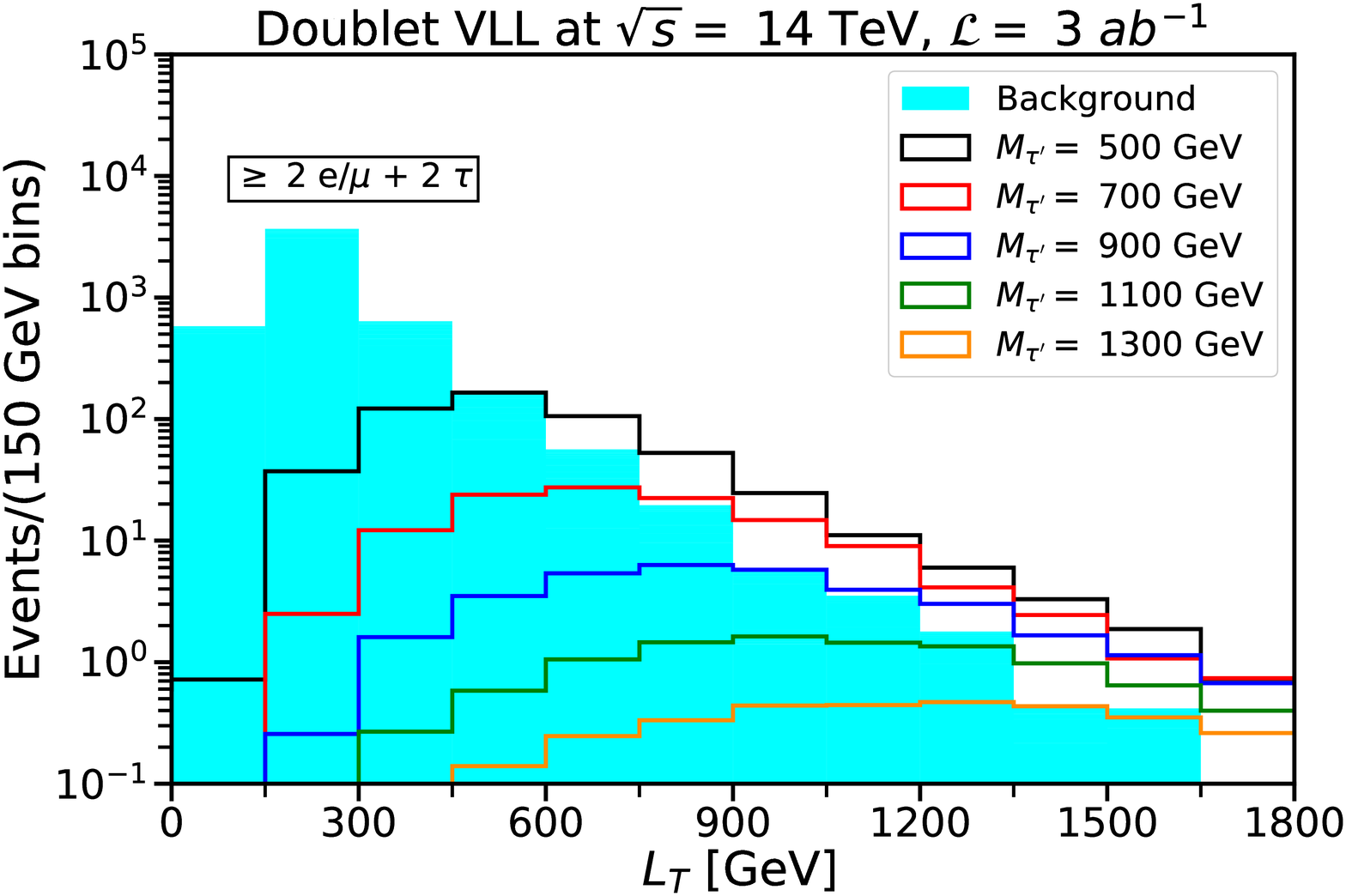}
  \end{minipage}
    \begin{minipage}[]{0.495\linewidth}
    \includegraphics[width=8.0cm,angle=0]{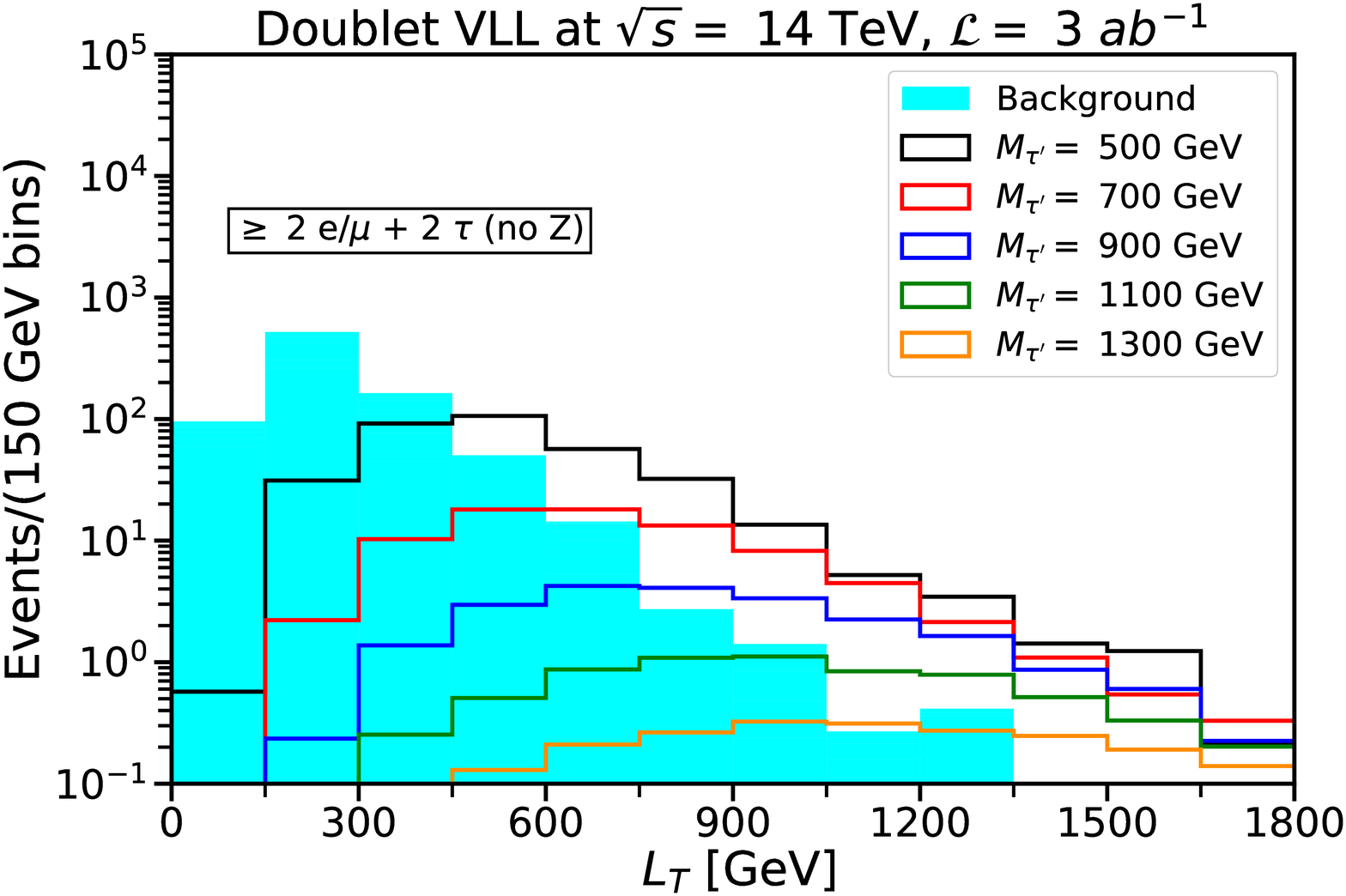}
  \end{minipage}
\begin{center}\begin{minipage}[]{0.95\linewidth}
\caption{\label{fig:LT_14TeV} 
$L_T$ event distributions for total background (shaded) and
Doublet VLL models (lines), for $pp$ collisions at $\sqrt{s} = 14$ TeV with an 
integrated luminosity $\mathcal{L} =$ 3 ab$^{-1}$. Five different masses
$M_{\tau'} = M_{\nu'} = 500,$ 700, 900, 1100, and 1300 GeV
are shown in each panel. The four panels show results for the four best signal regions, as labeled.}
\end{minipage}\end{center}
\end{figure}

In Figure \ref{fig:LT_14TeV}, we show the $L_T$ distributions for the best four
of these signal regions, for five different choices of $M_{\tau'}$ as labeled, 
and for the total of all backgrounds shown as the shaded histogram.
For $pp$ collisions with $\sqrt{s}=14$ TeV with 3 ab$^{-1}$, 
we found that the expected 
reach is approximately maximized if we then choose a cut 
$L_T > 800$ GeV. 

\begin{figure}[!tb]
  \begin{minipage}[]{0.495\linewidth}
    \includegraphics[width=8.0cm,angle=0]{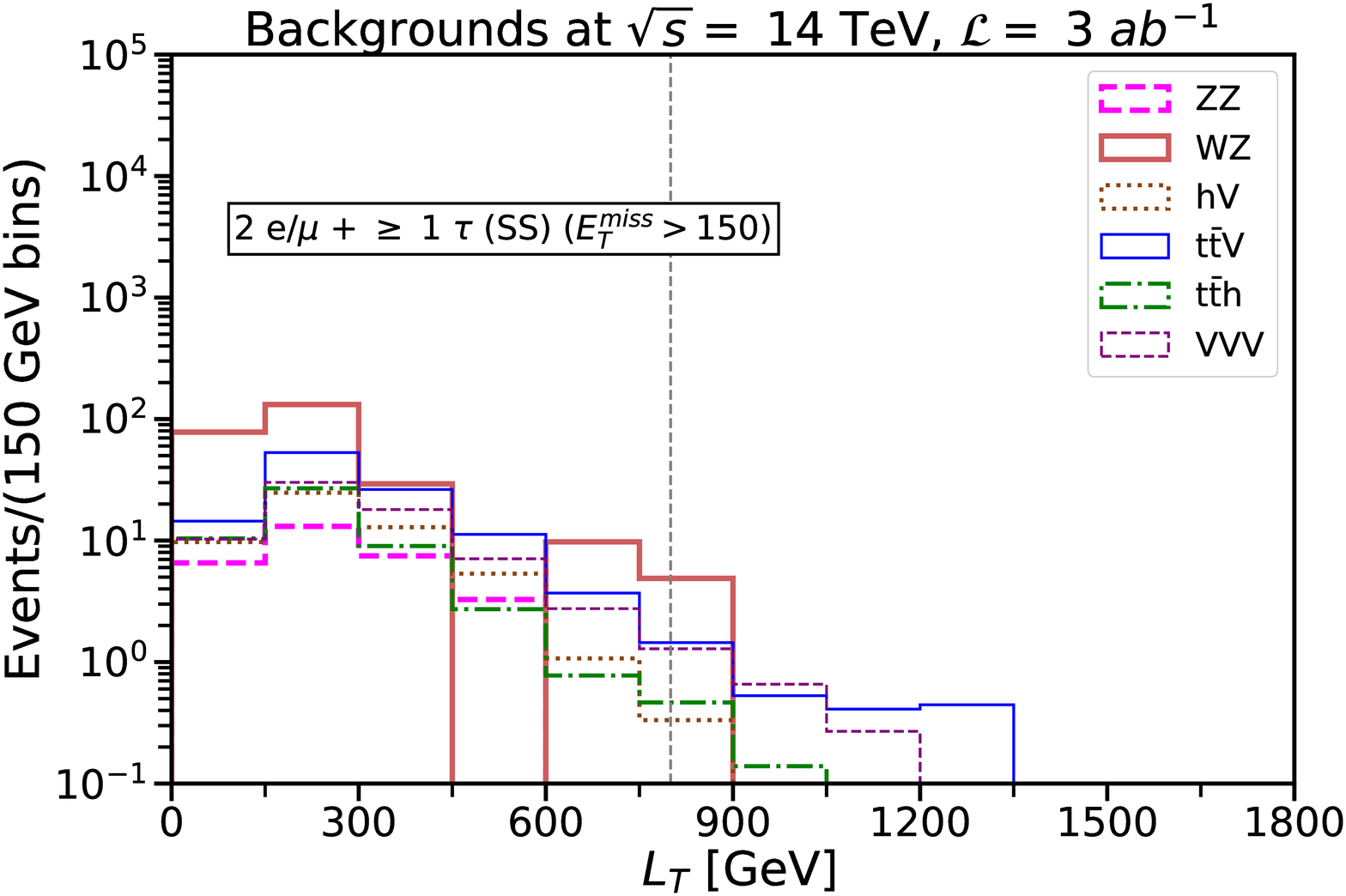}
  \end{minipage}
    \begin{minipage}[]{0.495\linewidth}
    \includegraphics[width=8.0cm,angle=0]{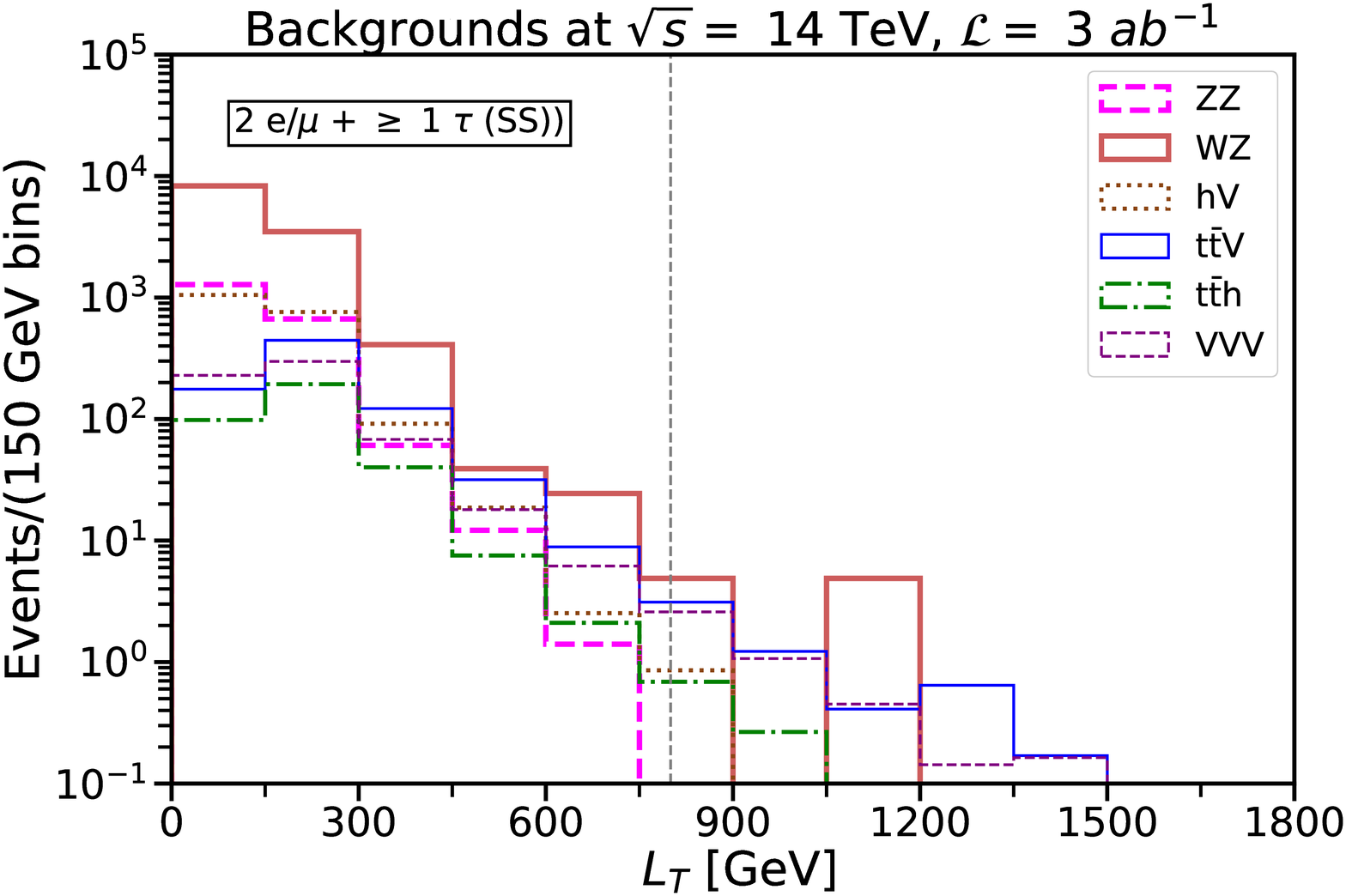}
  \end{minipage}
  \begin{minipage}[]{0.495\linewidth}
    \includegraphics[width=8.0cm,angle=0]{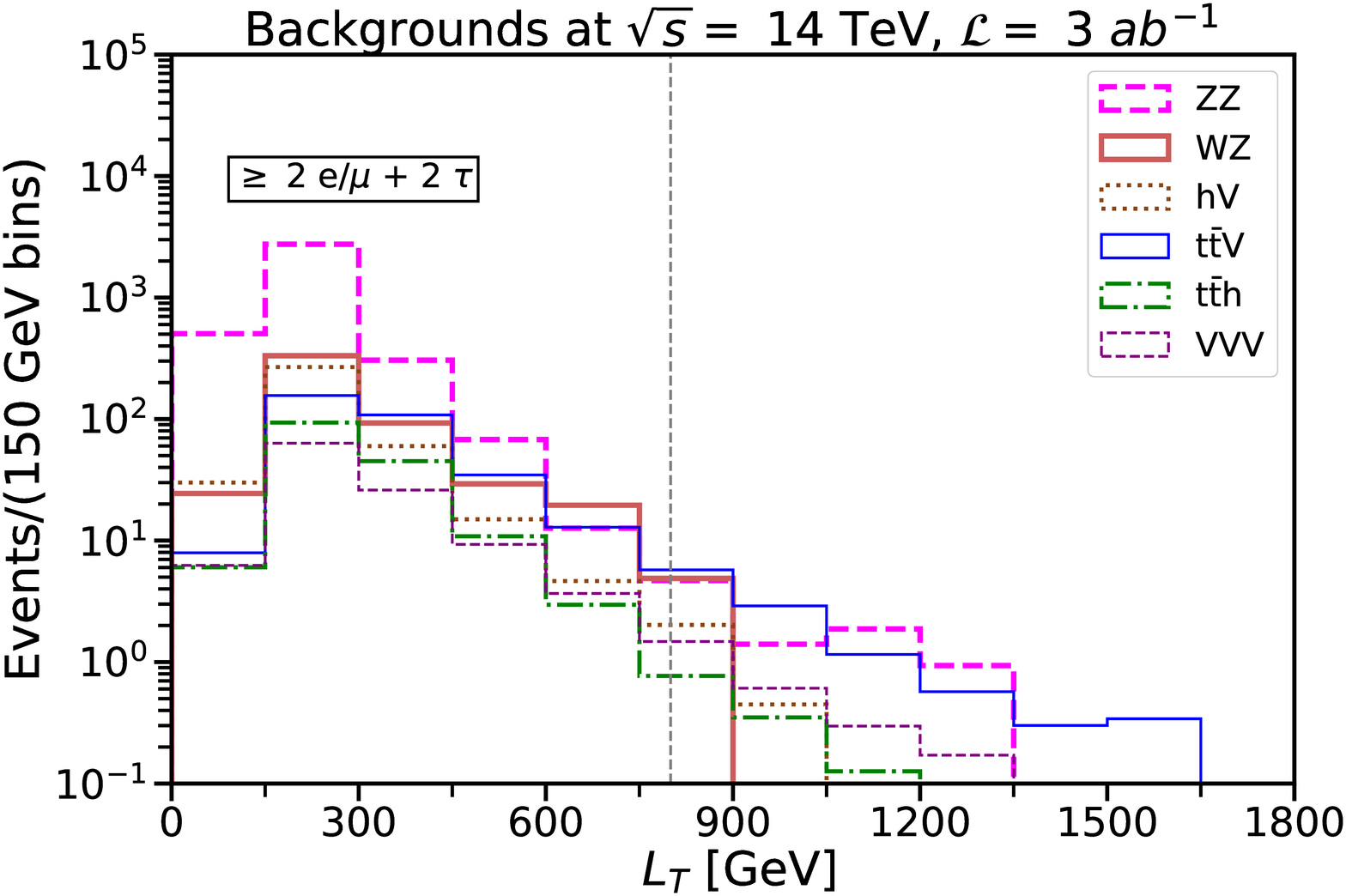}
  \end{minipage}
    \begin{minipage}[]{0.495\linewidth}
    \includegraphics[width=8.0cm,angle=0]{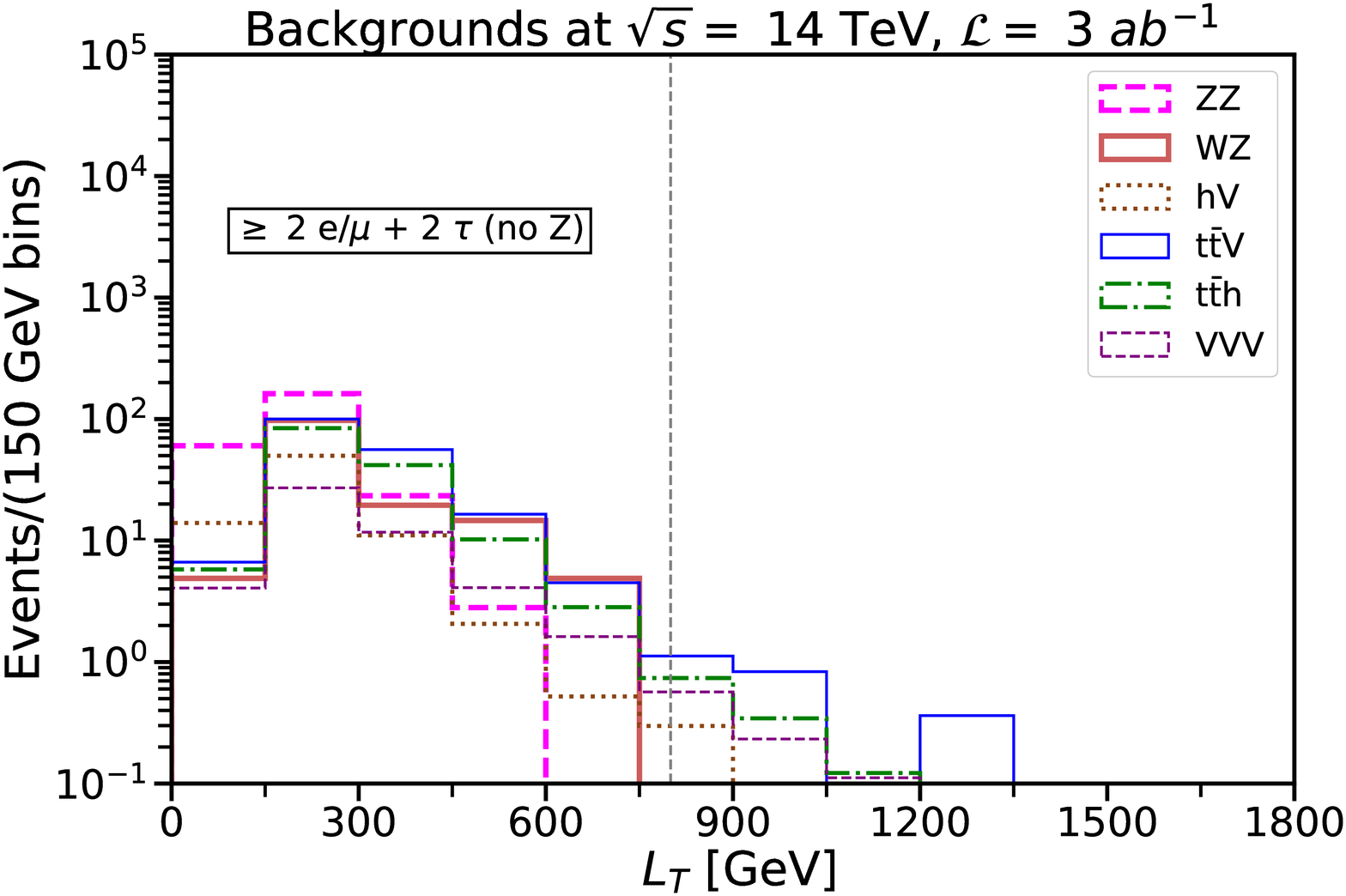}
  \end{minipage}
\begin{center}\begin{minipage}[]{0.95\linewidth}
\caption{\label{fig:LT_bg_14TeV} 
$L_T$ event distributions for all processes contributing to total SM background, 
for $pp$ collisions at $\sqrt{s} = 14$ TeV with an 
integrated luminosity $\mathcal{L} =$ 3 ab$^{-1}$. The four panels show results for 
the four best signal regions, as labeled. The vertical dashed line in all four panels shows our 
choice of $L_T$ cut.}
\end{minipage}\end{center}
\end{figure}

Figure \ref{fig:LT_bg_14TeV} shows the $L_T$ distributions for all background components, 
for the four best signal regions as labeled. The $L_T$ cut is shown in the figure 
as a vertical dashed line. After imposing the $L_T$ cut, the dominant SM backgrounds are 
$WZ$, $t\bar{t}V$ and $VVV$ (where $V=W, Z$) in the two signal regions with 
2 SS $e/\mu\> +\!\geq 1 \tau$, while the dominant SM backgrounds are $t\bar{t}V$ and $ZZ$ 
in the signal region with $\geq 2e/\mu + 2 \tau$, and $t\bar{t}V$ and $t\bar{t}h$ 
in the signal region with $\geq 2e/\mu + 2 \tau$ (no-$Z$).

\begin{figure}[!tb]
  \begin{minipage}[]{0.495\linewidth}
    \includegraphics[width=8.0cm,angle=0]{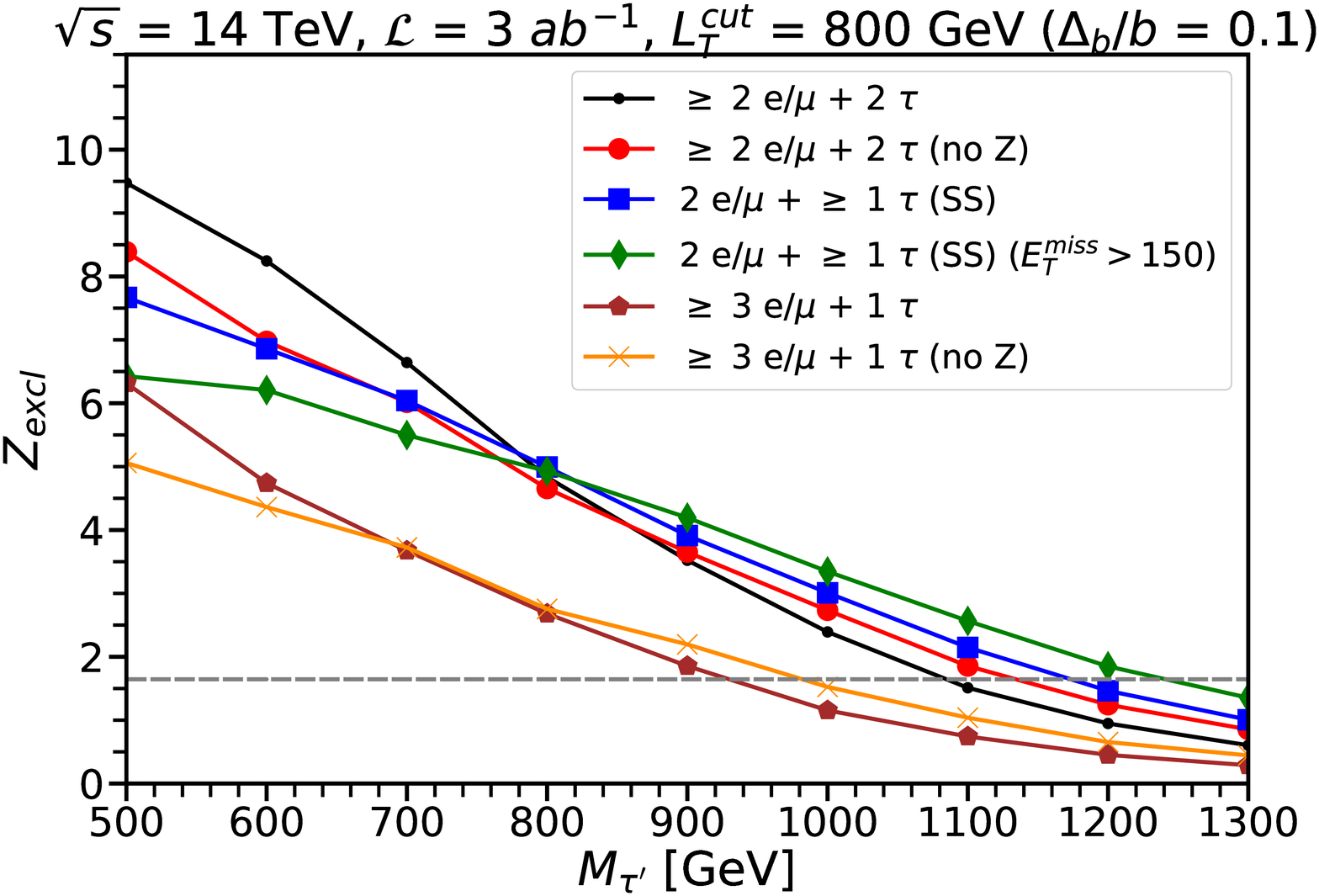}
  \end{minipage}
    \begin{minipage}[]{0.495\linewidth}
    \includegraphics[width=8.0cm,angle=0]{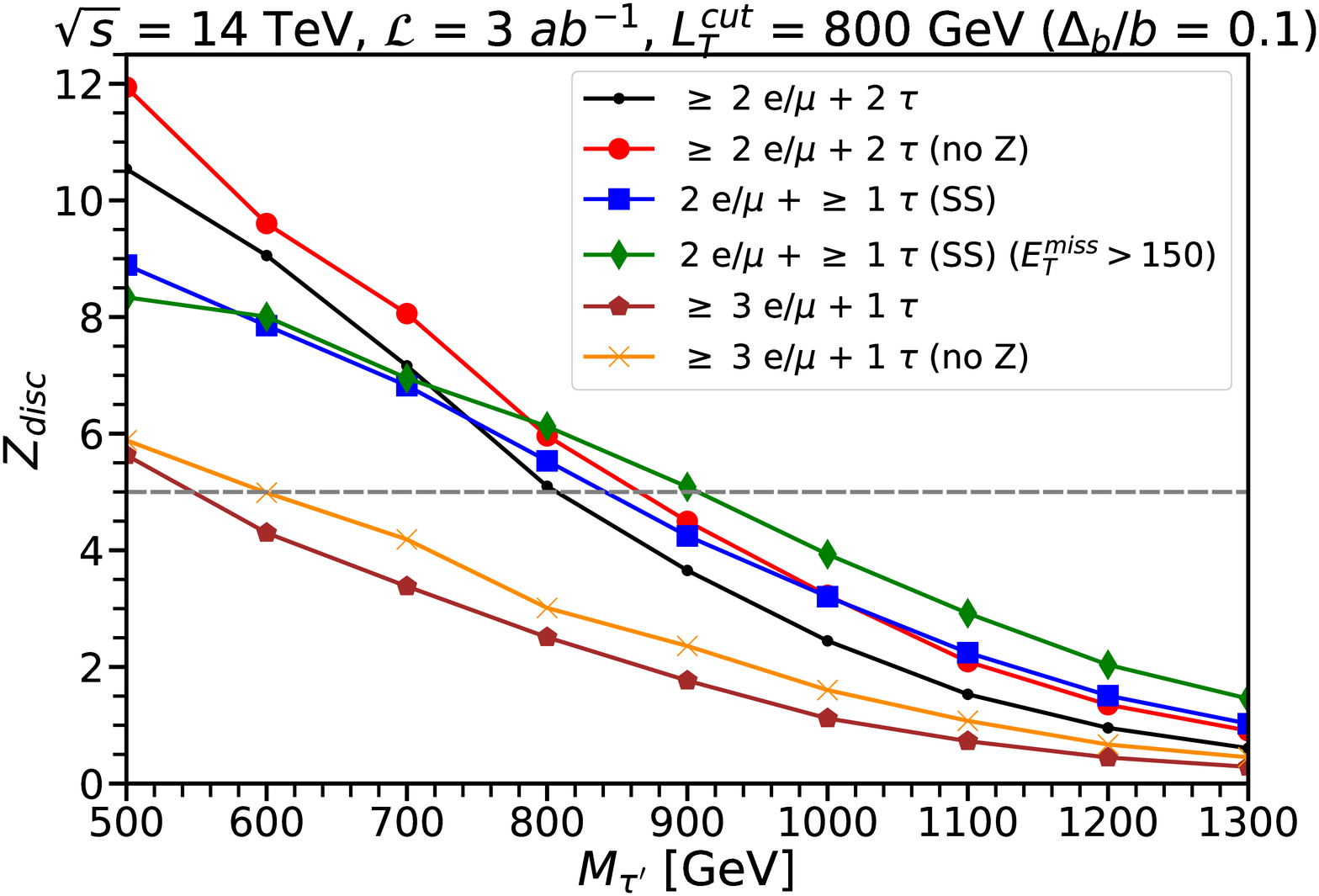}
  \end{minipage}
  \begin{minipage}[]{0.495\linewidth}
    \includegraphics[width=8.0cm,angle=0]{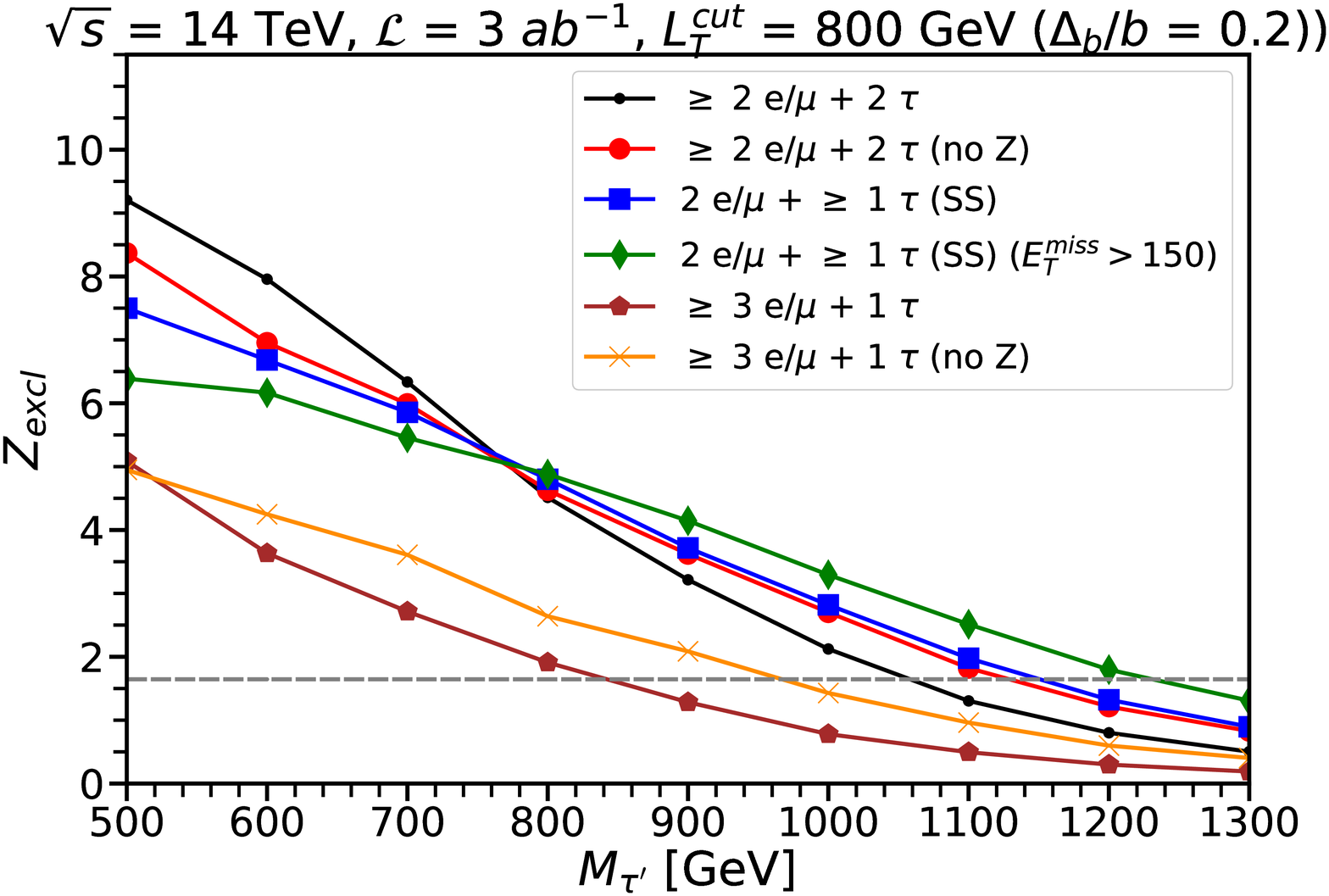}
  \end{minipage}
    \begin{minipage}[]{0.495\linewidth}
    \includegraphics[width=8.0cm,angle=0]{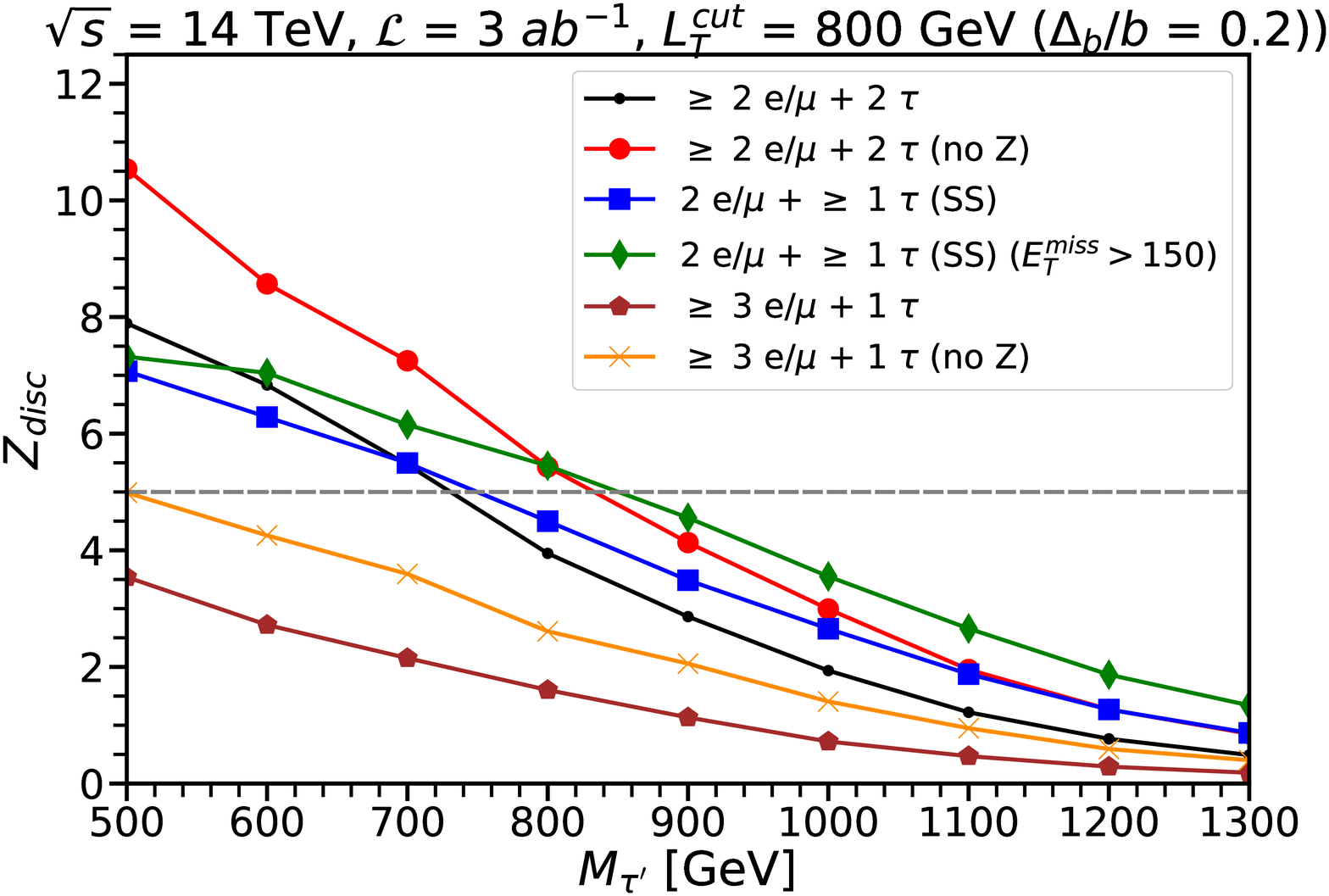}
  \end{minipage}
  \begin{minipage}[]{0.495\linewidth}
    \includegraphics[width=8.0cm,angle=0]{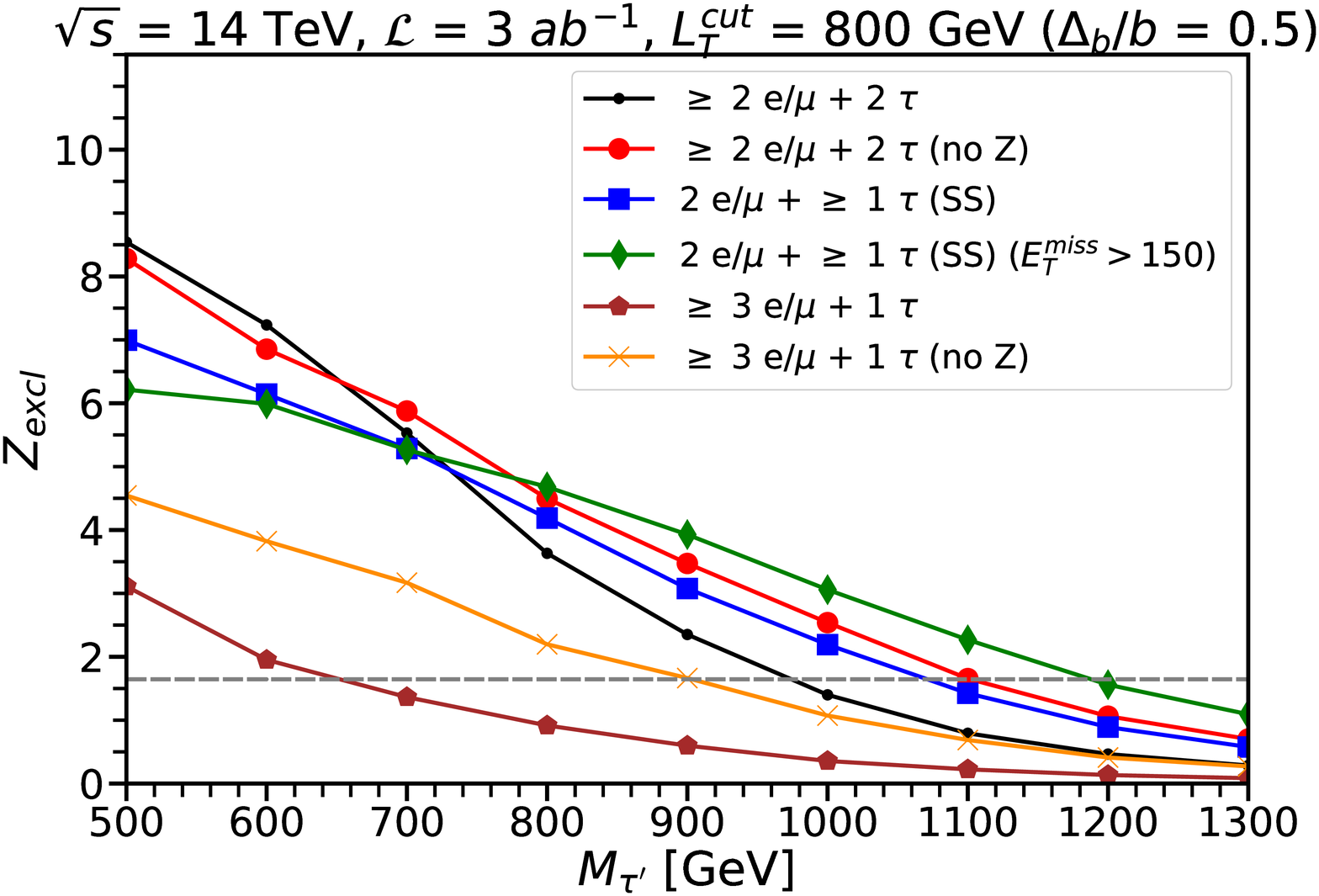}
  \end{minipage}
    \begin{minipage}[]{0.495\linewidth}
    \includegraphics[width=8.0cm,angle=0]{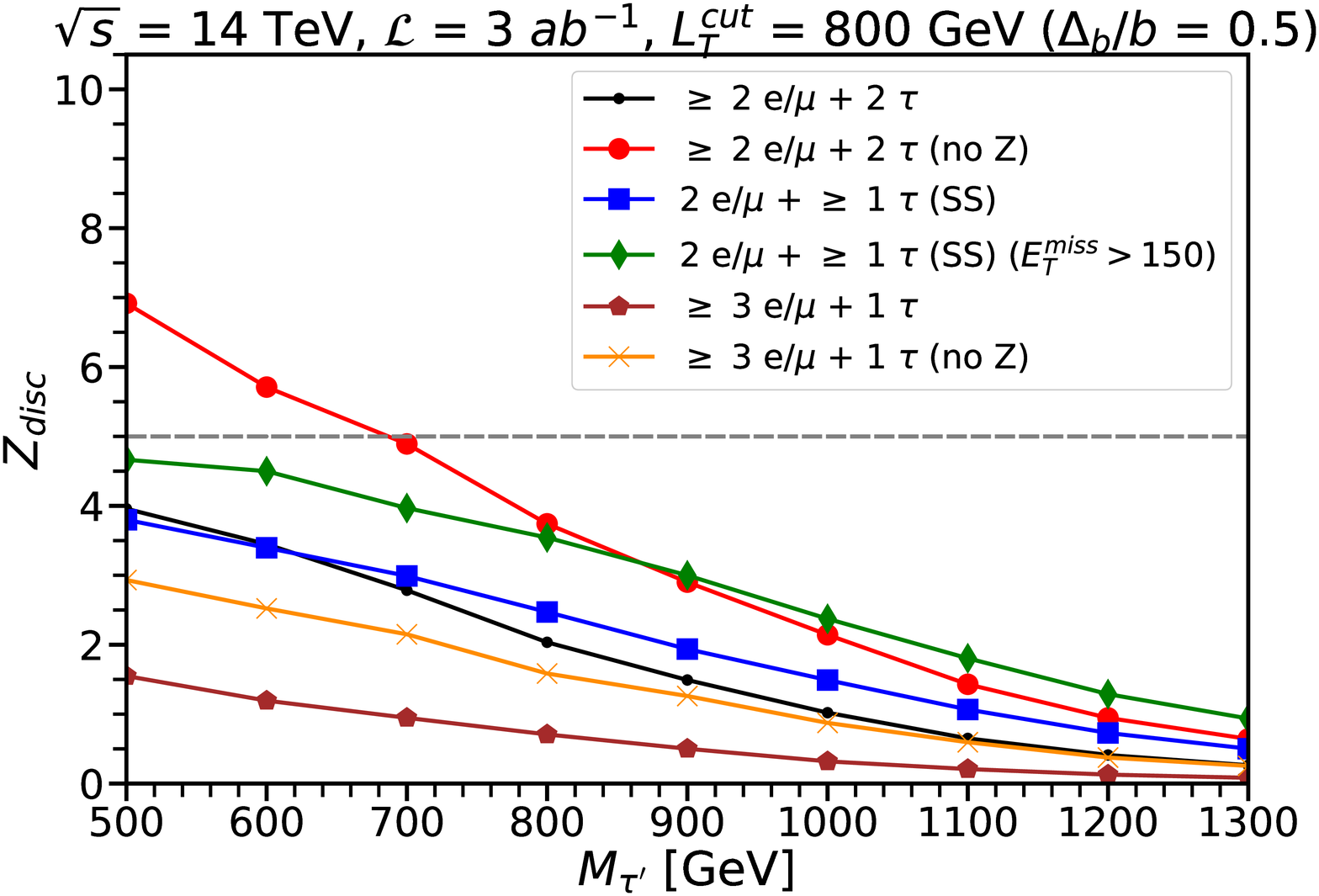}
  \end{minipage}
\begin{center}\begin{minipage}[]{0.95\linewidth}
\caption{\label{fig:Z_14TeV} 
The median expected significances for exclusion $Z_{\rm excl}$ (left panels) 
and discovery $Z_{\rm disc}$ (right panels) as a function of $M_{\tau^{\prime}}$ 
in the Doublet VLL model, for $pp$ collisions at 
$\sqrt{s} = 14$ TeV with integrated luminosity $\mathcal{L} = 3$ ab$^{-1}$, 
for six different signal regions as described in the text, each including a cut $L_T>$ 800 GeV.
The fractional uncertainty in the background is assumed to be $\Delta_b/b = 0.1$ (top row),
$0.2$ (middle row), and $0.5$ (bottom row).}
\end{minipage}\end{center}
\end{figure}

Figure \ref{fig:Z_14TeV} shows the resulting median expected significances 
for exclusion (left panels) and discovery (right panels), for 
$\Delta_b/b = 0.1$ (top row), $0.2$ (middle row), and $0.5$ (bottom row),
with the cut requirement $L_T > 800$ GeV imposed.
In all cases, the best signal regions for both exclusion and discovery
scenarios are the ones with 
2 SS $e/\mu\> +\!\geq 1 \tau$, with the additional requirement 
$E_T^{\rm miss} > 150$ GeV providing slightly more reach. The two signal
regions with $\geq 2e/\mu + 2 \tau$ give slightly less reach at higher masses,
but could actually provide comparable exclusion significances for lower masses.

From Figure \ref{fig:Z_14TeV} we conclude that a 14 TeV high-luminosity 
LHC with 3 ab$^{-1}$ should be able to exclude Doublet VLLs with 
$M_{\tau'}$ up to about 1250 GeV if they are indeed absent, or discover 
them if the mass is less than about 900 GeV, assuming that the future 
background determination from data is subject to uncertainties of order 10\%
or less. The figure also shows that prospects for exclusion are much 
less sensitive to uncertainty in the background than the prospects for discovery.
For $\Delta_b/b = 0.5$, one can still expect to exclude Doublet VLLs up to about
$M_{\tau'} = 1190$ GeV, or discover them if the mass is less than about 690 GeV, 
but the latter has already been excluded at 95\% confidence level by CMS \cite{CMS:2018cgi,Sirunyan:2019ofn}.

\begin{figure}[!tb]
  \begin{minipage}[]{0.505\linewidth}
  \begin{flushleft}
    \includegraphics[width=8.0cm,angle=0]{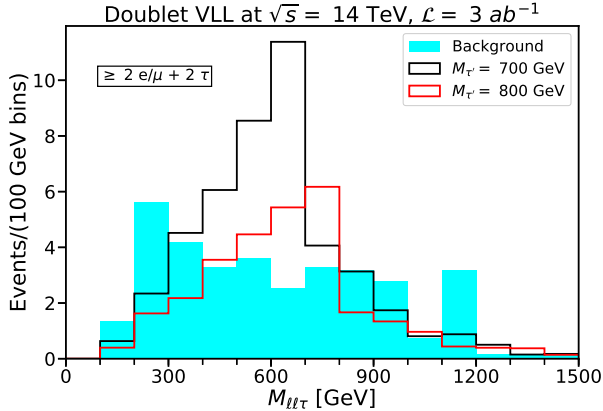}
  \end{flushleft}
  \end{minipage}
\begin{minipage}[]{0.445\linewidth}
\caption{\label{fig:inv_mass_ztau_14tev} 
The event distributions for 3-body invariant mass of $\tau^\pm e^+ e^-$ or $\tau^\pm \mu^+\mu^-$, for total background (shaded) and Doublet VLLs (lines), such that the $e^{+}e^{-}$ or $\mu^{+}\mu^{-}$ pair have an invariant mass within 10 GeV of $M_Z$ in a signal region with $\geq 2e/\mu + 2 \tau$, with the cut $L_T > 800$ GeV imposed. Two different masses $M_{\tau'} = M_{\nu'} = 700$ and $800$ GeV are shown.}   
  \end{minipage}
\end{figure}

In the case that there are enough events for a clear discovery, one can also hope to measure the mass
of the $\tau'$. In
figure \ref{fig:inv_mass_ztau_14tev}, we show the event distributions for the 3-body invariant mass of 
$\tau^\pm e^+ e^-$ or $\tau^\pm \mu^+\mu^-$, for the signal region with $\geq 2e/\mu + 2 \tau$, 
for two different choices of input $M_{\tau'}$, and for the total of all backgrounds shown 
as the shaded histogram. (Combinatorial backgrounds from wrong associations of the 
lepton pair and tau in the signal
sample are of course also present and included.) Here, we require the 2-body invariant mass of the 
$e^{+}e^{-}$ or $\mu^{+}\mu^{-}$ pair to be within 10 GeV of $M_Z$. Additionally, 
we also impose the cut $L_T > 800$ GeV. From Figure \ref{fig:inv_mass_ztau_14tev}, 
we observe that the distributions for Doublet VLLs are peaked just below their respective masses, 
which gives a possibility to measure the masses of Doublet VLLs, if they are indeed 
discovered. At $\sqrt{s} = 14$ TeV, the range of masses that one can hope to measure 
in this way is limited by the present exclusion up to $M_{\tau'} = 790$ GeV
by CMS \cite{Sirunyan:2019ofn} and by the
fact that the cross-section decreases rapidly for higher masses.

\subsection{Singlet VLL models}

Singlet VLLs are much more challenging than Doublet VLL, 
due to their much smaller production 
cross sections. For the minimal Singlet model, 
we find no possible exclusion or discovery at $\sqrt{s} = 14$ TeV 
with integrated luminosity of 3 $ab^{-1}$. And, for the non-minimal 
Singlet models, we find some exclusion possibility, 
but with no prospects for discovery. In view of the difficulties involved, below we consider only the
case that the uncertainty in the background is $\Delta_b/b = 0.1$.

\begin{figure}[!tb]
  \begin{minipage}[]{0.505\linewidth}
  \begin{flushleft}
    \includegraphics[width=8.0cm,angle=0]{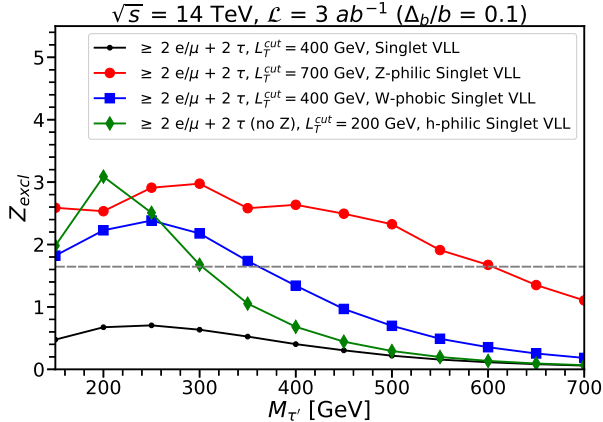}
  \end{flushleft}
  \end{minipage}
\begin{minipage}[]{0.445\linewidth}
\caption{\label{fig:Z_14TeV_Singlet} 
The median expected significances for exclusion $Z_{\rm excl}$ as a function of $M_{\tau^{\prime}}$ 
in the Singlet VLL models, for $pp$ collisions at 
$\sqrt{s} = 14$ TeV with integrated luminosity $\mathcal{L} = 3$ ab$^{-1}$, 
for the best signal region for each of the Singlet VLL models, including a cut on $L_T$ as shown in the plot.
The fractional uncertainty in the background is assumed to be $\Delta_b/b = 0.1$.
}   
  \end{minipage}
\end{figure}

To maximize the exclusion reach in the non-minimal cases, we chose a cut $L_T > 700$ GeV for the $Z$-philic Singlet VLL, $L_T > 400$ GeV for the $W$-phobic Singlet VLL and $L_T > 200$ GeV for the Higgs-philic Singlet VLL. Figure \ref{fig:Z_14TeV_Singlet} shows the resulting median expected significances 
for exclusion, for 
$\Delta_b/b = 0.1$, for the best signal region for each of the Singlet VLL models,
with the cuts on $L_T$ imposed. In all the models, the best signal region for exclusion is the one which requires $\geq 2e/\mu + 2 \tau$. However, for the Higgs-philic Singlet VLL model, the additional requirement of no-$Z$ proved to be beneficial.
 
From Figure \ref{fig:Z_14TeV_Singlet}, we conclude that assuming 
$\Delta_b/b = 0.1$, a 14 TeV $pp$ collider with 3 ab$^{-1}$ 
should be able to exclude Singlet VLLs with masses up to about 600 GeV 
if they are $Z$-philic, or exclude masses up to about 360 GeV 
if they are $W$-phobic, or exclude masses up to about 300 GeV 
if they are Higgs-philic. We find that there is no possible exclusion 
of Singlet VLLs in the minimal model. Furthermore, there are no 
discovery prospects in minimal or non-minimal Singlet VLL models.

\baselineskip=15.pt

\section{Results for the HE-LHC collider\label{sec:pp27TeV}}
\setcounter{equation}{0}
\setcounter{figure}{0}
\setcounter{table}{0}
\setcounter{footnote}{1}

In this section, we discuss the prospects for exclusion and discovery 
of VLLs at $\sqrt{s} = 27$ TeV with 15 ab$^{-1}$ of $pp$ collisions 
in the six signal regions mentioned in eqs.~(\ref{eq:signal1})-(\ref{eq:signal6}). 
All leptons including hadronic tau candidates are required to satisfy:
\beq
p_T^\ell > \mbox{25 GeV}.
\eeq 
along with the same pseudo-rapidity, isolation and other requirements of eqs.~(\ref{eta})-(\ref{nbjets}), with a trigger requirement of:
\beq
p_T^{e_1}\>\,\mbox{or}\>\, p_T^{\mu_1} > 50 \textrm{ GeV}.
\eeq

\subsection{Doublet VLL model}

\begin{figure}[!tb]
  \begin{minipage}[]{0.495\linewidth}
    \includegraphics[width=8.0cm,angle=0]{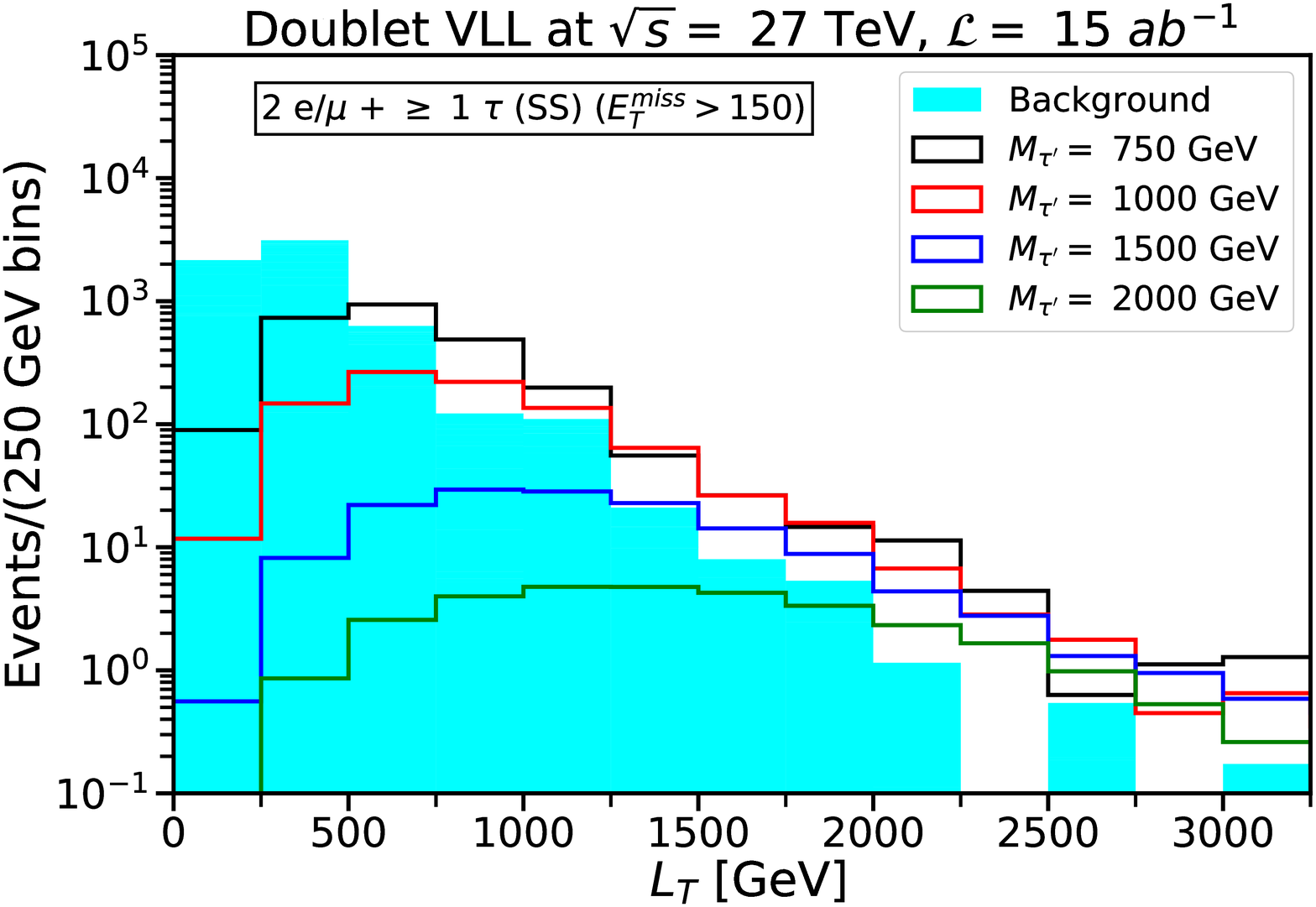}
  \end{minipage}
    \begin{minipage}[]{0.495\linewidth}
    \includegraphics[width=8.0cm,angle=0]{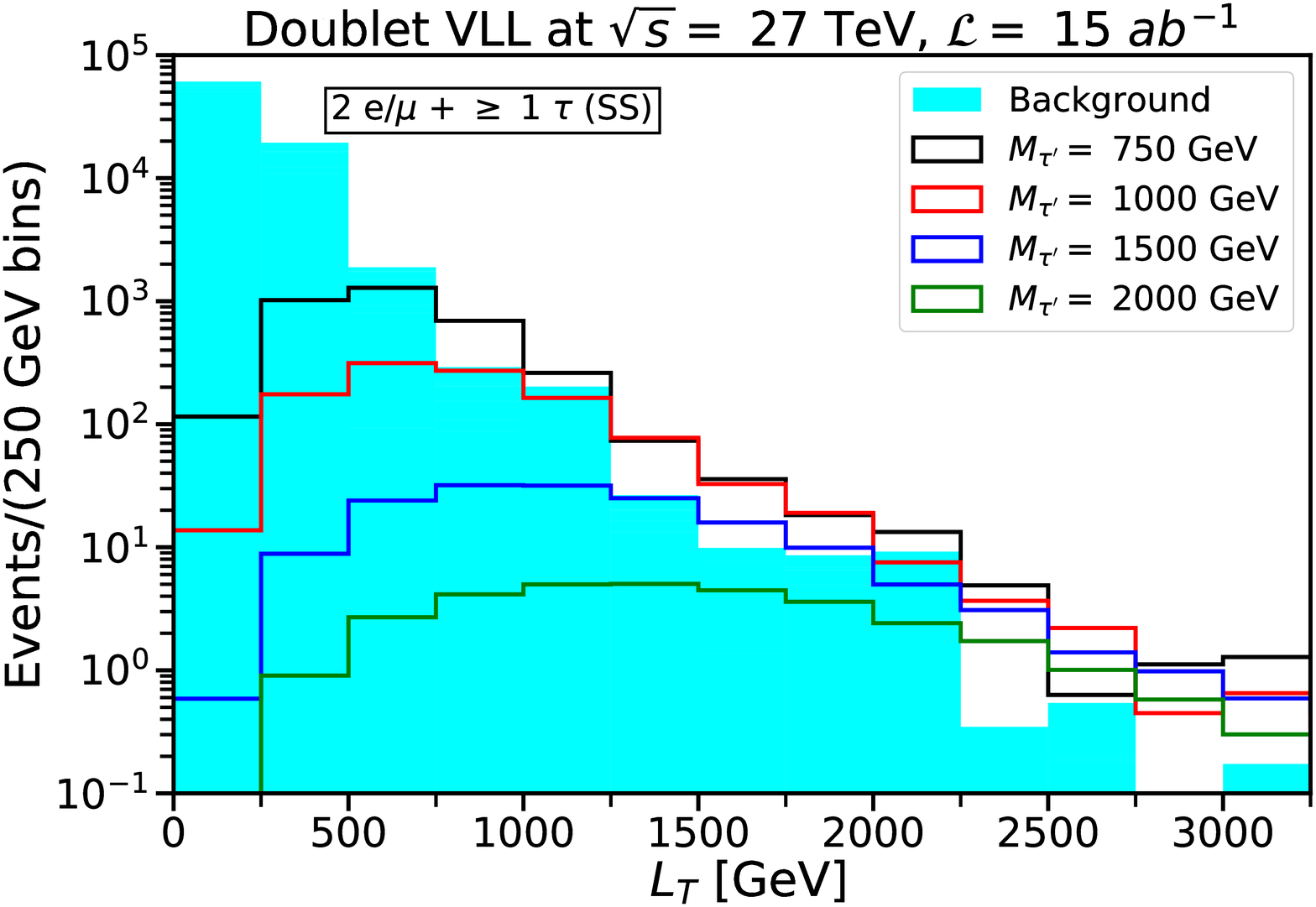}
  \end{minipage}
  \begin{minipage}[]{0.495\linewidth}
    \includegraphics[width=8.0cm,angle=0]{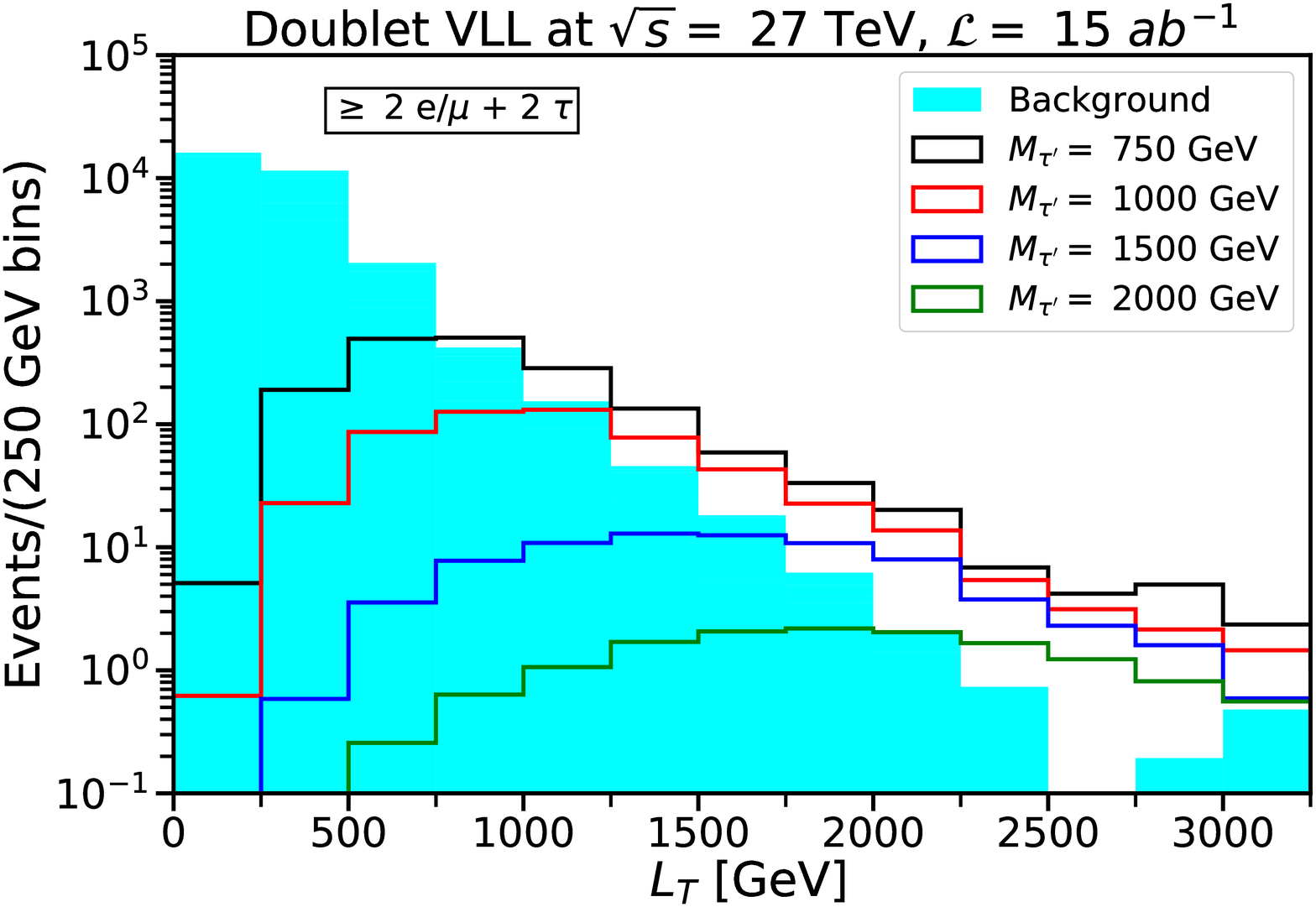}
  \end{minipage}
    \begin{minipage}[]{0.495\linewidth}
    \includegraphics[width=8.0cm,angle=0]{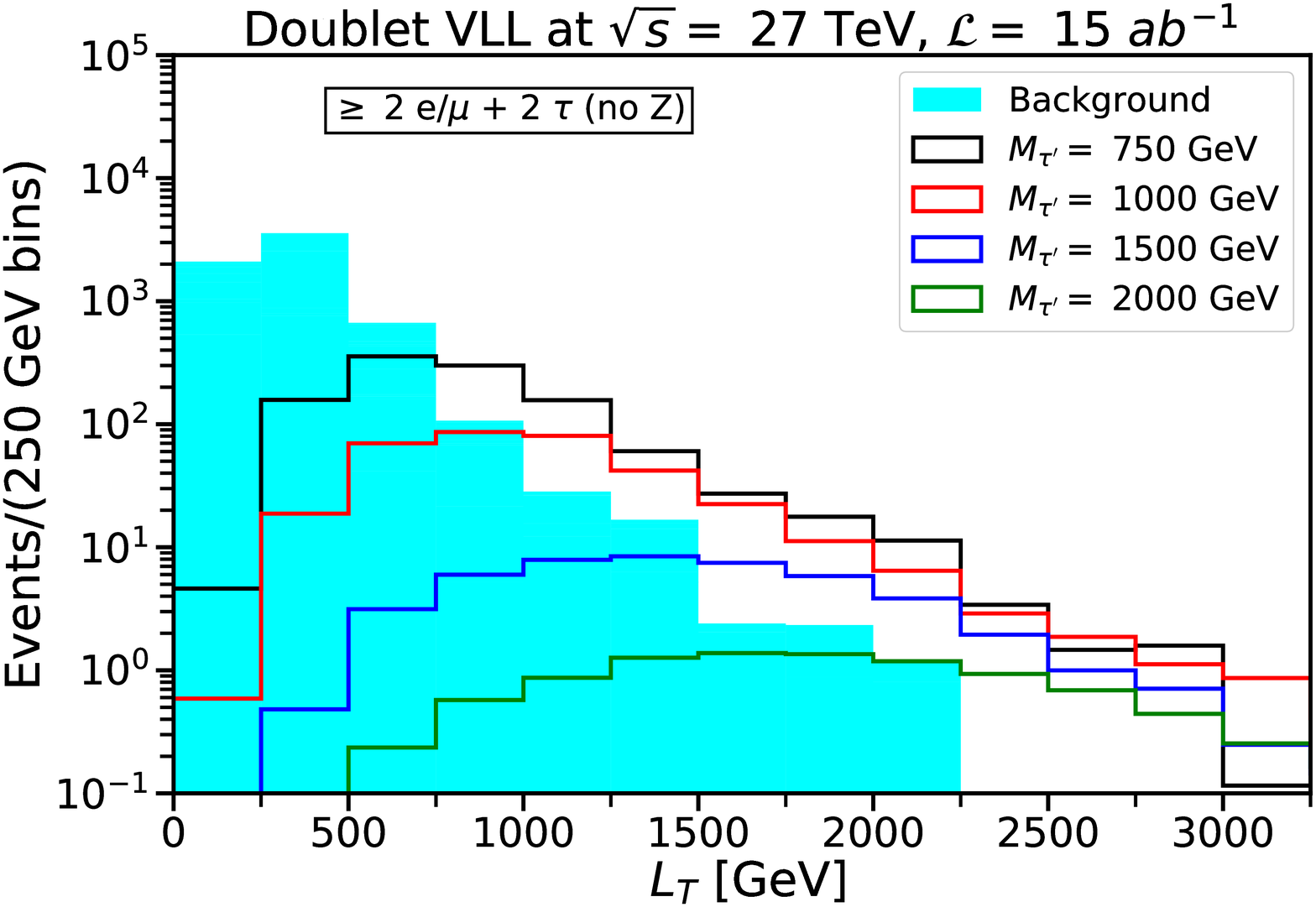}
  \end{minipage}
\begin{center}\begin{minipage}[]{0.95\linewidth}
\caption{\label{fig:LT_27TeV} 
$L_T$ event distributions for total background (shaded) and Doublet VLL
models (lines), for $pp$ collisions at $\sqrt{s} = 27$ TeV with an
integrated luminosity $\mathcal{L} =$ 15 ab$^{-1}$. Four different masses
$M_{\tau'} = M_{\nu'} = 750,$ 1000, 1500, and 2000 GeV are shown in
each panel. The four panels show results for the four best signal
regions, as labeled.}
\end{minipage}\end{center}
\end{figure}

In Figure \ref{fig:LT_27TeV}, we show the $L_T$ distributions for the best four of these signal regions, 
for four different choices of $M_{\tau'}$ as labeled, and for the total of all backgrounds shown as the shaded histogram. 
The cut chosen to enhance the reach for exclusion and discovery in this case 
was $L_T > 1500$ GeV.

\begin{figure}[!tb]
  \begin{minipage}[]{0.495\linewidth}
    \includegraphics[width=8.0cm,angle=0]{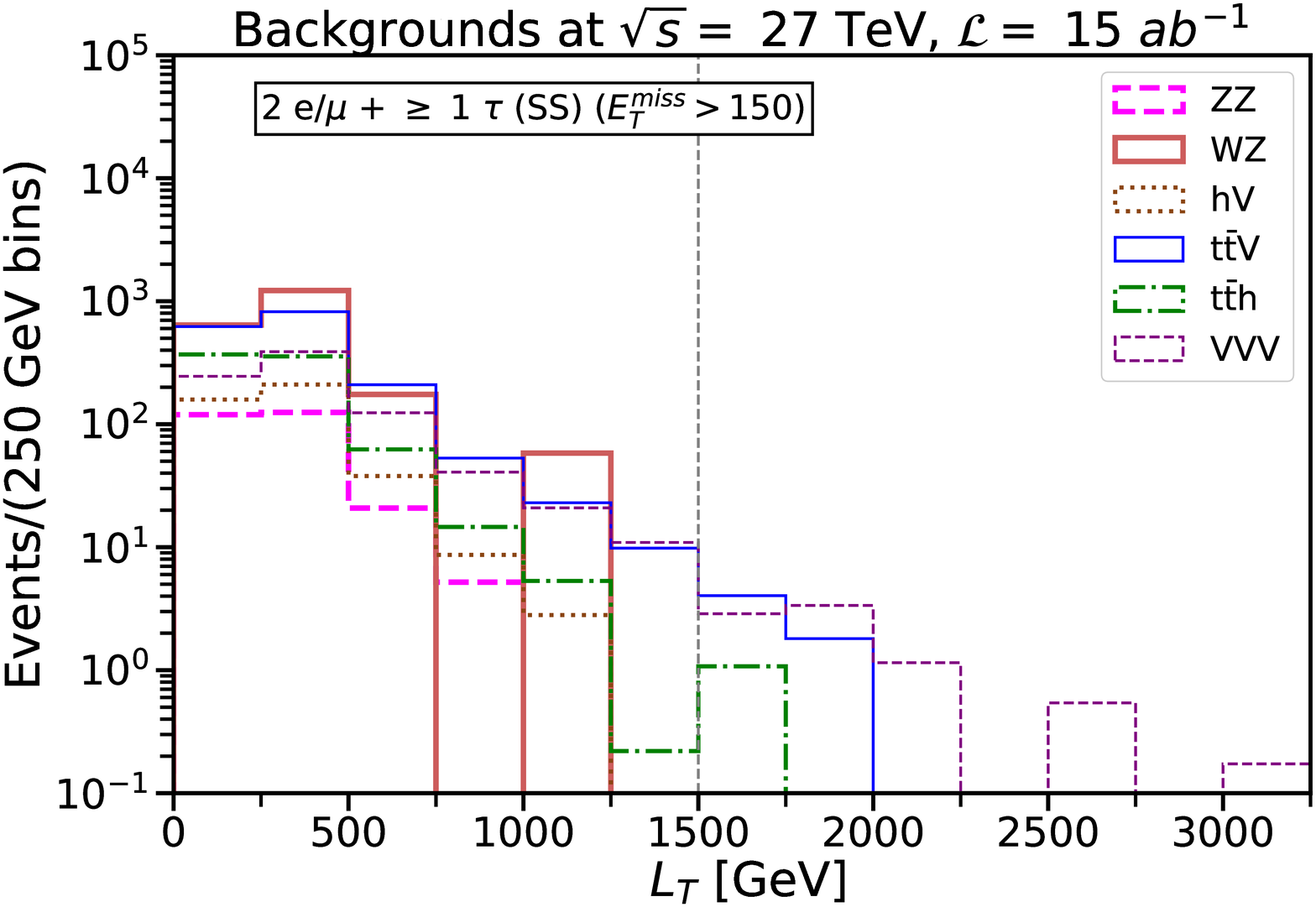}
  \end{minipage}
    \begin{minipage}[]{0.495\linewidth}
    \includegraphics[width=8.0cm,angle=0]{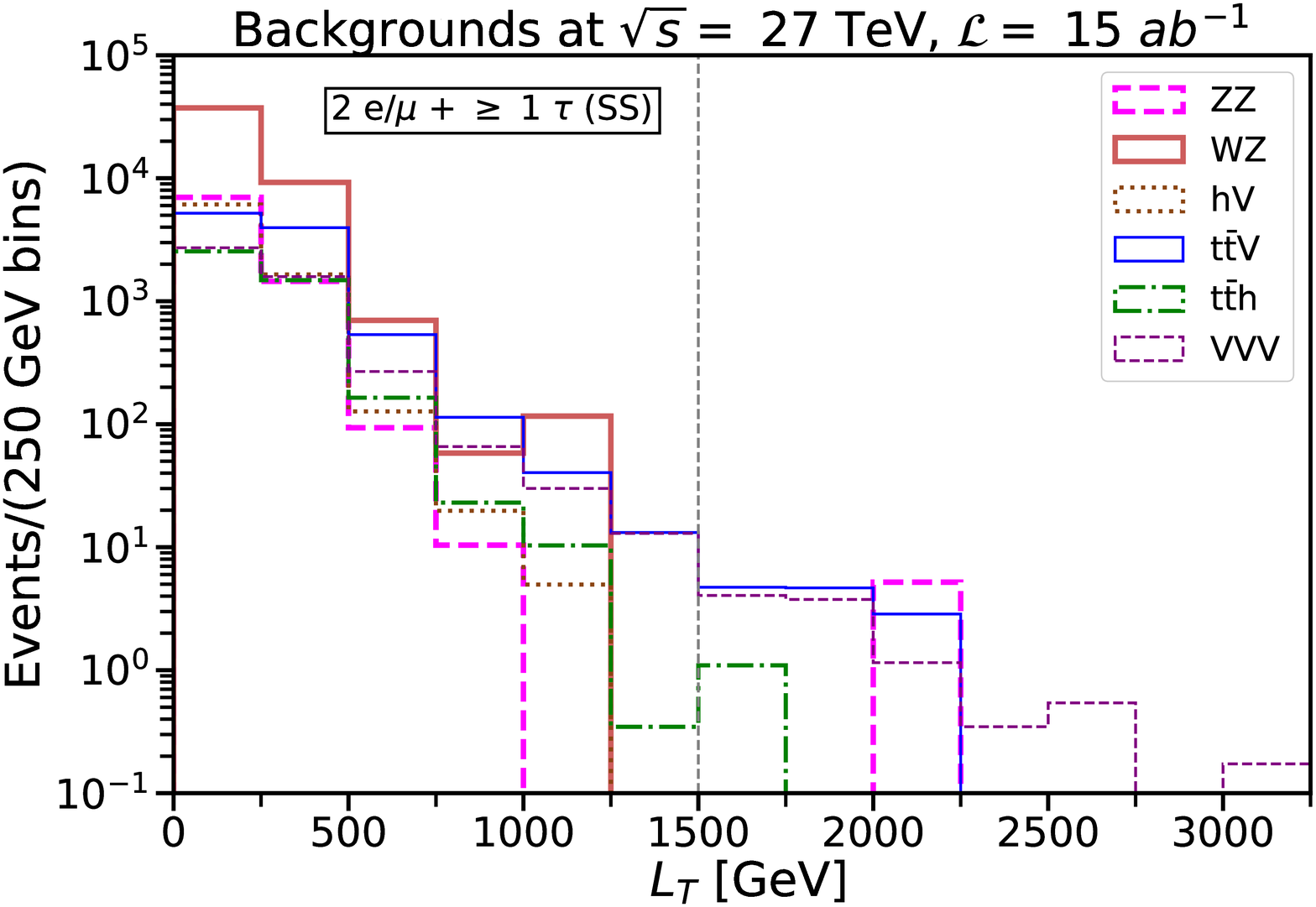}
  \end{minipage}
  \begin{minipage}[]{0.495\linewidth}
    \includegraphics[width=8.0cm,angle=0]{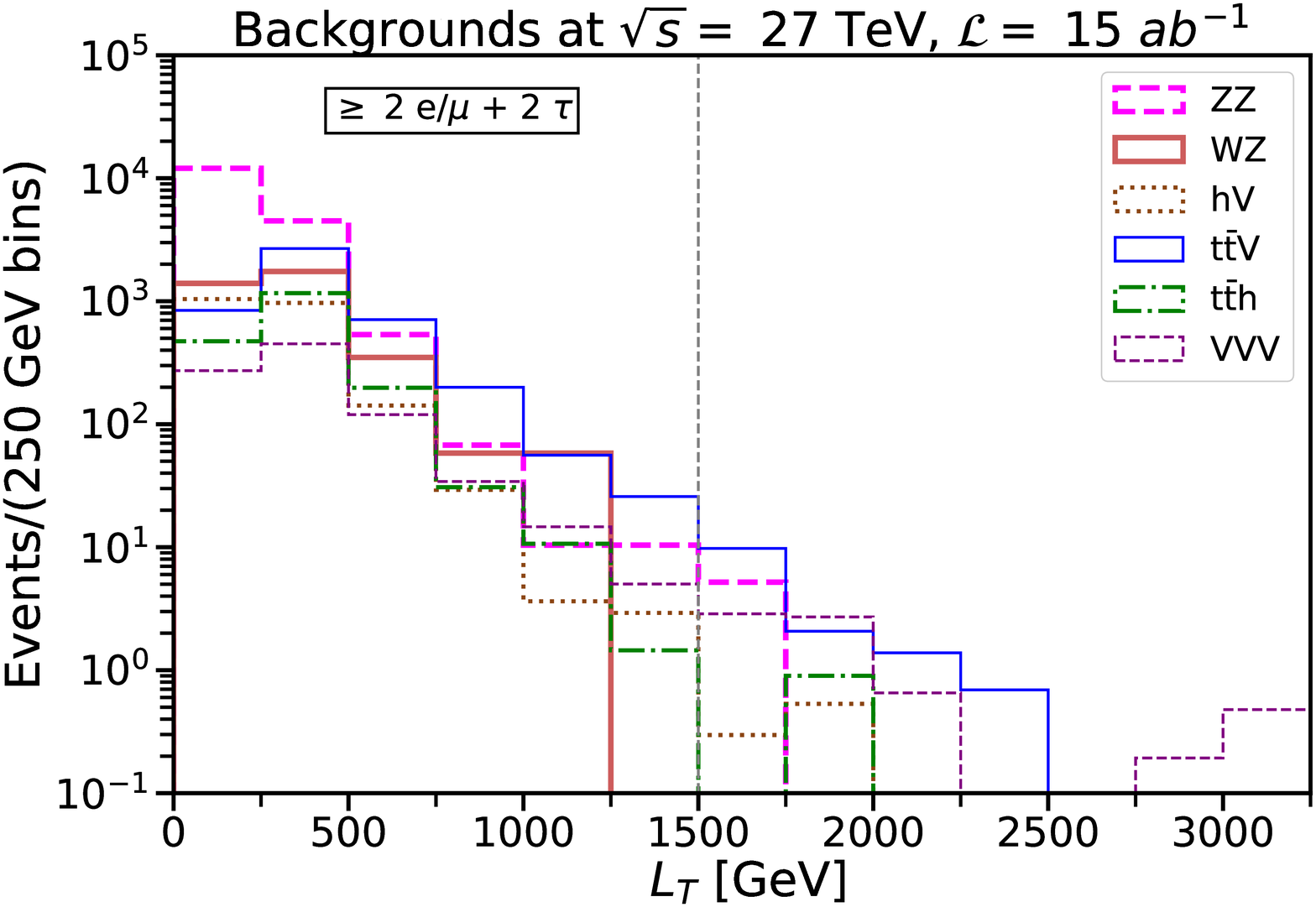}
  \end{minipage}
    \begin{minipage}[]{0.495\linewidth}
    \includegraphics[width=8.0cm,angle=0]{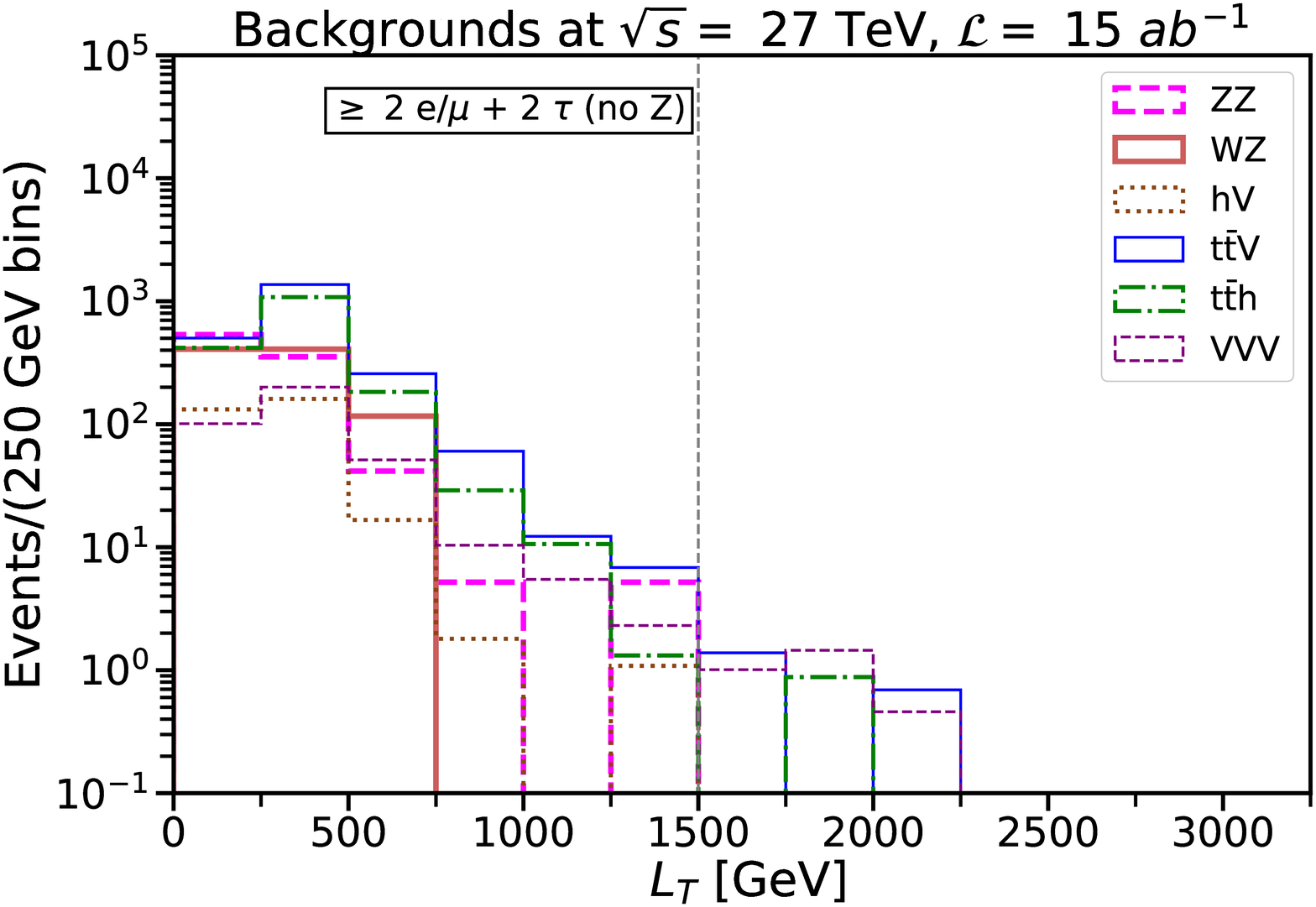}
  \end{minipage}
\begin{center}\begin{minipage}[]{0.95\linewidth}
\caption{\label{fig:LT_bg_27TeV} 
$L_T$ event distributions for all processes contributing to total SM background, 
for $pp$ collisions at $\sqrt{s} = 27$ TeV with an 
integrated luminosity $\mathcal{L} =$ 15 ab$^{-1}$. The four panels show results for the 
four best signal regions, as labeled. The vertical dashed line in all four panels 
shows our choice of $L_T$ cut of 1500 GeV.}
\end{minipage}\end{center}
\end{figure}

Figure \ref{fig:LT_bg_27TeV} shows the $L_T$ distributions for all background components, for the four best signal regions as labeled. The $L_T$ cut is shown in the figure as a vertical dashed line. After imposing the $L_T$ cut, the most important SM backgrounds 
are $t\bar{t}V$ and $VVV$ in all four of these signal regions.

\begin{figure}[!tb]
  \begin{minipage}[]{0.495\linewidth}
    \includegraphics[width=8.0cm,angle=0]{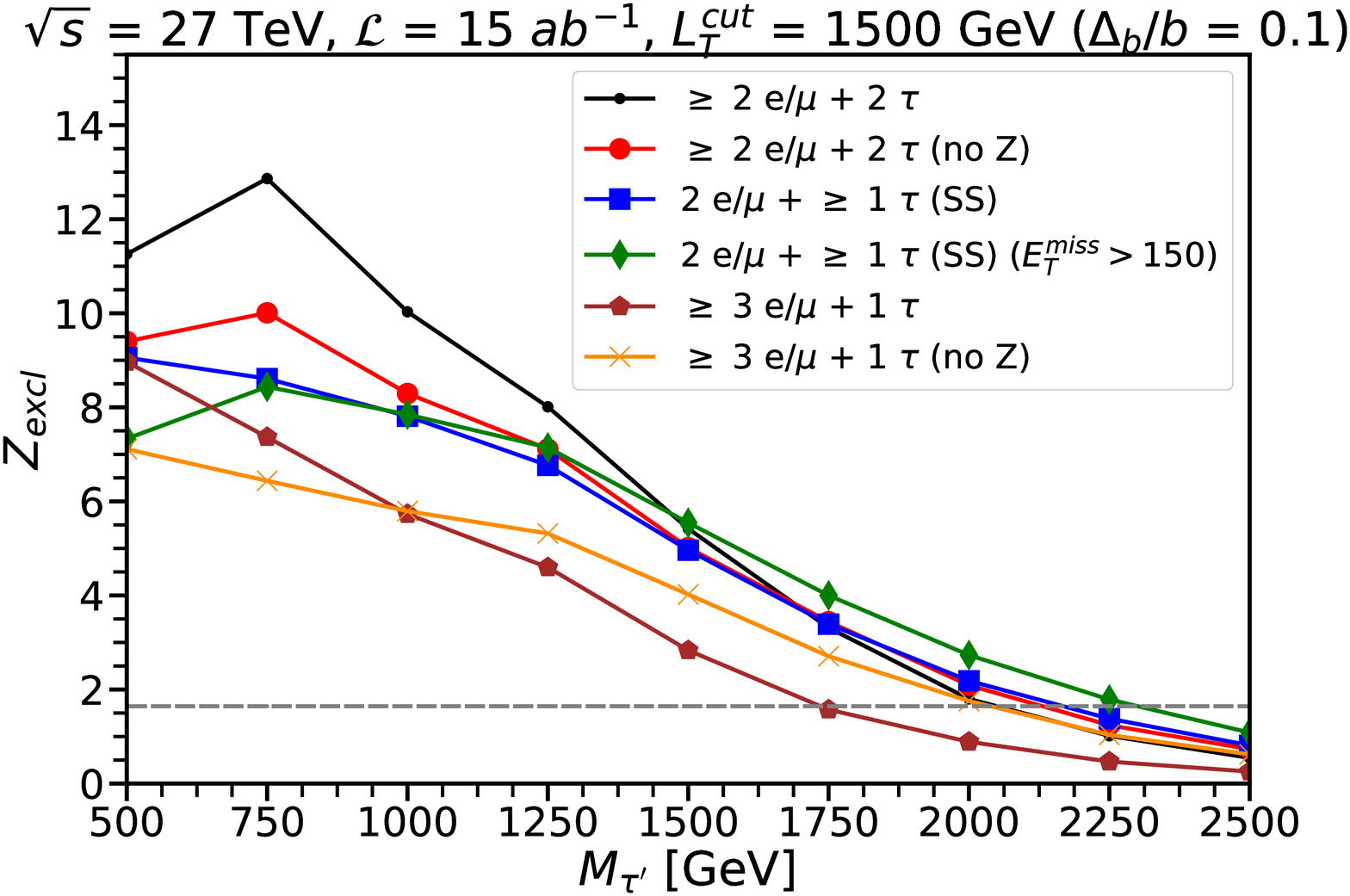}
  \end{minipage}
    \begin{minipage}[]{0.495\linewidth}
    \includegraphics[width=8.0cm,angle=0]{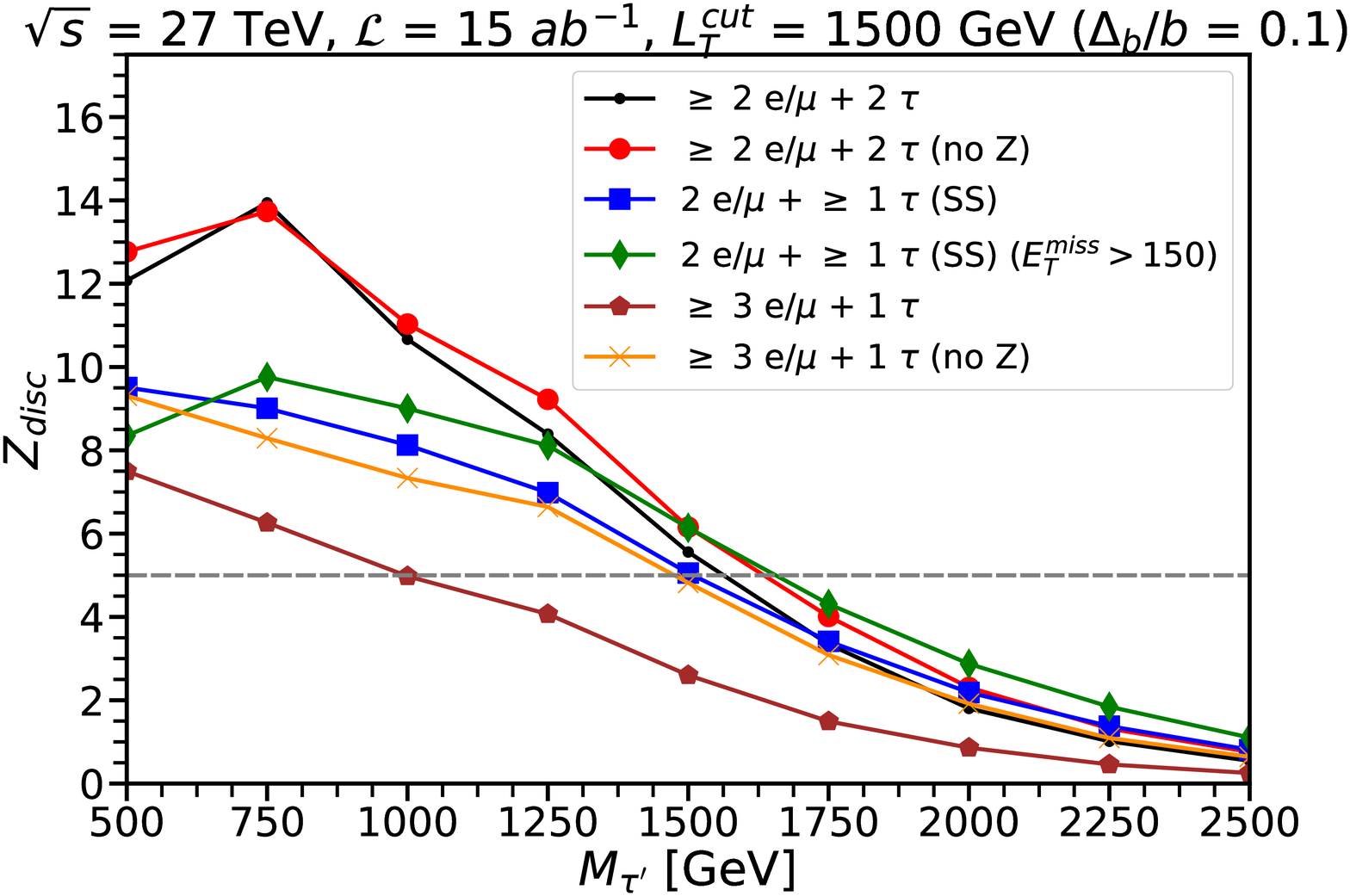}
  \end{minipage}
    \begin{minipage}[]{0.495\linewidth}
    \includegraphics[width=8.0cm,angle=0]{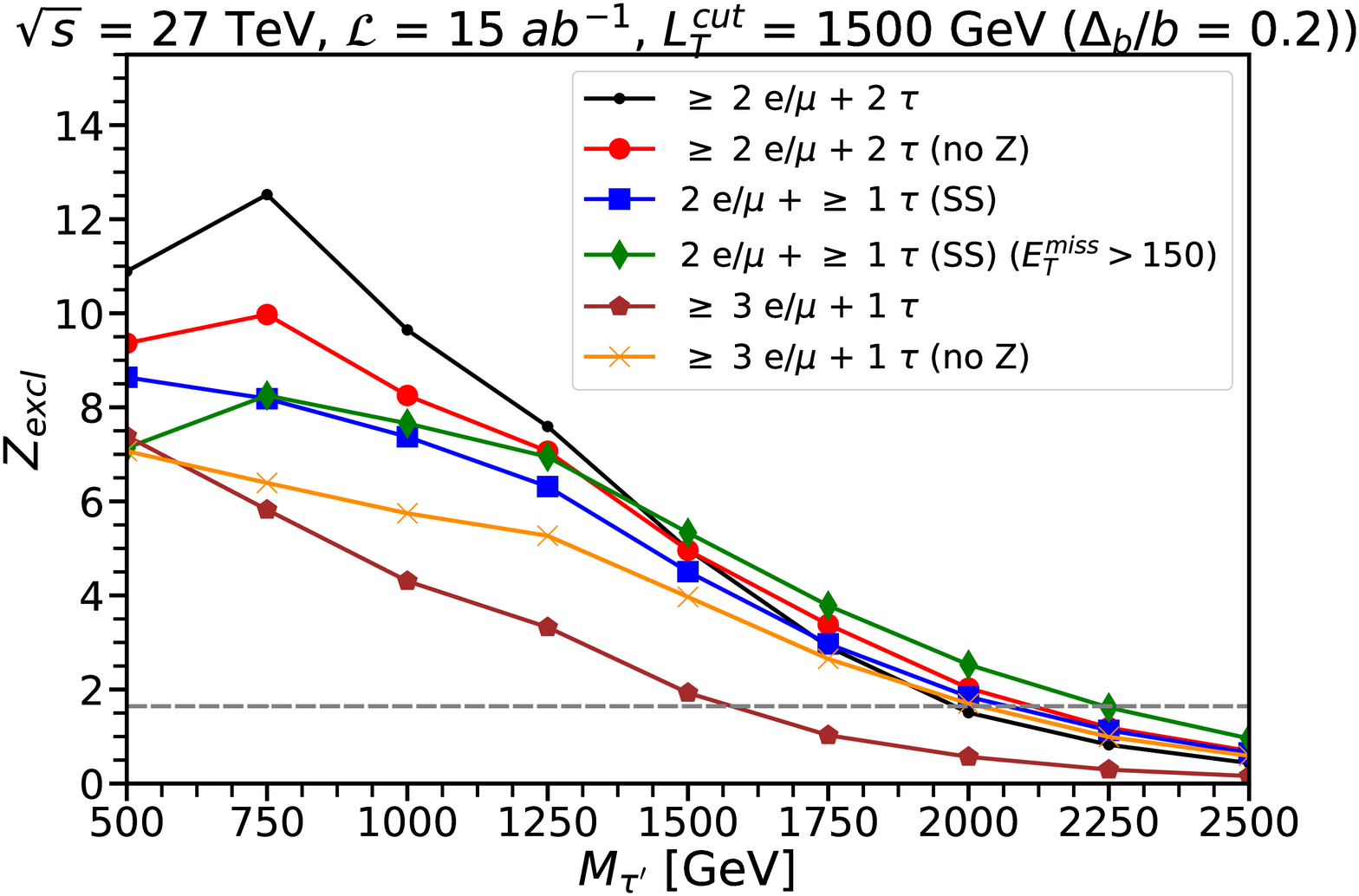}
  \end{minipage}
    \begin{minipage}[]{0.495\linewidth}
    \includegraphics[width=8.0cm,angle=0]{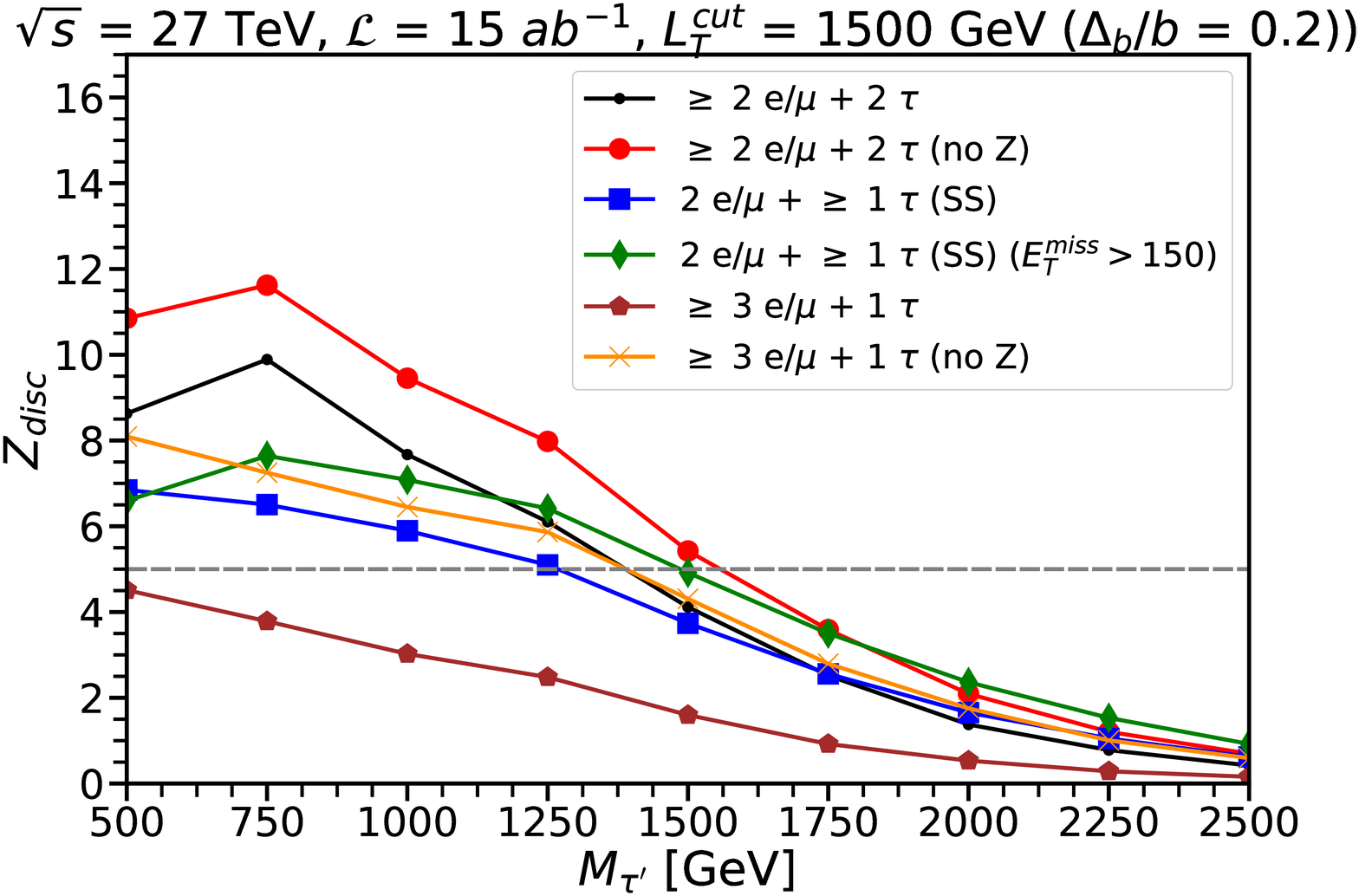}
  \end{minipage}
  \begin{minipage}[]{0.495\linewidth}
    \includegraphics[width=8.0cm,angle=0]{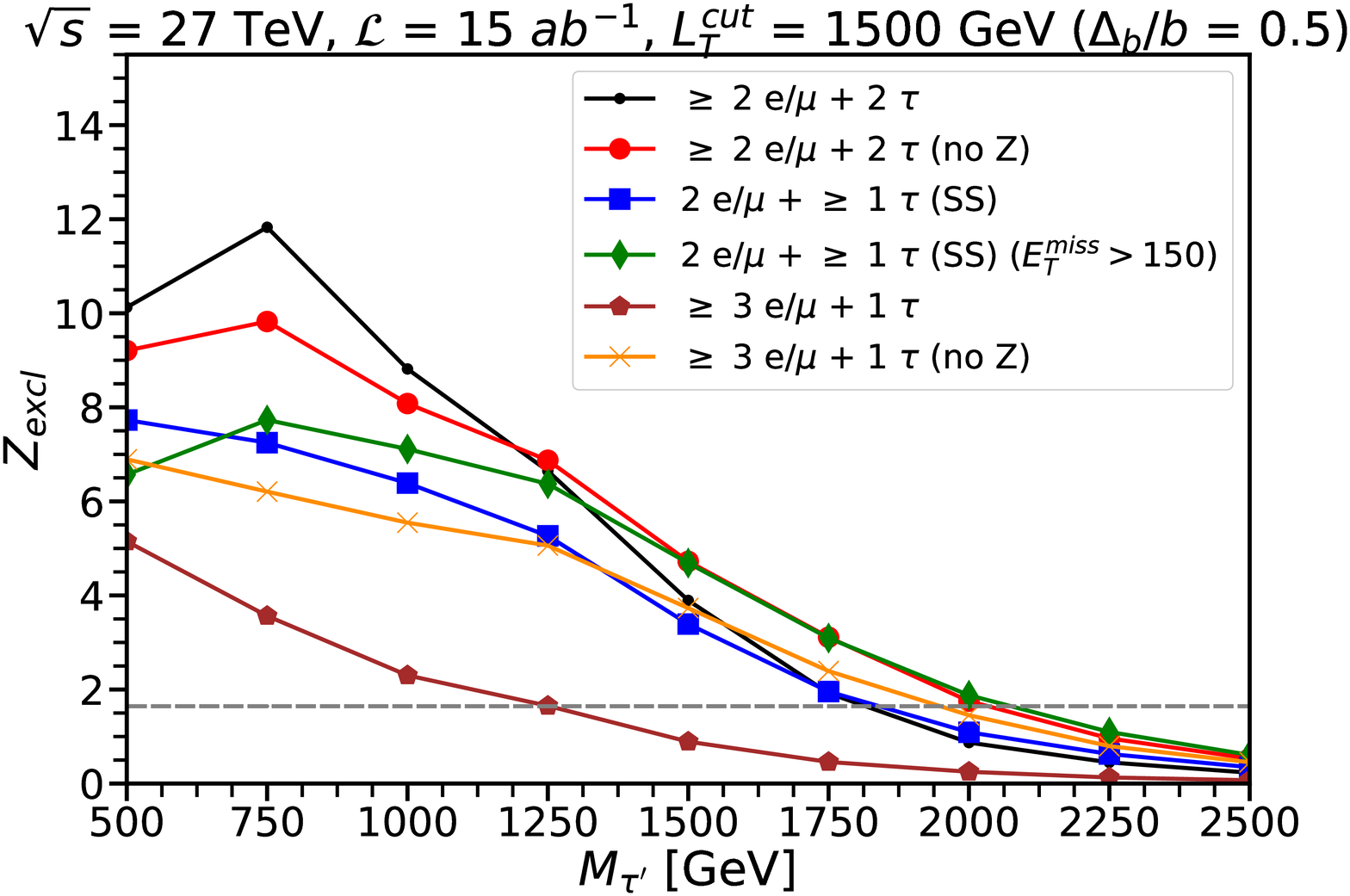}
  \end{minipage}
    \begin{minipage}[]{0.495\linewidth}
    \includegraphics[width=8.0cm,angle=0]{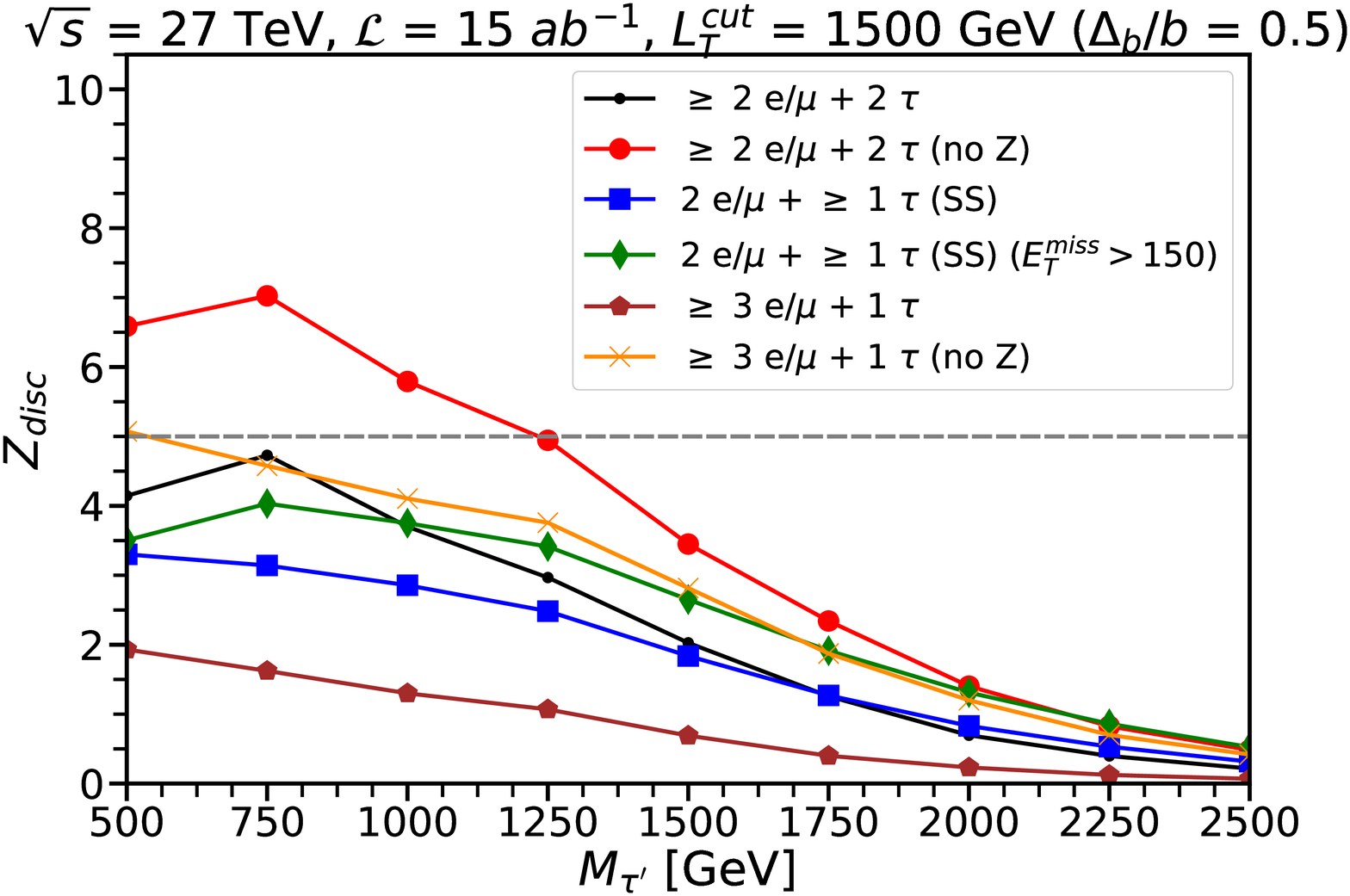}
  \end{minipage}
\begin{center}\begin{minipage}[]{0.95\linewidth}
\caption{\label{fig:Z_27TeV} 
The median expected significances for exclusion $Z_{\rm excl}$ 
(left panels) and discovery $Z_{\rm disc}$ (right panels) as a function of
$M_{\tau^{\prime}}$ in the Doublet VLL model, for $pp$ collisions at
$\sqrt{s} = 27$ TeV with integrated luminosity $\mathcal{L} = 15$ 
ab$^{-1}$, for six different signal regions as described in the text, each including a cut $L_T>$ 1500 GeV.
The fractional uncertainty in the background is assumed to be $\Delta_b/b = 0.1$ (top row),
$0.2$ (middle row), and $0.5$ (bottom row).
}
\end{minipage}\end{center}
\end{figure}

Figure \ref{fig:Z_27TeV} shows the $Z_{\rm excl}$ (left panels) and 
$Z_{\rm disc}$ (right panels) as a function of $M_{\tau'}$, for 
$\Delta_b/b = 0.1$ (top row), $0.2$ (middle row), and $0.5$ (bottom row),
with the cut $L_T > 1500$ GeV imposed. The signal region with 
2 SS $e/\mu\> +\!\geq 1 \tau$ with $E_T^{\rm miss} > 150$ GeV has 
the farthest mass reach. However, at lower masses the two signal regions 
with $\geq 2e/\mu + 2 \tau$ give slightly better exclusion and discovery significances. 

By looking at Figure \ref{fig:Z_27TeV}, we conclude that a 
27 TeV high-energy LHC with 15 ab$^{-1}$ could 
exclude Doublet VLLs with masses up to about 2300 GeV 
or discover them if the mass is less than about 1700 GeV, 
assuming the fractional uncertainty in the background to be 0.1. 
On the other hand, if we assume $\Delta_b/b = 0.5$, 
we would still be able to exclude Doublet VLLs up to 
$M_{\tau'} = 2050$ GeV, or discover them if the mass is less than about 1250 GeV. 
Just as in the case of the HL-LHC, 
from Figure \ref{fig:Z_27TeV}, a larger uncertainty in the background has a 
moderate effect on the prospects for exclusion, 
but a much larger impact on the prospects for discovery. 

\begin{figure}[!tb]
  \begin{minipage}[]{0.505\linewidth}
  \begin{flushleft}
    \includegraphics[width=8.0cm,angle=0]{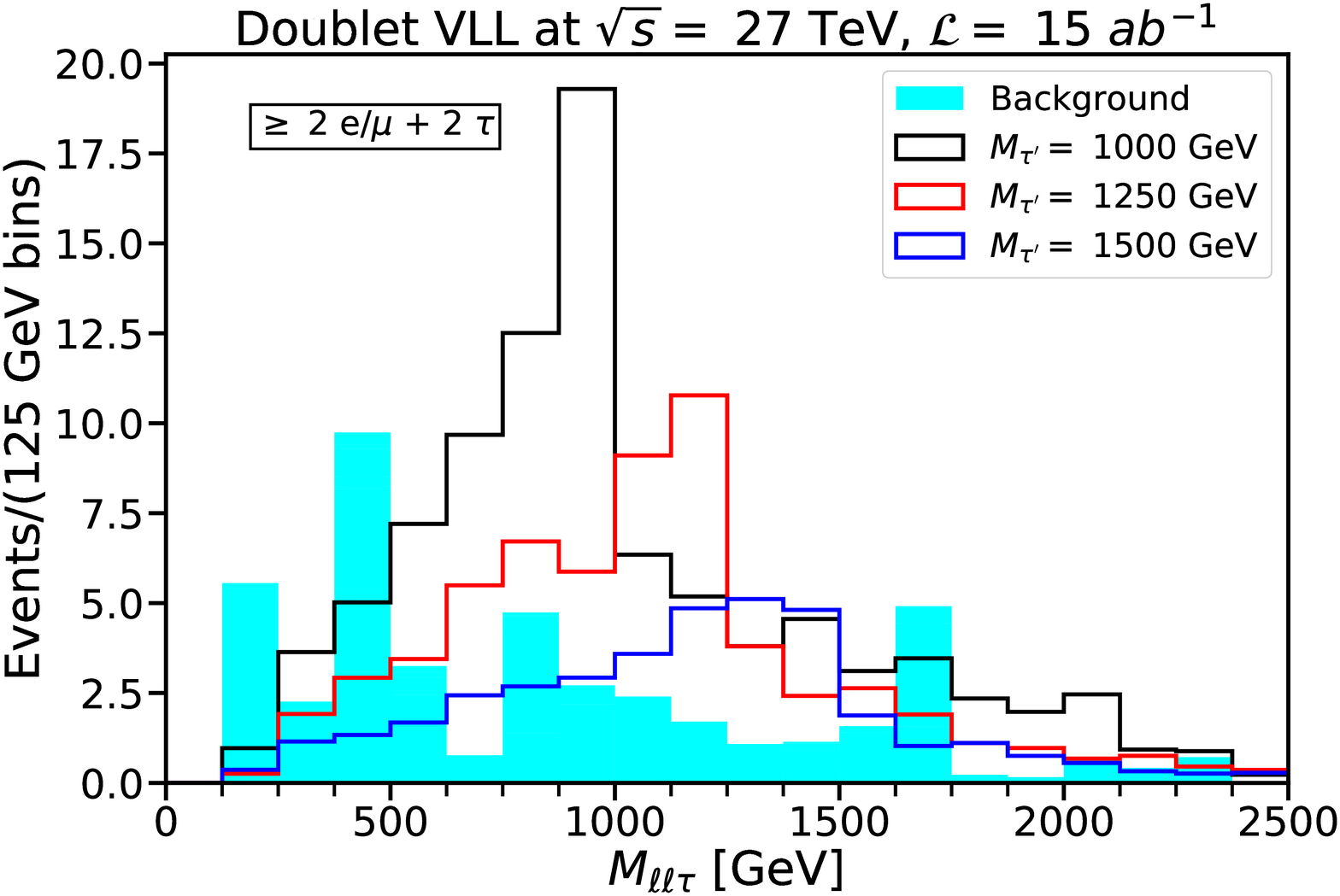}
  \end{flushleft}
  \end{minipage}
\begin{minipage}[]{0.445\linewidth}
\caption{\label{fig:inv_mass_ztau_27tev} 
The event distributions for 3-body invariant mass of $\tau^\pm e^+ e^-$ or $\tau^\pm \mu^+\mu^-$, for total background (shaded) and Doublet VLLs (lines), such that the $e^{+}e^{-}$ or $\mu^{+}\mu^{-}$ pair have an invariant mass within 10 GeV of $M_Z$ in a signal region with $\geq 2e/\mu + 2 \tau$, with the cut $L_T > 1500$ GeV imposed. Three different masses $M_{\tau'} = M_{\nu'} = 1000,$ 1250, and 1500 GeV are shown in the plot.
}   
  \end{minipage}
\end{figure}

We again consider the prospects for observing a mass peak in the case that there
are enough events for a clear discovery. In
figure \ref{fig:inv_mass_ztau_27tev}, we show the event distributions for the 3-body invariant 
mass of $\tau^\pm e^+ e^-$ or $\tau^\pm \mu^+\mu^-$, for the signal region with 
$\geq 2e/\mu + 2 \tau$, for three different choices of $M_{\tau'}$. The total of all backgrounds 
is shown as the shaded histogram. We require the 2-body invariant mass of the $e^{+}e^{-}$ 
or $\mu^{+}\mu^{-}$ pair to be within 10 GeV of $M_Z$. Additionally, we also impose the cut 
$L_T > 1500$ GeV. From Figure \ref{fig:inv_mass_ztau_27tev}, we observe that the 
distributions for Doublet VLLs are peaked just below
their respective masses, which gives a possibility to measure the masses of Doublet VLLs, 
if they are indeed discovered. For higher masses, statistical limitations 
and the combinatorial background mean that the mass determinations will be problematic. 


\subsection{Singlet VLL models}

At $\sqrt{s} = 27$ TeV with integrated luminosity of 15 $ab^{-1}$, we 
find that there is no exclusion or discovery reach for the minimal 
and the Higgs-philic Singlet VLL models. For the other two 
non-minimal Singlet models, we find some exclusion possibility, 
but again with no prospects for discovery. 

\begin{figure}[!tb]
  \begin{minipage}[]{0.505\linewidth}
  \begin{flushleft}
    \includegraphics[width=8.0cm,angle=0]{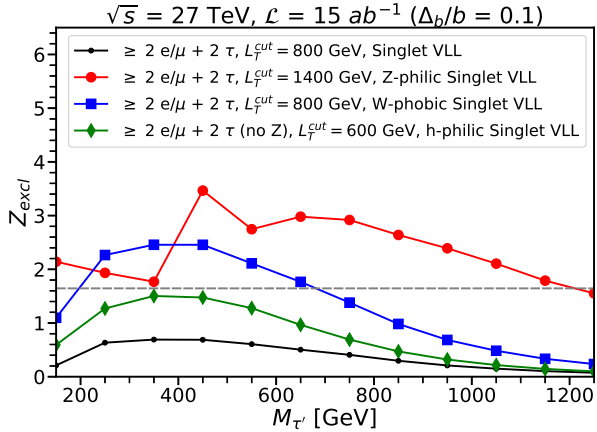}
  \end{flushleft}
  \end{minipage}
\begin{minipage}[]{0.445\linewidth}
\caption{\label{fig:Z_27TeV_Singlet} 
The median expected significances for exclusion $Z_{\rm excl}$ as a function of $M_{\tau^{\prime}}$ 
in the Singlet VLL models, for $pp$ collisions at 
$\sqrt{s} = 27$ TeV with integrated luminosity $\mathcal{L} = 15$ ab$^{-1}$, 
for the best signal region for each of the Singlet VLL models, including a cut on $L_T$ as shown in the plot.
The fractional uncertainty in the background is assumed to be $\Delta_b/b = 0.1$.
}   
  \end{minipage}
\end{figure}

We chose a cut $L_T > 1400$ GeV for the $Z$-philic Singlet VLL, $L_T > 800$ GeV for the $W$-phobic Singlet VLL, to maximize the exclusion reach. A cut $L_T > 600$ GeV for the Higgs-philic Singlet VLL and $L_T > 800$ GeV for the minimal Singlet VLL model were chosen for the best possible exclusion significance. Figure \ref{fig:Z_27TeV_Singlet} shows the resulting median expected significances 
for exclusion, for 
$\Delta_b/b = 0.1$, for the best signal region for each of the Singlet VLL models,
with the cuts on $L_T$ imposed, as mentioned above. The best signal region for exclusion, for all the Singlet VLL models is the one which requires 
$\geq 2e/\mu + 2 \tau$, except that 
an additional requirement of no-$Z$ gave better results for the Higgs-philic model.

From Figure \ref{fig:Z_27TeV_Singlet}, we can conclude that a 27 TeV $pp$ 
collider with 15 ab$^{-1}$ could possibly exclude Singlet VLLs 
with masses up to about 1200 GeV in the $Z$-philic model, 
or exclude masses up to about 670 GeV in the $W$-phobic model, 
but with no prospects for discovery. 
These results assume a fractional uncertainty in the background
of $\Delta_b/b = 0.1$.
In both minimal and the Higgs-philic Singlet VLL models, there is no 
possibility for exclusion or discovery. From the results for Singlet VLL analyses 
for 14 and 27 TeV colliders, we can note that 
the exclusion reach scaled approximately linearly with $\sqrt{s}$, 
for both $Z$-philic and the $W$-phobic Singlet VLL models.

\section{Results for a $pp$ collider with 
$\sqrt{s} = 70$ TeV\label{sec:pp70TeV}}
\setcounter{equation}{0}
\setcounter{figure}{0}
\setcounter{table}{0}
\setcounter{footnote}{1}

In this section, we turn our attention to prospects for exclusion and discovery of VLLs at possible future $pp$ collider 
at $\sqrt{s} =$ 70 TeV, with integrated luminosity of 30 ab$^{-1}$ in the six signal regions mentioned in eqs.~(\ref{eq:signal1})-(\ref{eq:signal6}). All the leptons 
including hadronic tau candidates are required to satisfy:
\beq
p_T^\ell &>& \mbox{75 GeV}.
\eeq 
along with the same pseudo-rapidity, isolation and other requirements of eqs.~(\ref{eta})-(\ref{nbjets}), with at least one $e$ or $\mu$ satisfying a trigger requirement:
\beq
p_T^{e_1}\>\,\mbox{or}\>\, p_T^{\mu_1} &>& 150 \textrm{ GeV}.
\eeq

\subsection{Doublet VLL model}

\begin{figure}[!tb]
  \begin{minipage}[]{0.495\linewidth}
    \includegraphics[width=8.0cm,angle=0]{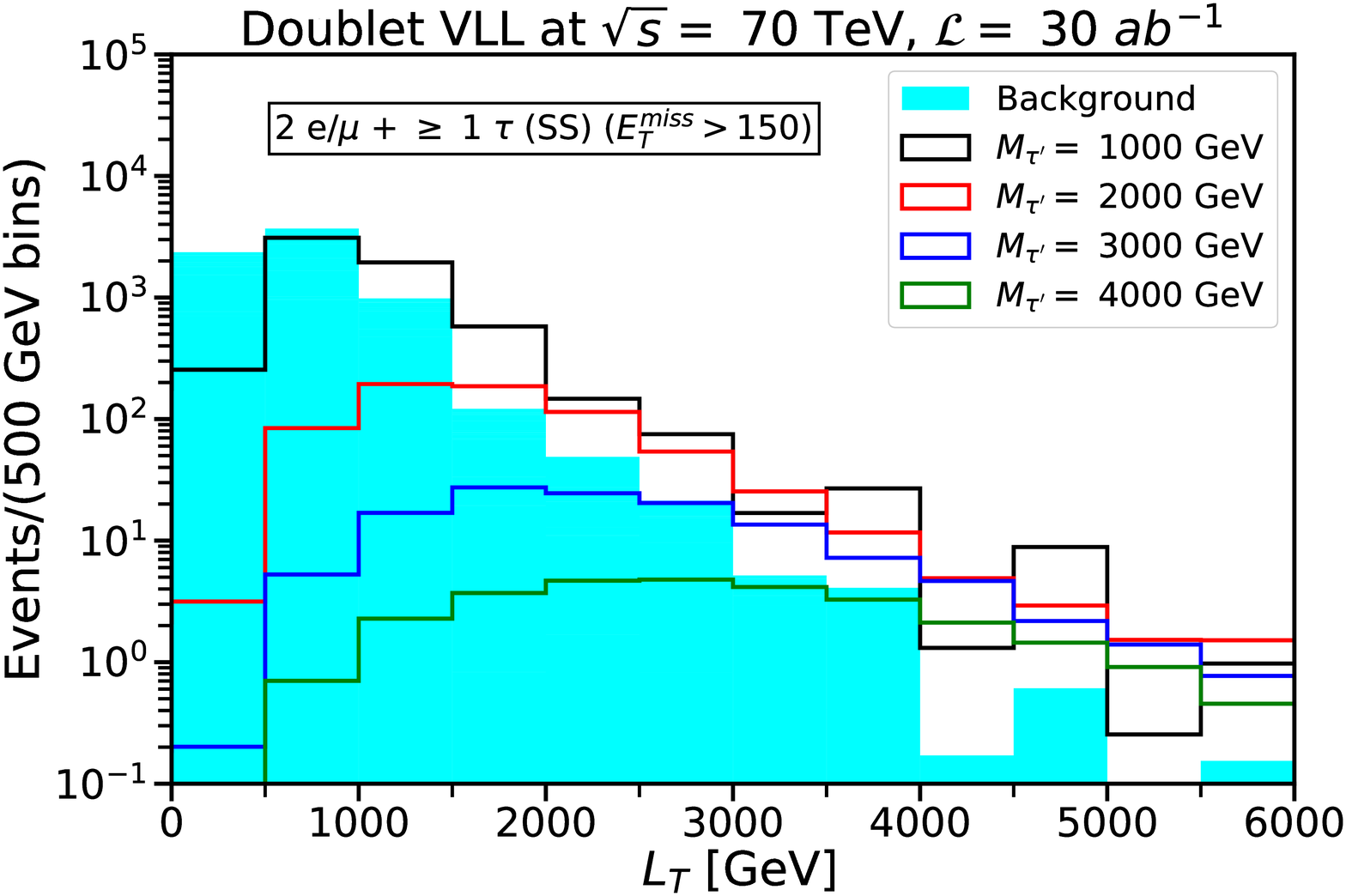}
  \end{minipage}
    \begin{minipage}[]{0.495\linewidth}
    \includegraphics[width=8.0cm,angle=0]{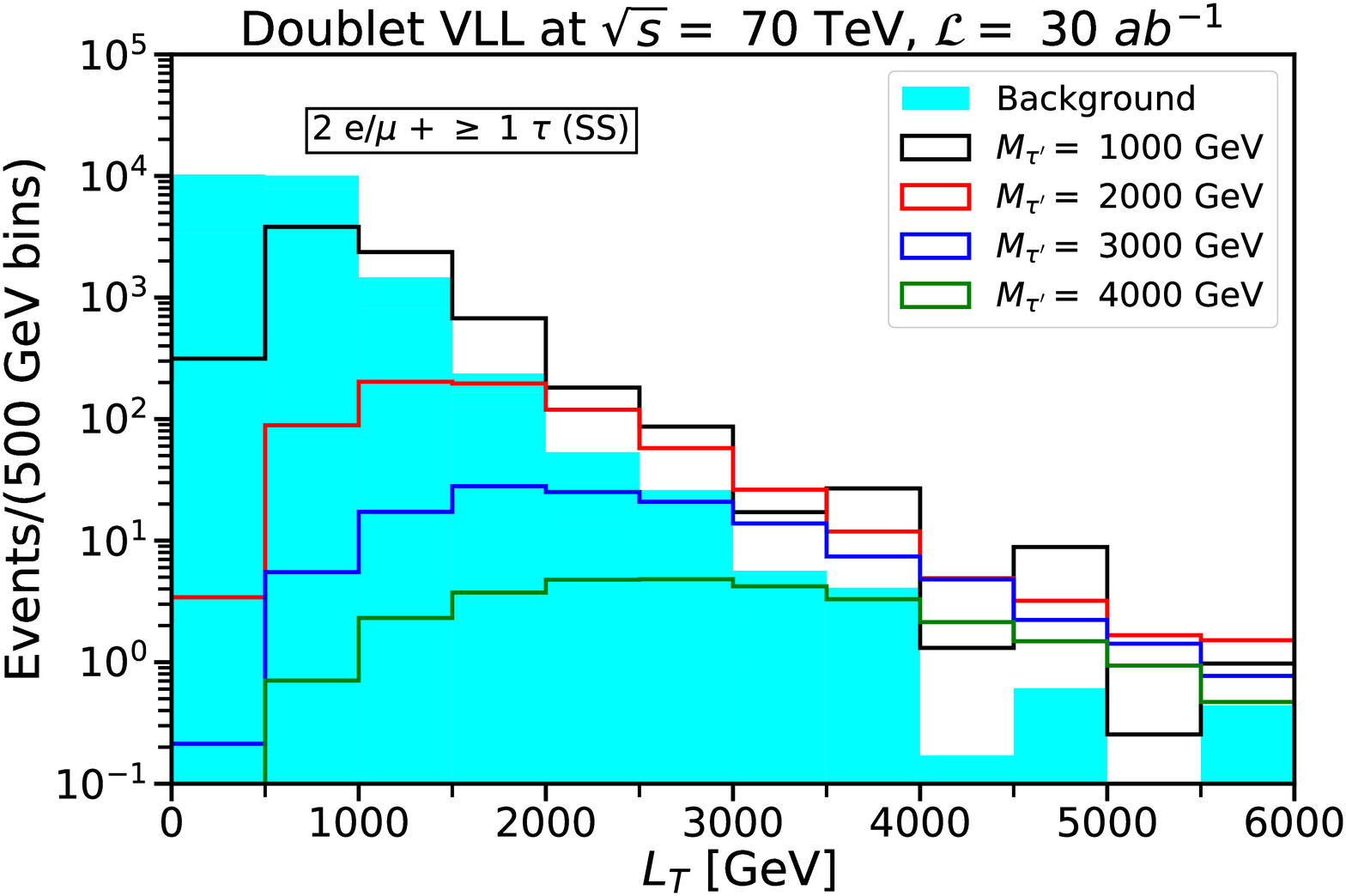}
  \end{minipage}
  \begin{minipage}[]{0.495\linewidth}
    \includegraphics[width=8.0cm,angle=0]{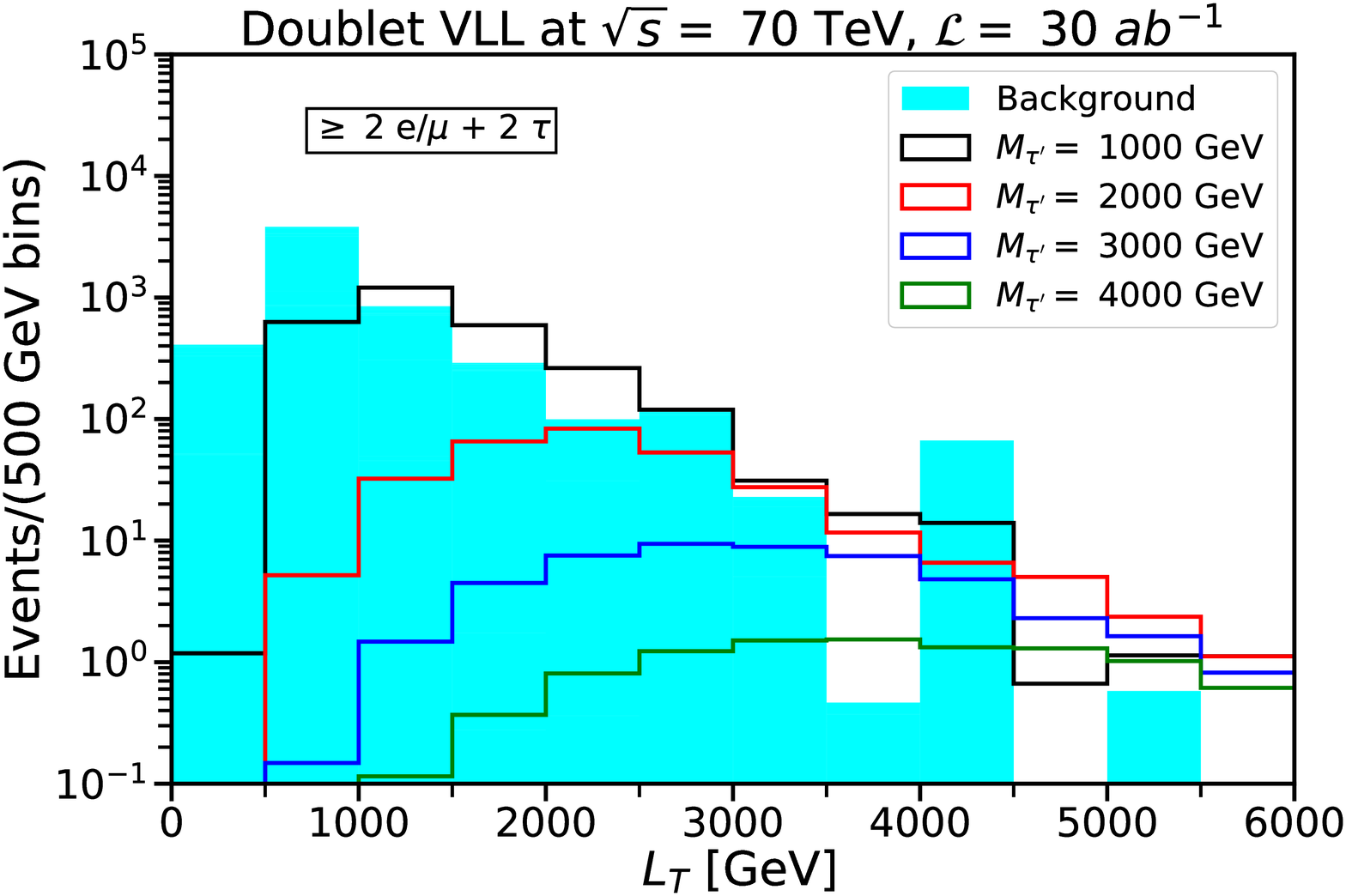}
  \end{minipage}
    \begin{minipage}[]{0.495\linewidth}
    \includegraphics[width=8.0cm,angle=0]{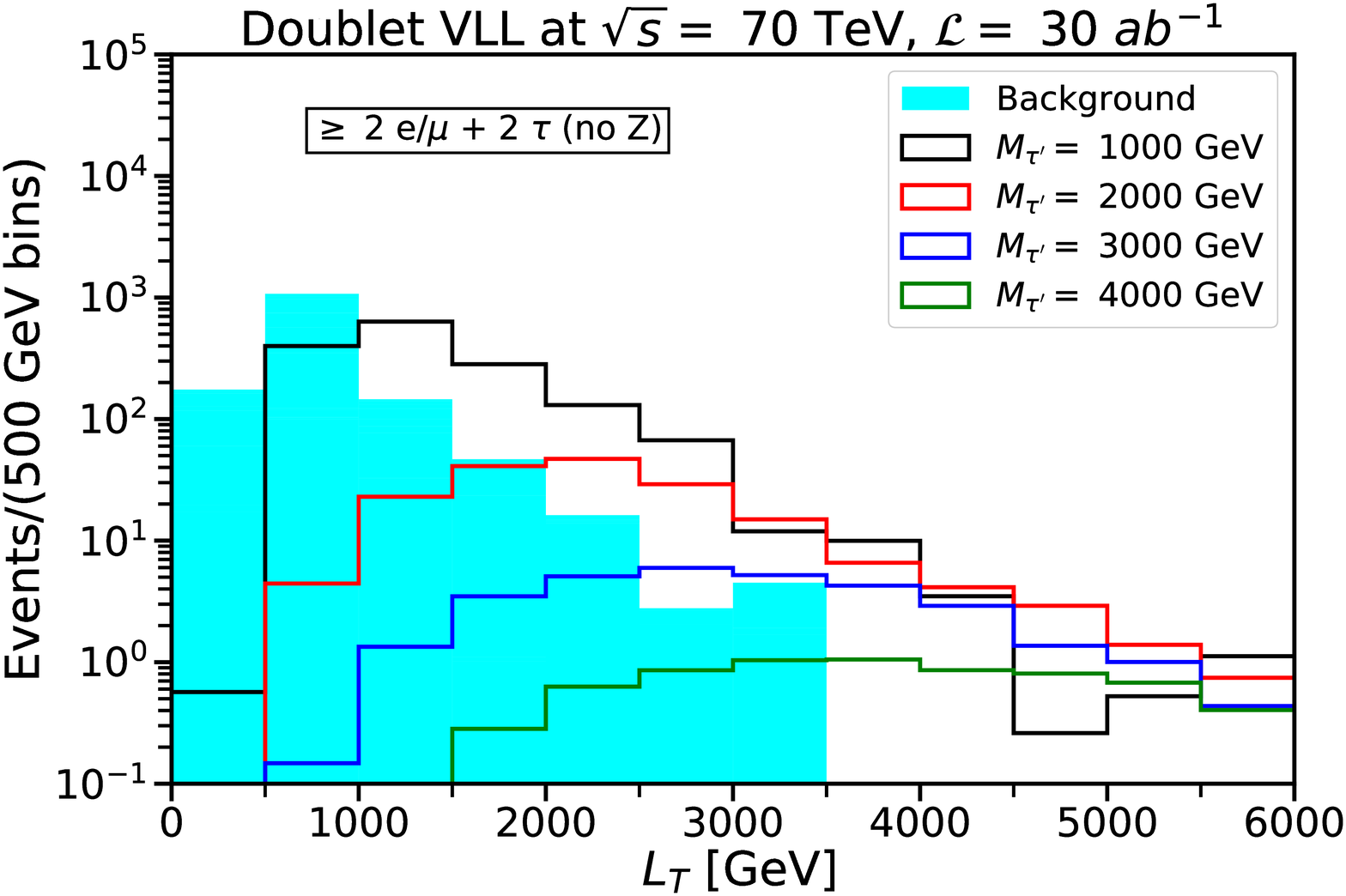}
  \end{minipage}
\begin{center}\begin{minipage}[]{0.95\linewidth}
\caption{\label{fig:LT_70TeV} 
$L_T$ event distributions for total background (shaded) and Doublet VLL
models (lines), for $pp$ collisions at $\sqrt{s} = 70$ TeV with an
integrated luminosity $\mathcal{L} =$ 30 ab$^{-1}$. Four different masses
$M_{\tau'} = M_{\nu'} = 1000,$ 2000, 3000, and 4000 GeV are shown in
each panel. The four panels show results for the four best signal
regions, as labeled.
}
\end{minipage}\end{center}
\end{figure}

In Figure \ref{fig:LT_70TeV}, we show the $L_T$ distributions 
for the best four of the 
signal regions, for four different choices of $M_{\tau'}$ as labeled, and for 
the total of all backgrounds shown as the shaded histogram. We found that choosing 
the cut $L_T > 2800$ GeV provides a good reach for both exclusion and discovery in this case.

\begin{figure}[!tb]
  \begin{minipage}[]{0.495\linewidth}
    \includegraphics[width=8.0cm,angle=0]{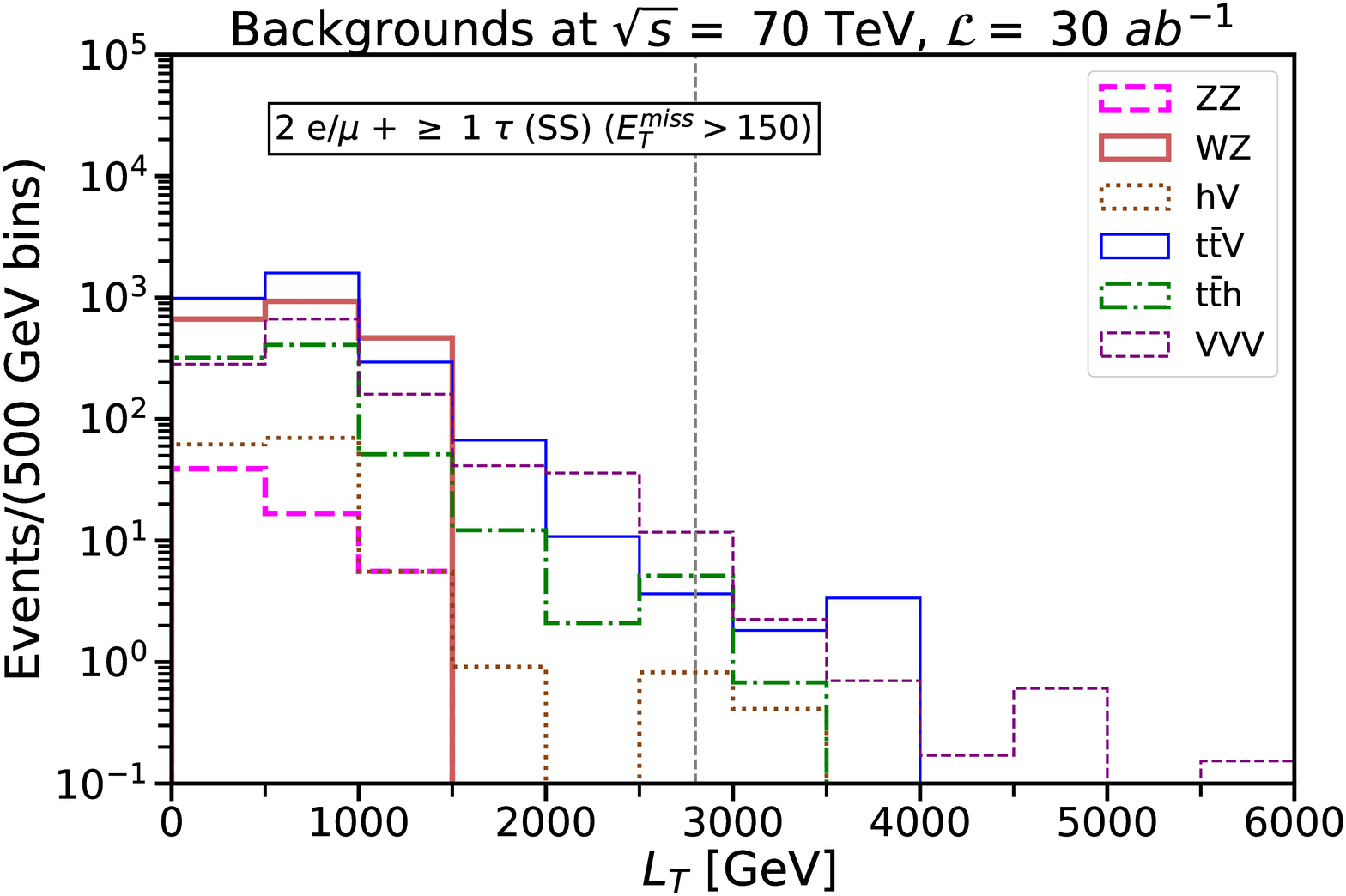}
  \end{minipage}
    \begin{minipage}[]{0.495\linewidth}
    \includegraphics[width=8.0cm,angle=0]{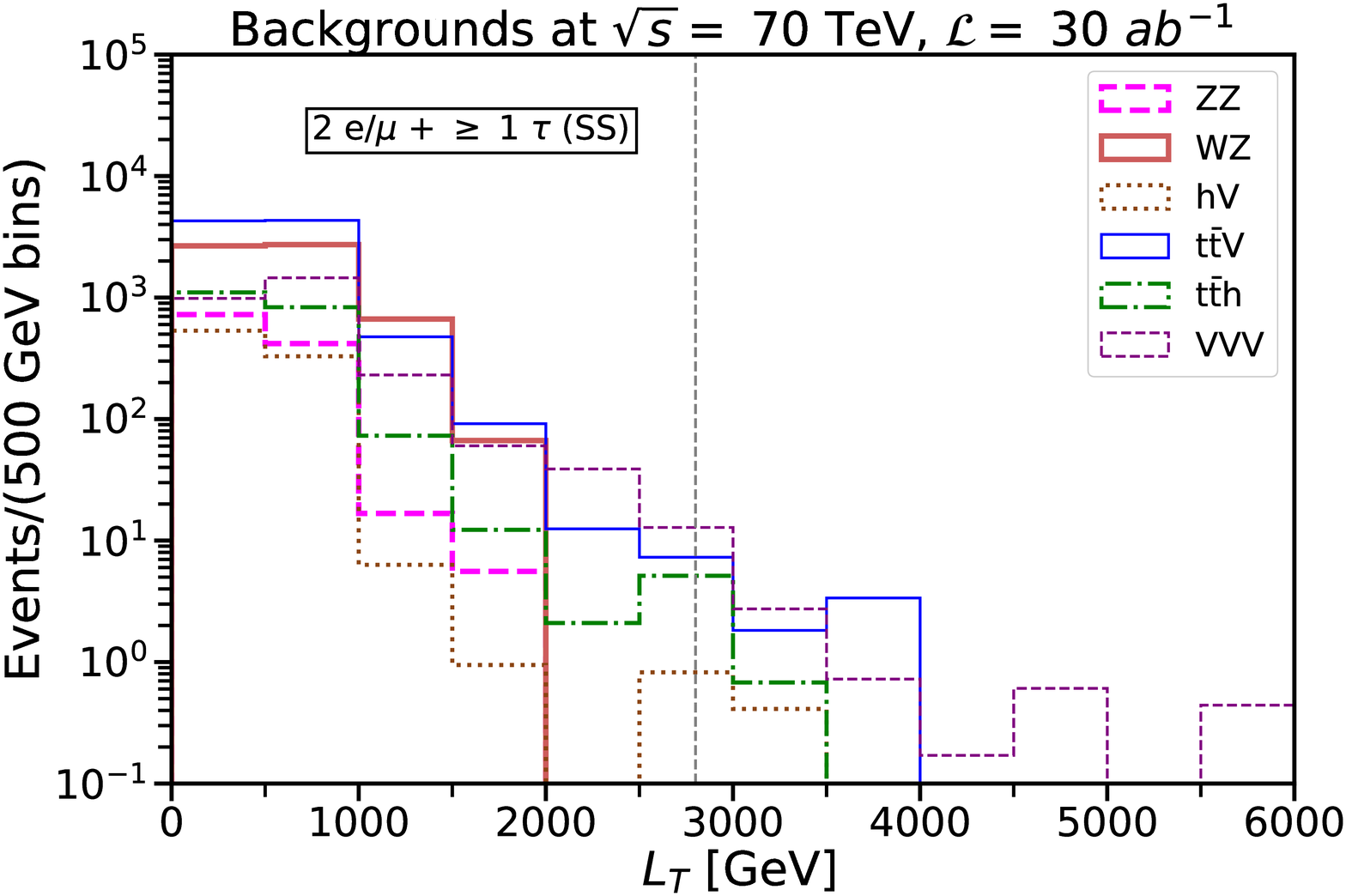}
  \end{minipage}
  \begin{minipage}[]{0.495\linewidth}
    \includegraphics[width=8.0cm,angle=0]{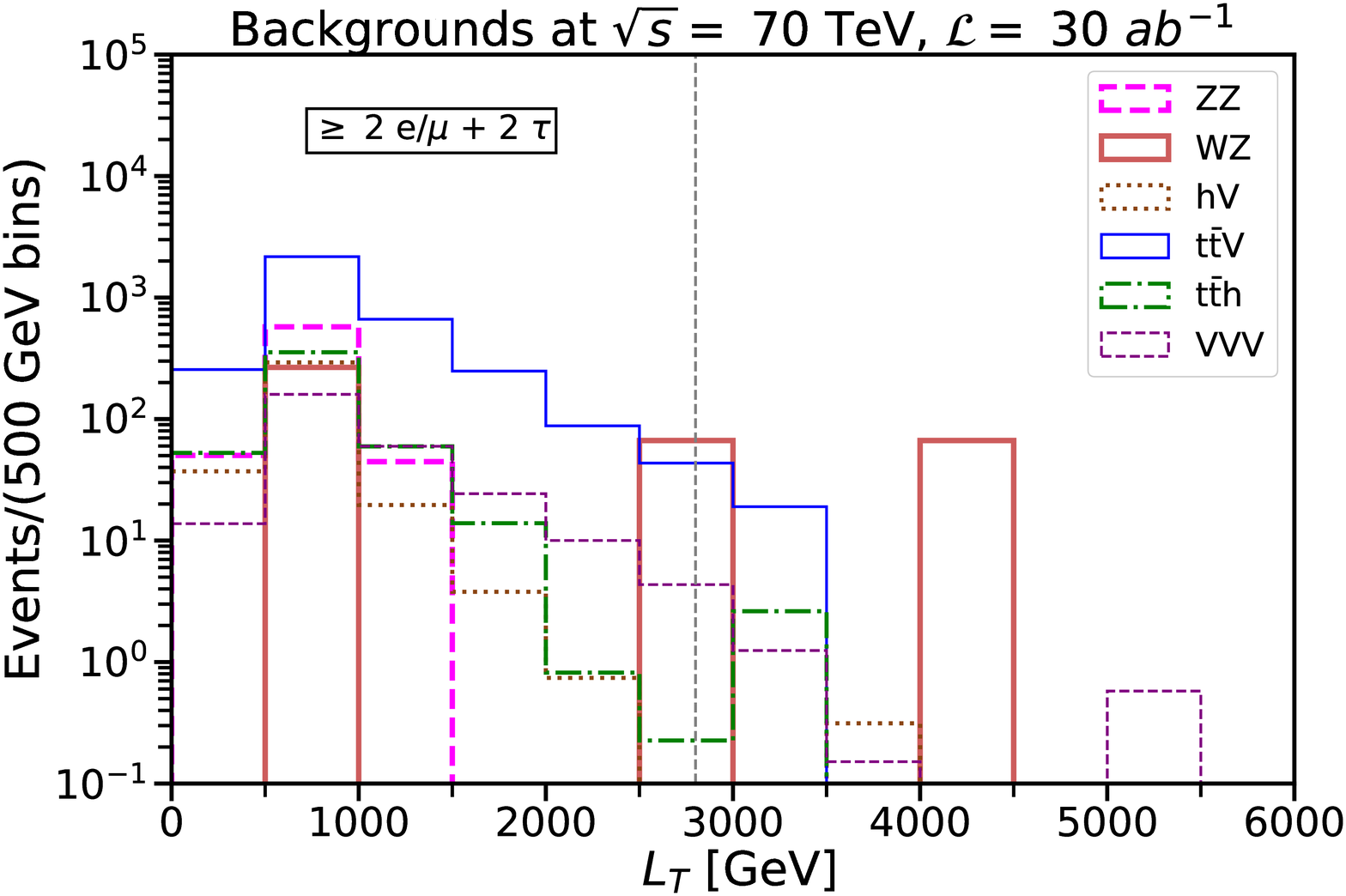}
  \end{minipage}
    \begin{minipage}[]{0.495\linewidth}
    \includegraphics[width=8.0cm,angle=0]{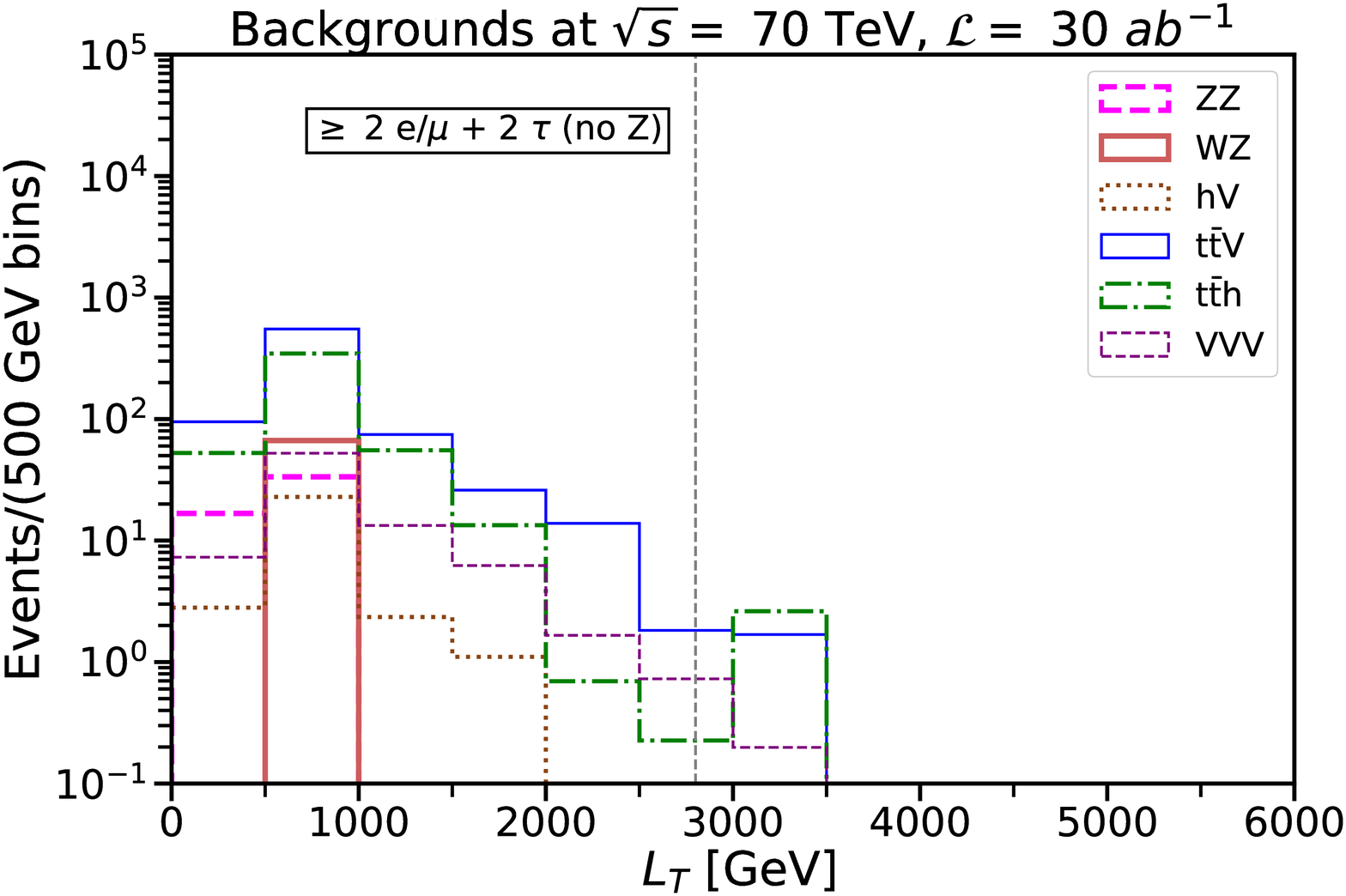}
  \end{minipage}
\begin{center}\begin{minipage}[]{0.95\linewidth}
\caption{\label{fig:LT_bg_70TeV} 
$L_T$ event distributions for all processes contributing to total SM background, for $pp$ collisions at $\sqrt{s} = 70$ TeV with an 
integrated luminosity $\mathcal{L} =$ 30 ab$^{-1}$. The four panels show results for the four best signal regions, as labeled. The vertical dashed line in all four panels shows our choice of 
$L_T$ cut of 2800 GeV.}
\end{minipage}\end{center}
\end{figure}

Figure \ref{fig:LT_bg_70TeV} shows the $L_T$ distributions 
for all background components, for the four best signal regions as labeled. 
The $L_T$ cut is shown in the figure as a vertical dashed line. 
After imposing the $L_T$ cut, the most important SM backgrounds are $t\bar{t}V$ 
and $VVV$ in the two signal regions with 2 SS $e/\mu\> +\!\geq 1 \tau$, 
while the most important SM backgrounds are $WZ$ and $t\bar{t}V$ in the 
signal region with $\geq 2e/\mu + 2 \tau$, and $t\bar{t}h$ and $t\bar{t}V$ 
in the signal region with $\geq 2e/\mu + 2 \tau$ (no-$Z$). We note that 
the fluctuation of about 66 normalized events in the bin from 4000 to 4500 GeV 
in the lower left panel (i.e. signal region 
with $\geq 2e/\mu + 2 \tau$) of Figure \ref{fig:LT_70TeV} is due to 
a single simulated event of $WZ$ background.
This is because of the large cross-section but extremely low yield for this 
background component in this signal region even after forcing decays to leptons.
This is an unavoidable source of uncertainty for our analysis; given that our sample
size was already $5.5 \times 10^6$ simulated events for this component, no practically feasible
increase in sample size would yield significantly better statistics. However, in the real world
the background can perhaps be determined more accurately from data.

\begin{figure}[!tb]
  \begin{minipage}[]{0.495\linewidth}
    \includegraphics[width=8.0cm,angle=0]{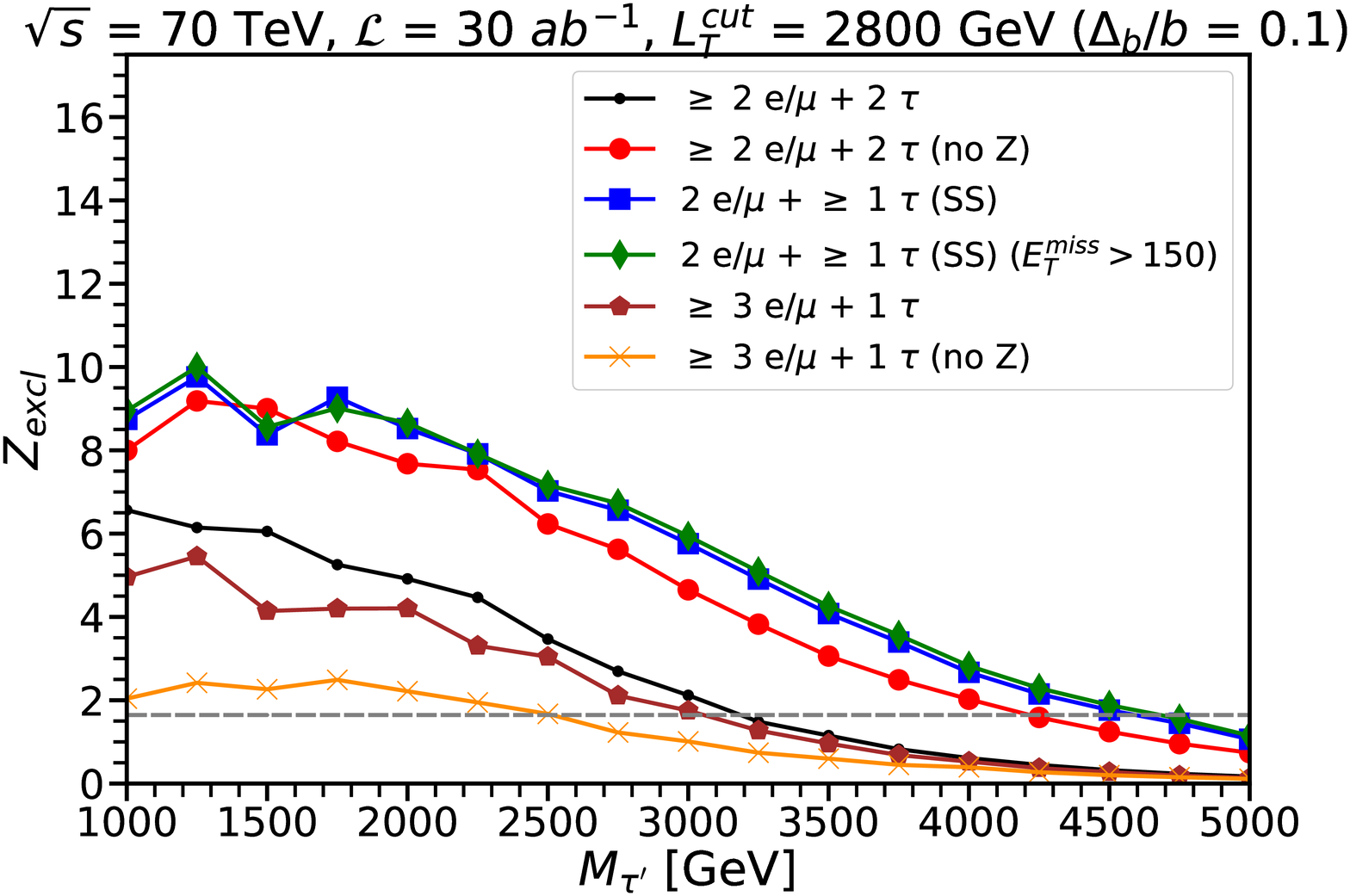}
  \end{minipage}
    \begin{minipage}[]{0.495\linewidth}
    \includegraphics[width=8.0cm,angle=0]{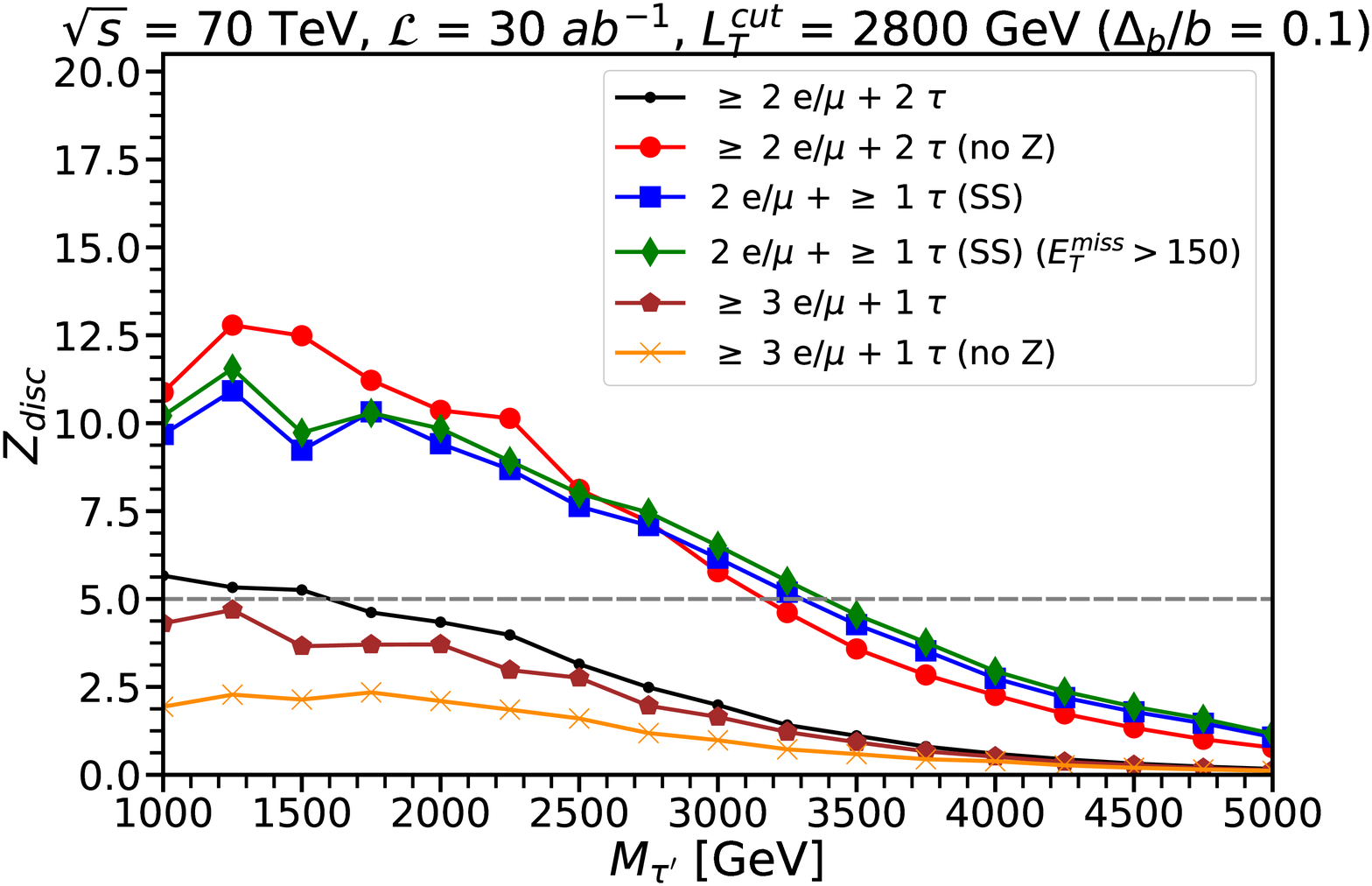}
  \end{minipage}
    \begin{minipage}[]{0.495\linewidth}
    \includegraphics[width=8.0cm,angle=0]{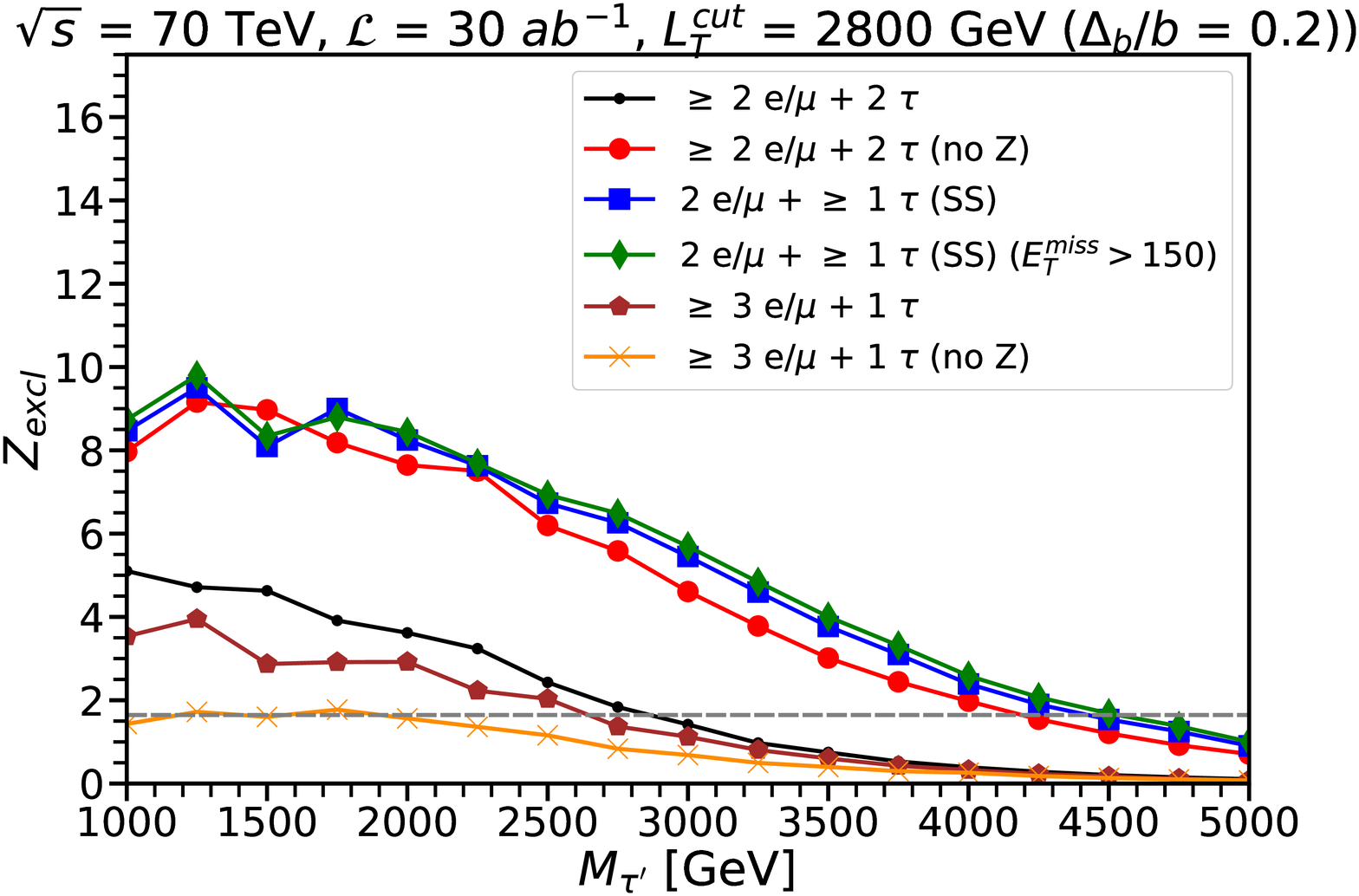}
  \end{minipage}
    \begin{minipage}[]{0.495\linewidth}
    \includegraphics[width=8.0cm,angle=0]{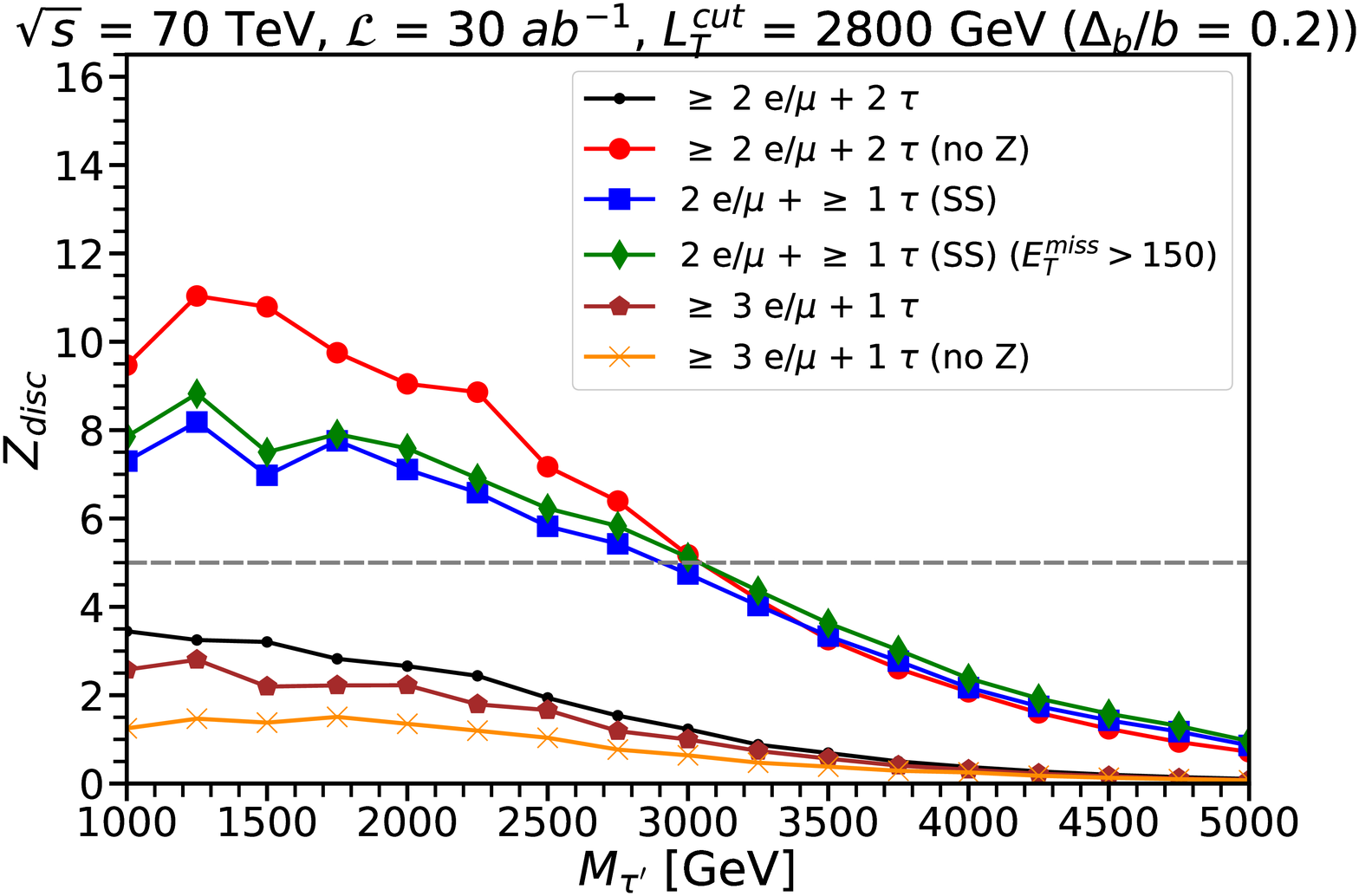}
  \end{minipage}
  \begin{minipage}[]{0.495\linewidth}
    \includegraphics[width=8.0cm,angle=0]{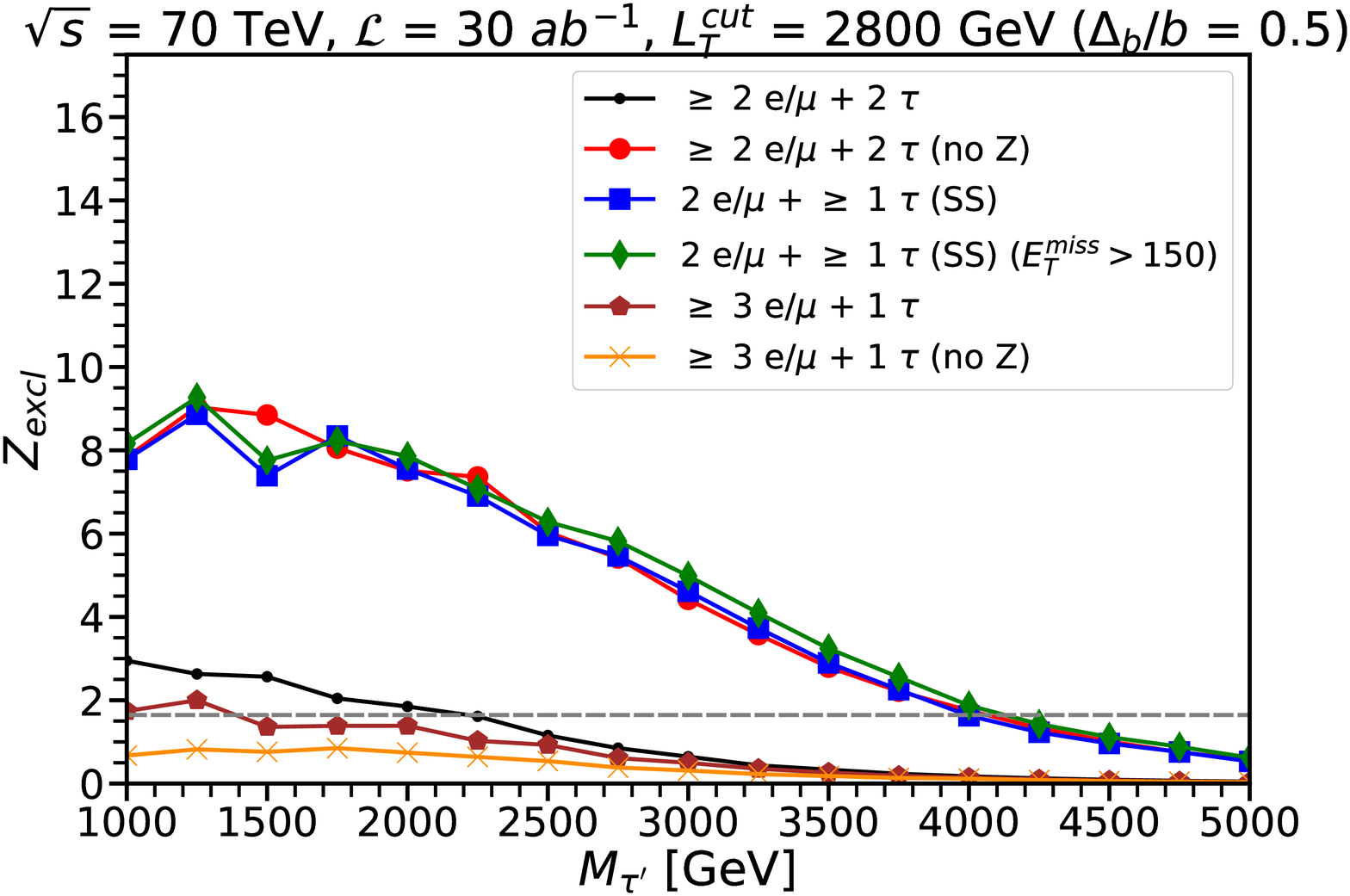}
  \end{minipage}
    \begin{minipage}[]{0.495\linewidth}
    \includegraphics[width=8.0cm,angle=0]{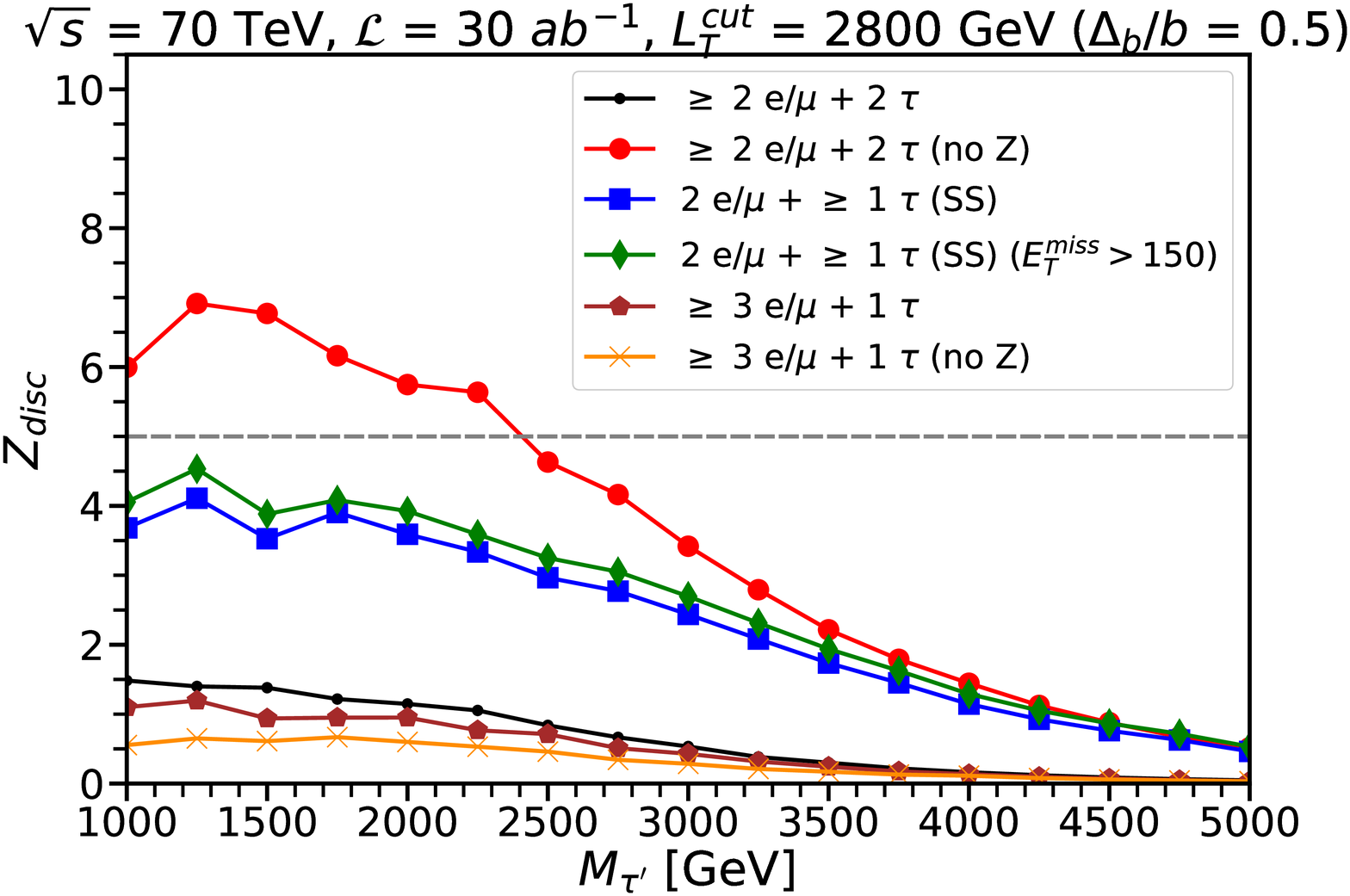}
  \end{minipage}
  
\begin{center}\begin{minipage}[]{0.95\linewidth}
\caption{\label{fig:Z_70TeV} 
The median expected significances for exclusion $Z_{\rm excl}$ 
(left panels) and discovery $Z_{\rm disc}$ (right panels) as a function of
$M_{\tau^{\prime}}$ in the Doublet VLL model, for $pp$ collisions at
$\sqrt{s} = 70$ TeV with integrated luminosity $\mathcal{L} = 30$ 
ab$^{-1}$, for six different signal regions as described in the text, each including a cut $L_T>$
2800 GeV. The fractional uncertainty in the background is assumed to be
$\Delta_b/b = 0.1$ (top row), $0.2$ (middle row), and $0.5$ (bottom
row).}
\end{minipage}\end{center}
\end{figure}

Figure \ref{fig:Z_70TeV} shows the median expected significances for exclusion $Z_{\rm excl}$ (left panels) and discovery 
$Z_{\rm disc}$ (right panels) as a function of $M_{\tau'}$, for 
$\Delta_b/b = 0.1$ (top row), $0.2$ (middle row), and $0.5$ (bottom row),
with the cut $L_T > 2800$ GeV imposed. The two signal regions with 2 SS $e/\mu\> +\!\geq 1 \tau$ and the 
signal region with $\geq 2e/\mu + 2 \tau$ (no-$Z$) have comparable exclusion significances and reaches. However, the latter has higher discovery significances at lower masses, as well as at higher fractional uncertainties in the background, 
e.g. $\Delta_b/b = 0.5$ .

Figure \ref{fig:Z_70TeV} shows a possibility of excluding Doublet VLLs of masses up to about 
4700 GeV or discovering them if mass is less than about 3400 GeV with a 70 TeV $pp$ collider 
with 30 ab$^{-1}$, assuming the fractional uncertainty in the background to be 
$\Delta_b/b = 0.1$. If $\Delta_b/b = 0.5$, we can still expect to exclude Doublet VLLs up to 
$M_{\tau'} = 4150$ GeV, or discover them if the mass is less than about 2400 GeV. Again, a larger uncertainty in the background has a moderate effect on the prospects for exclusion, 
but a much larger impact on the prospects for discovery. 

\clearpage

\subsection{Singlet VLL models}
\begin{figure}[!tb]
  \begin{minipage}[]{0.505\linewidth}
  \begin{flushleft}
    \includegraphics[width=8.0cm,angle=0]{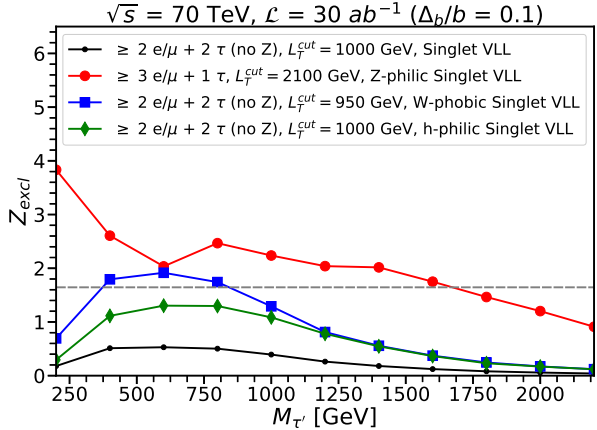}
  \end{flushleft}
  \end{minipage}
\begin{minipage}[]{0.445\linewidth}
\caption{\label{fig:Z_70TeV_Singlet} 
The median expected significances for exclusion $Z_{\rm excl}$ as a function of $M_{\tau^{\prime}}$ 
in the Singlet VLL models, for $pp$ collisions at 
$\sqrt{s} = 70$ TeV with integrated luminosity $\mathcal{L} = 30$ ab$^{-1}$, 
for the best signal region for each of the Singlet VLL models, including a cut on $L_T$ as shown in the plot.
The fractional uncertainty in the background is assumed to be $\Delta_b/b = 0.1$.
}   
  \end{minipage}
\end{figure}

For $pp$ collisions with $\sqrt{s} = 70$ TeV with 30 $ab^{-1}$, we again
find that there is no possible exclusion or discovery reach for the minimal 
and the Higgs-philic Singlet VLL models. While we find some exclusion possibility for the 
$Z$-philic and the $W$-phobic Singlet VLL models, there are again no prospects 
for discovery. 
Figure \ref{fig:Z_70TeV_Singlet} shows the resulting median expected significances for exclusion, for 
$\Delta_b/b = 0.1$, for the best signal region for each of the Singlet VLL models,
with appropriate cuts on $L_T$ imposed. 
To approximately maximize the exclusion reaches for the $Z$-philic and $W$-phobic 
Singlet VLL models, we chose a 
cut $L_T > 2100$ GeV for the former and $L_T > 950$ GeV for the latter. 
The best signal region for exclusion for the 
$Z$-philic Singlet VLL model is the one which requires $\geq 3 e/\mu + 1 \tau$, 
while it was $\geq 2 e/\mu + 2 \tau$, no-$Z$ for all other Singlet VLL models.

From Figure \ref{fig:Z_70TeV_Singlet}, we conclude that a 70 TeV $pp$ collider with 30 ab$^{-1}$ could possibly exclude Singlet VLLs with masses up to about 1700 GeV in the $Z$-philic model, or exclude masses up to  850 GeV in the $W$-phobic model, but with no prospects for $5\sigma$ discovery. 
In both minimal and the Higgs-philic Singlet VLL models, there is unfortunately no 
possibility for exclusion or discovery, at least with the signal regions we considered.

\section{Results for a $pp$ collider with 
$\sqrt{s} = 100$ TeV\label{sec:pp100TeV}}
\setcounter{equation}{0}
\setcounter{figure}{0}
\setcounter{table}{0}
\setcounter{footnote}{1}

Finally, we consider the possibility of excluding or discovering VLLs at a future 
very high energy $pp$ collider at $\sqrt{s} = 100$ TeV, with integrated luminosity 
of 30 $ab^{-1}$, in the six signal regions mentioned in 
eqs.~(\ref{eq:signal1})-(\ref{eq:signal6}). We require all leptons including 
hadronic tau candidates to satisfy: 
\beq
p_T^\ell > \mbox{100 GeV}.
\eeq 
along with the same pseudo-rapidity, isolation and other requirements of 
eqs.~(\ref{eta})-(\ref{nbjets}). We then require the leading $e$ or $\mu$ 
in each event to satisfy a trigger requirement:
\beq
p_T^{e_1}\>\,\mbox{or}\>\, p_T^{\mu_1} > 200 \textrm{ GeV}.
\eeq

\subsection{Doublet VLL model}

\begin{figure}[!tb]
  \begin{minipage}[]{0.495\linewidth}
    \includegraphics[width=8.0cm,angle=0]{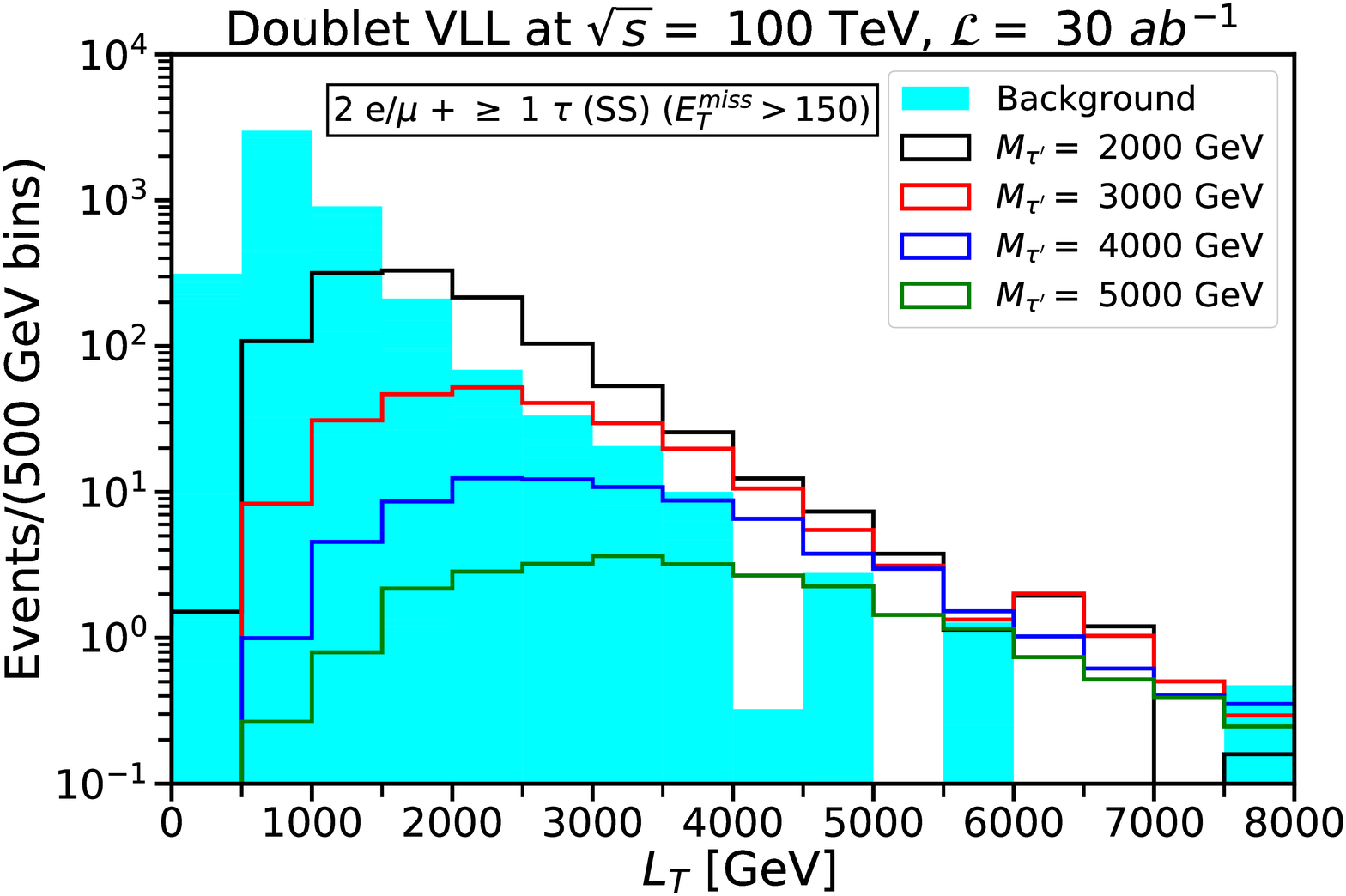}
  \end{minipage}
    \begin{minipage}[]{0.495\linewidth}
    \includegraphics[width=8.0cm,angle=0]{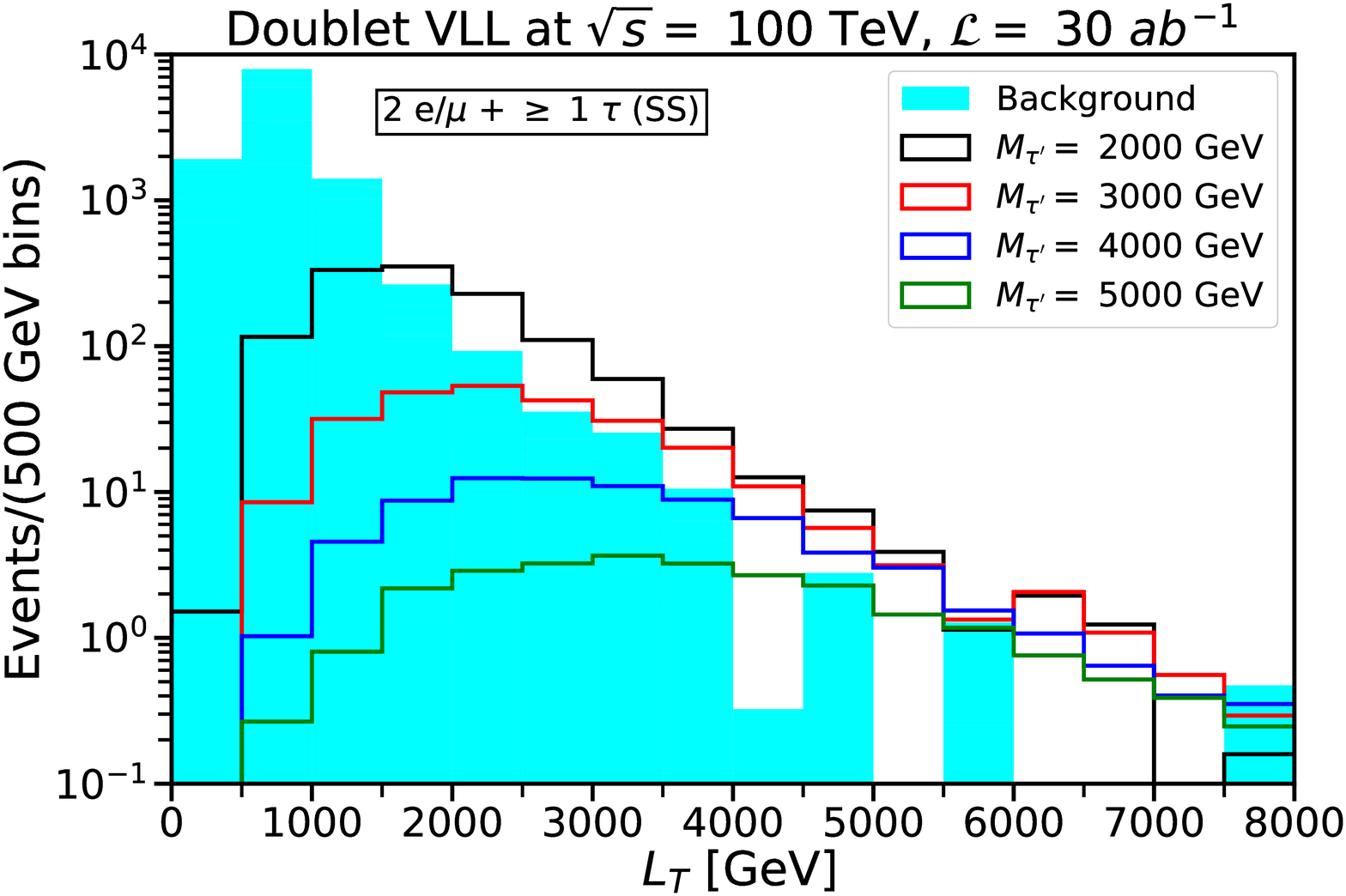}
  \end{minipage}
  \begin{minipage}[]{0.495\linewidth}
    \includegraphics[width=8.0cm,angle=0]{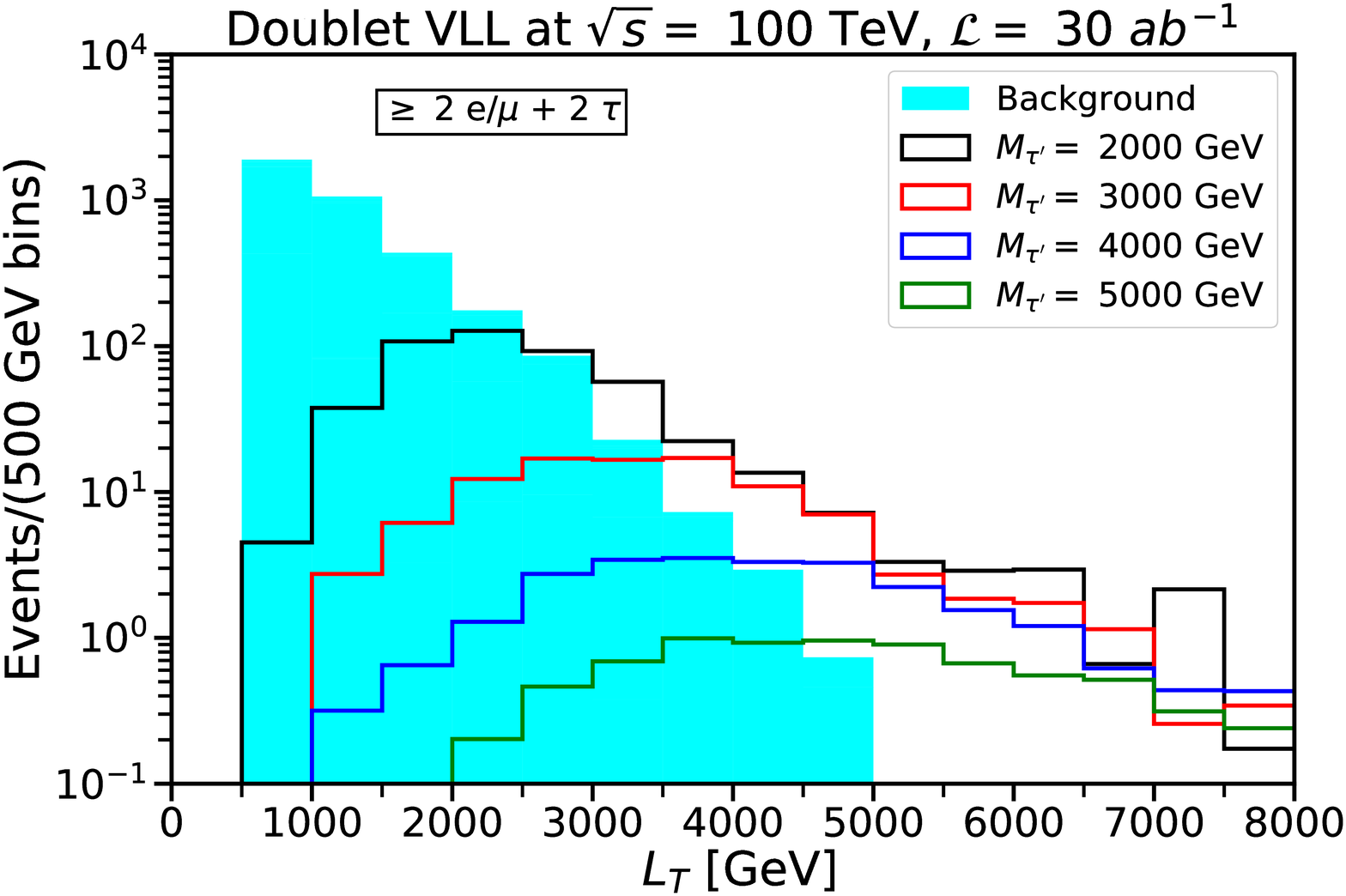}
  \end{minipage}
    \begin{minipage}[]{0.495\linewidth}
    \includegraphics[width=8.0cm,angle=0]{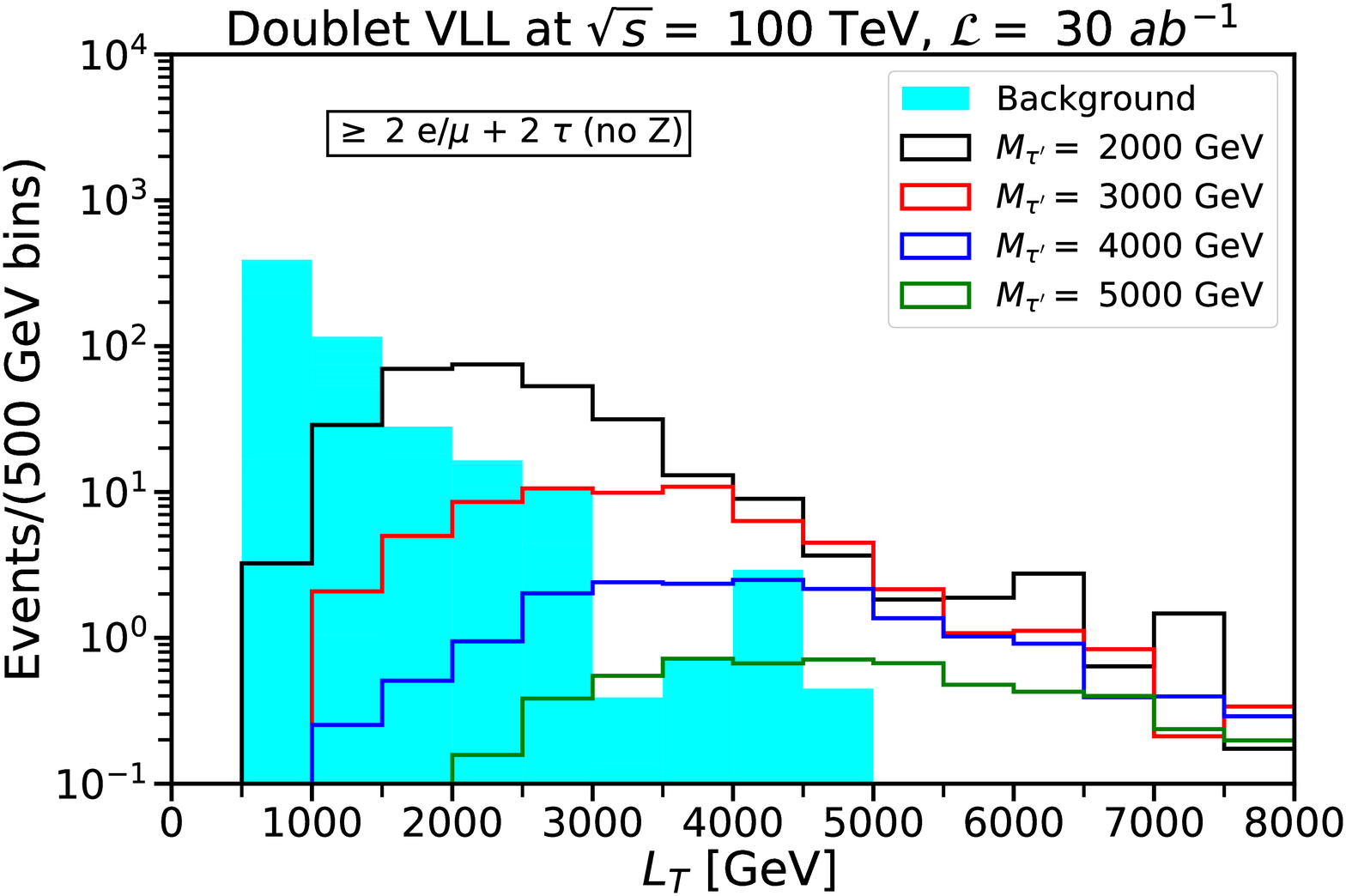}
  \end{minipage}
\begin{center}\begin{minipage}[]{0.95\linewidth}
\caption{\label{fig:LT_100TeV} 
$L_T$ event distributions for total background (shaded) and Doublet VLL
models (lines), for $pp$ collisions at $\sqrt{s} = 100$ TeV with an
integrated luminosity $\mathcal{L} =$ 30 ab$^{-1}$. Four different masses
$M_{\tau'} = M_{\nu'} = 2000,$ 3000, 4000, and 5000 GeV are shown in
each panel. The four panels show results for the four best signal
regions, as labeled.
}
\end{minipage}\end{center}
\end{figure}

In Figure \ref{fig:LT_100TeV}, we show the $L_T$ distributions 
for the best four of these signal regions, 
for four different choices of $M_{\tau'}$ as labeled, 
and for the total of all backgrounds 
shown as the shaded histogram. 
To obtain enhanced expected reaches for both exclusion and discovery, 
we then chose a cut  $L_T > 3500$ GeV.

\begin{figure}[!tb]
  \begin{minipage}[]{0.495\linewidth}
    \includegraphics[width=8.0cm,angle=0]{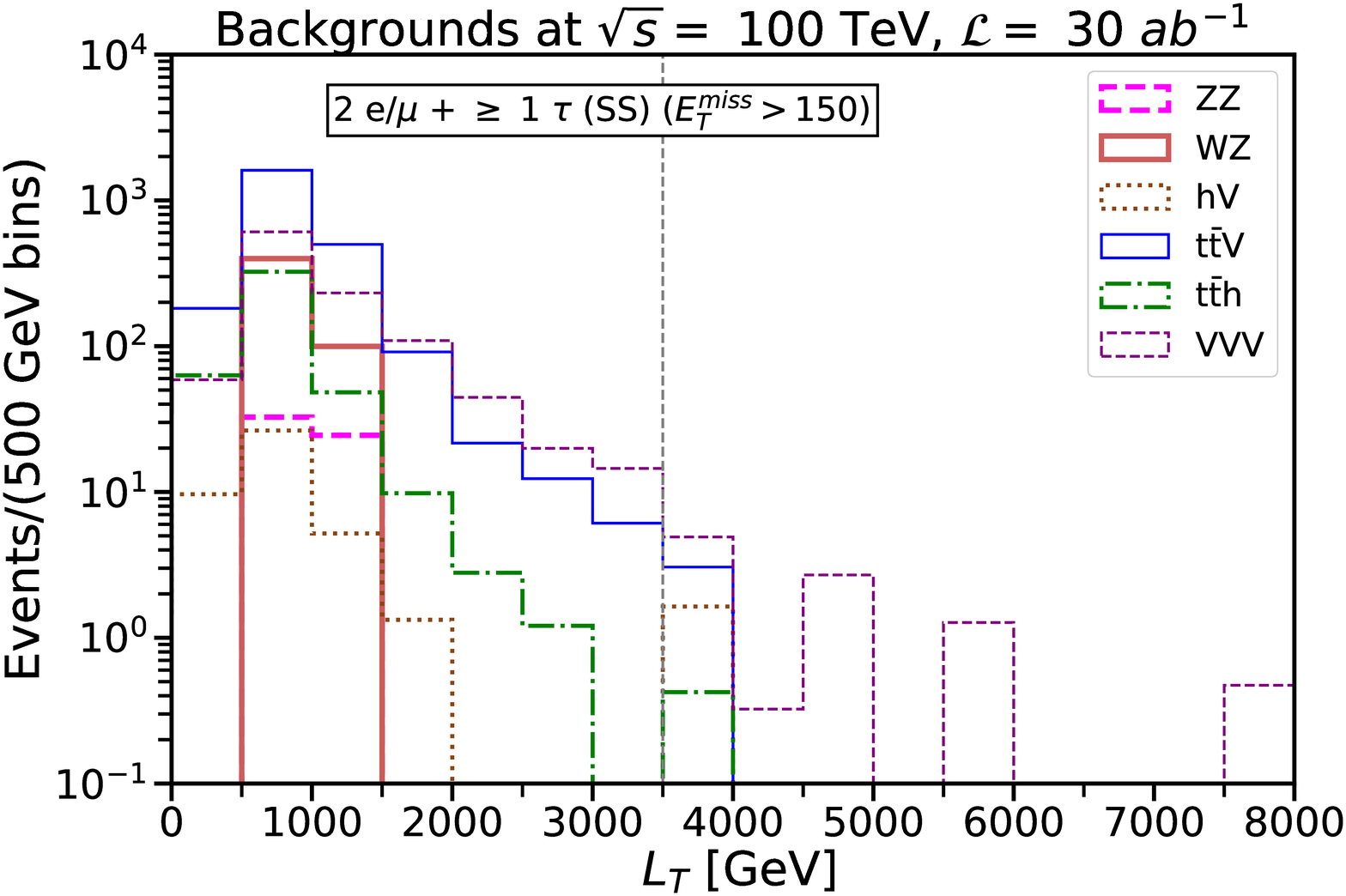}
  \end{minipage}
    \begin{minipage}[]{0.495\linewidth}
    \includegraphics[width=8.0cm,angle=0]{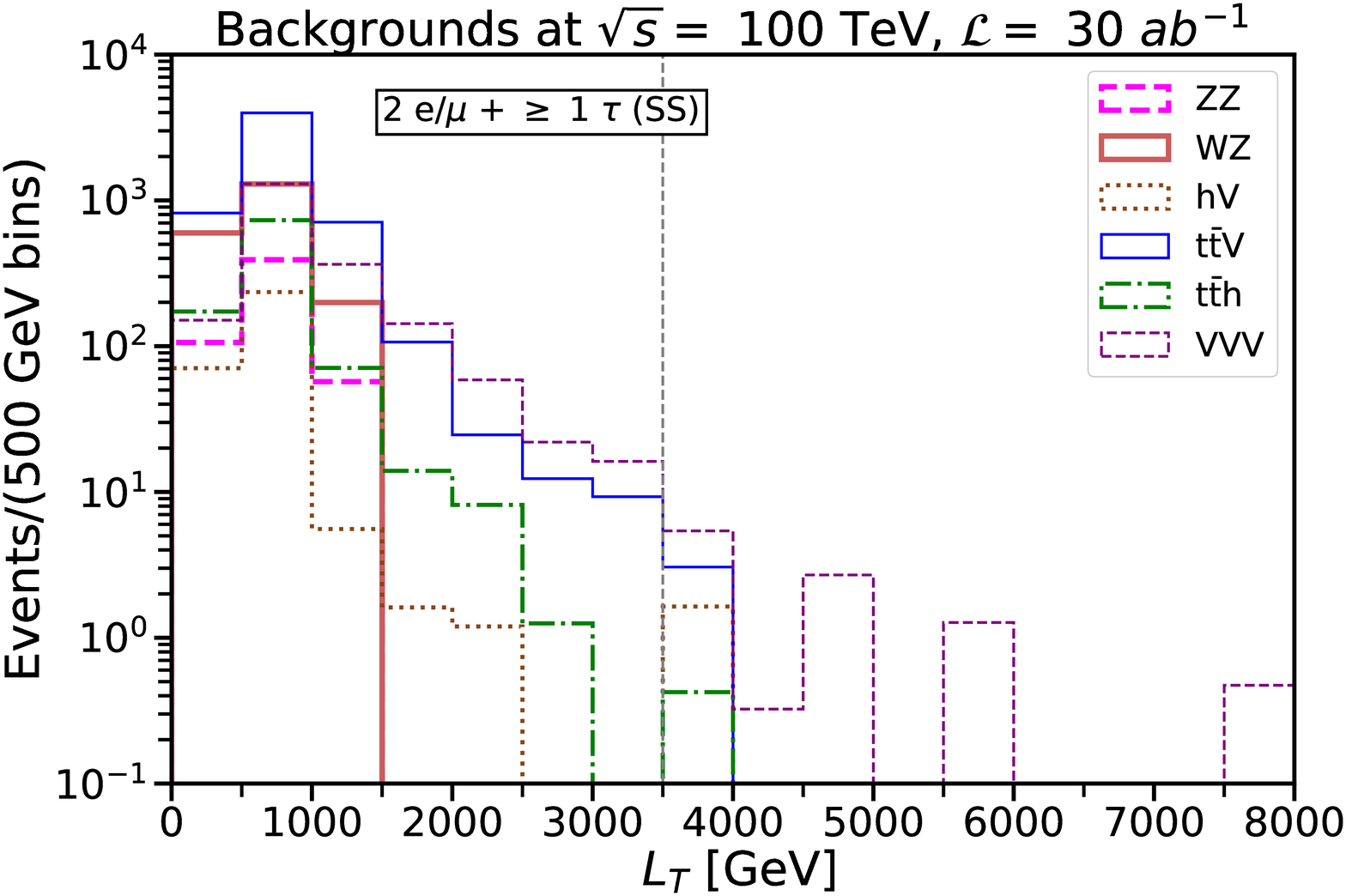}
  \end{minipage}
  \begin{minipage}[]{0.495\linewidth}
    \includegraphics[width=8.0cm,angle=0]{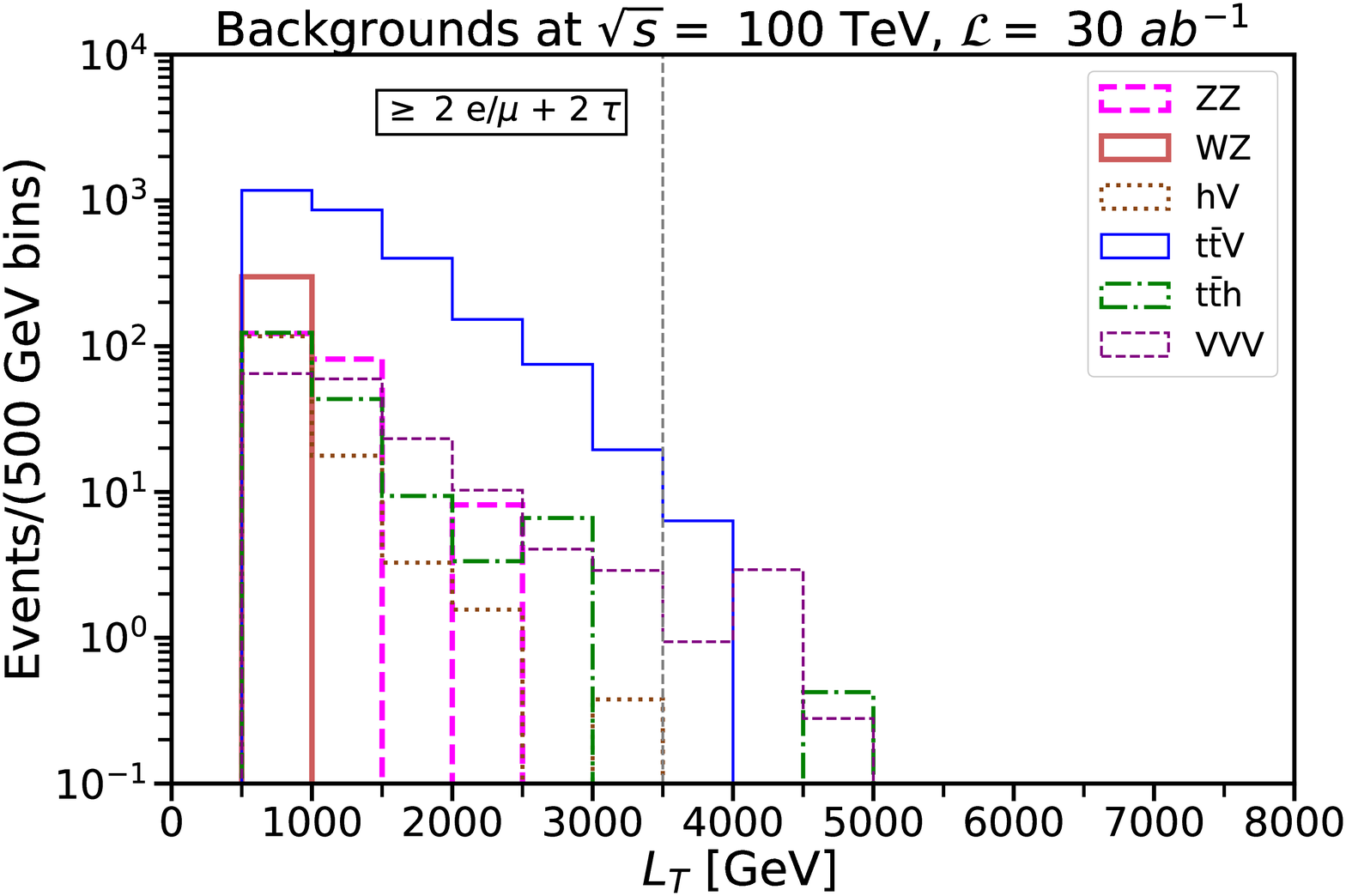}
  \end{minipage}
    \begin{minipage}[]{0.495\linewidth}
    \includegraphics[width=8.0cm,angle=0]{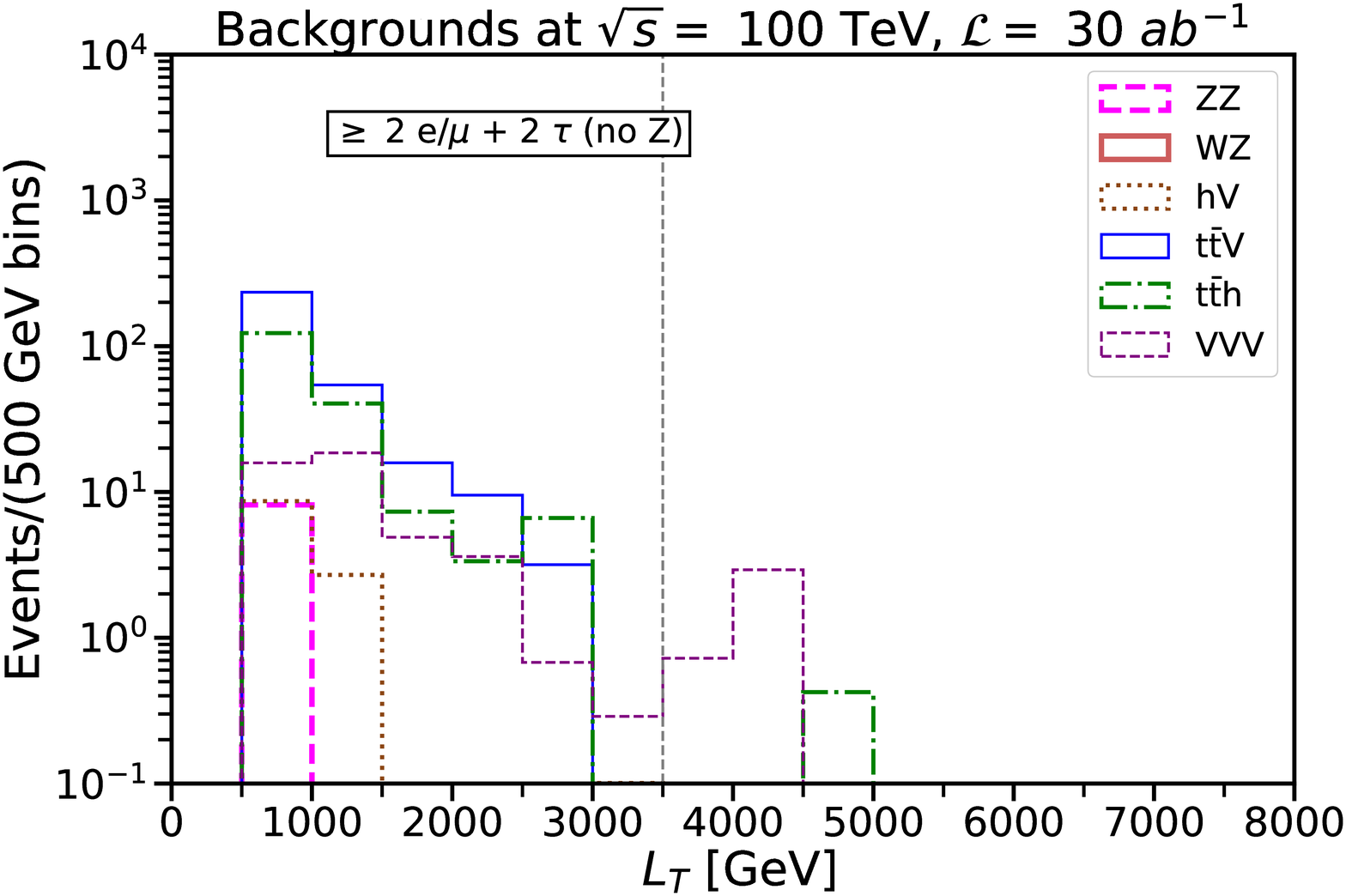}
  \end{minipage}
\begin{center}\begin{minipage}[]{0.95\linewidth}
\caption{\label{fig:LT_bg_100TeV} 
$L_T$ event distributions for all processes contributing to total SM background, 
for $pp$ collisions at $\sqrt{s} = 100$ TeV with an 
integrated luminosity $\mathcal{L} =$ 30 ab$^{-1}$. The 
four panels show results for the four best signal regions, as labeled. 
The vertical dashed line in all four panels shows our choice of $L_T$ cut of 3500 GeV.}
\end{minipage}\end{center}
\end{figure}

Figure \ref{fig:LT_bg_100TeV} shows the $L_T$ distributions for all background 
components, for the four best signal regions as labeled. The $L_T$ cut is shown 
in the figure as a vertical dashed line. After imposing the $L_T$ cut, the 
most important SM backgrounds are $t\bar{t}V$ and $VVV$ in the two signal regions 
with 2 SS $e/\mu\> +\!\geq 1 \tau$ and in the signal region with $\geq 2e/\mu + 2 \tau$, 
while the most important SM background in the signal region 
with $\geq 2e/\mu + 2 \tau$ (no-$Z$) is $VVV$.

\begin{figure}[!tb]
  \begin{minipage}[]{0.495\linewidth}
    \includegraphics[width=8.0cm,angle=0]{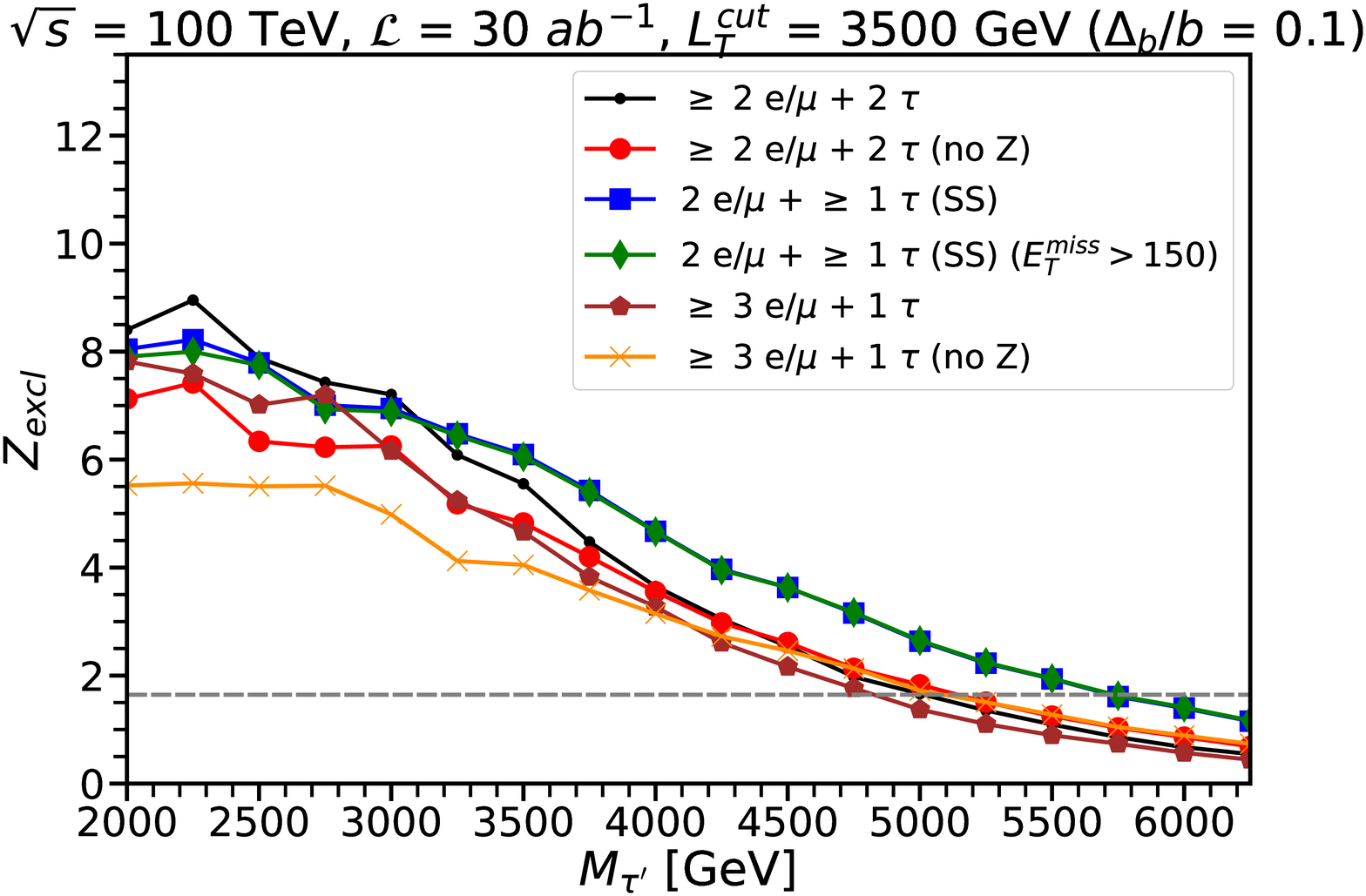}
  \end{minipage}
    \begin{minipage}[]{0.495\linewidth}
    \includegraphics[width=8.0cm,angle=0]{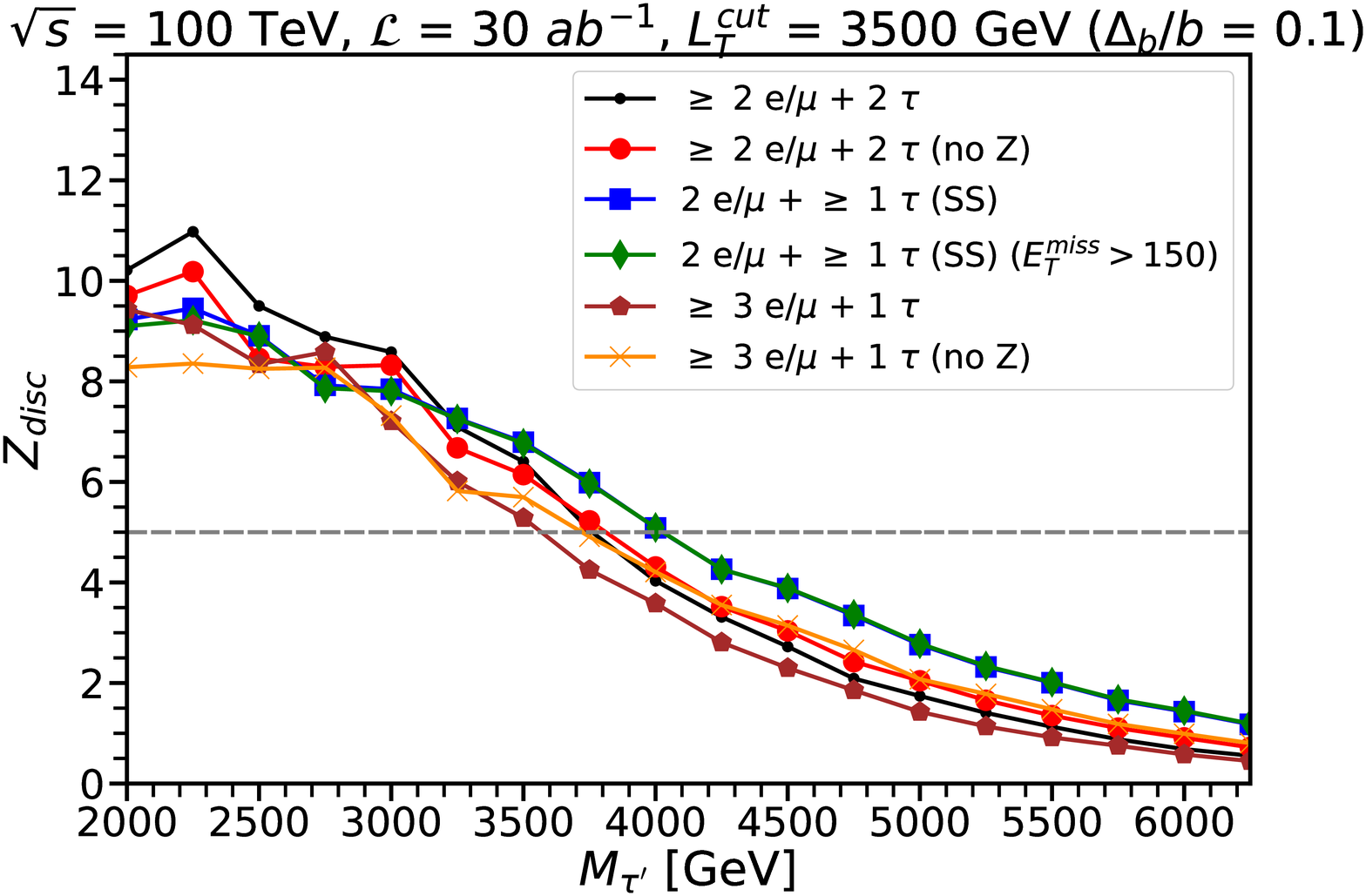}
  \end{minipage}
    \begin{minipage}[]{0.495\linewidth}
    \includegraphics[width=8.0cm,angle=0]{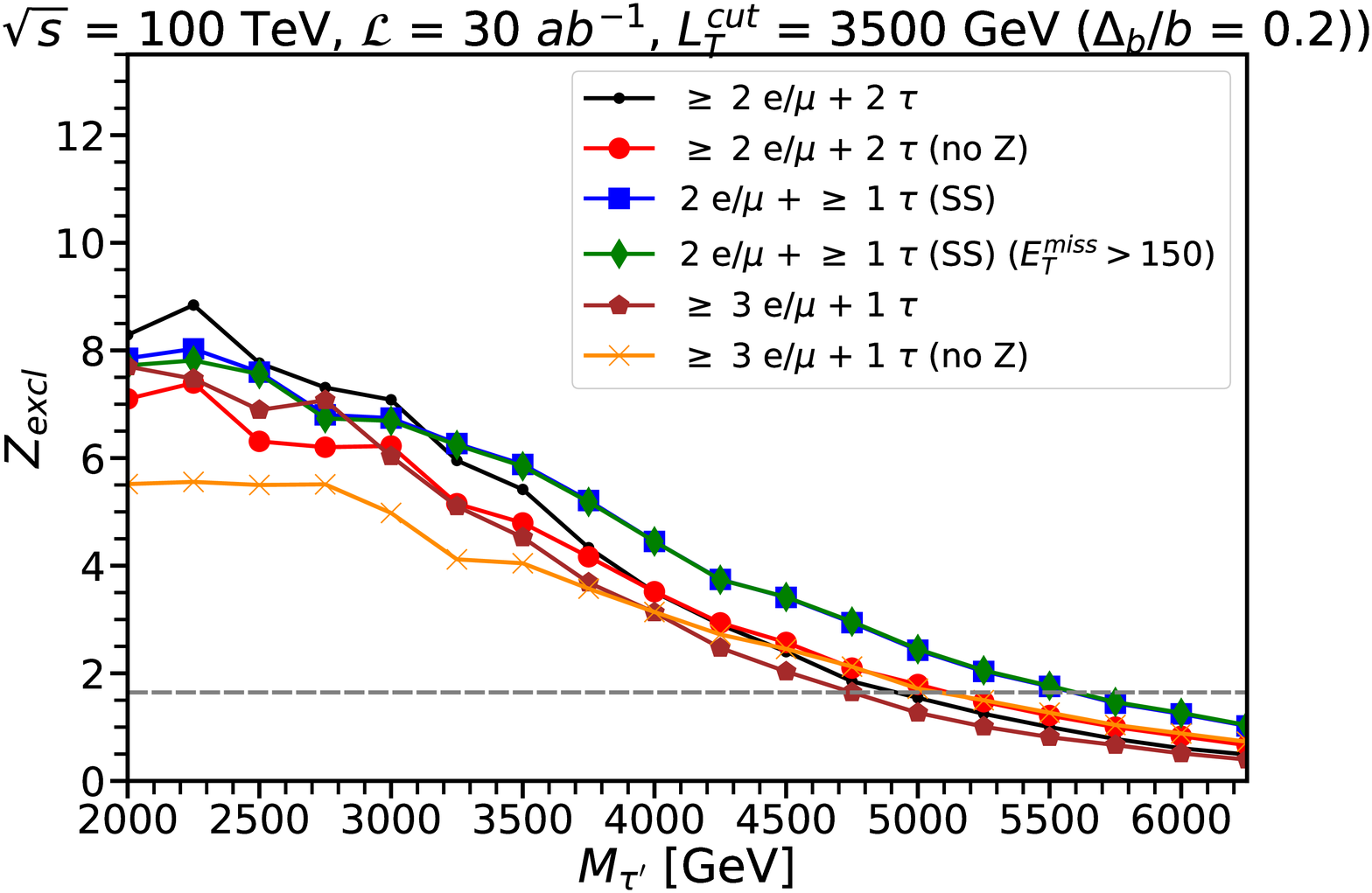}
  \end{minipage}
    \begin{minipage}[]{0.495\linewidth}
    \includegraphics[width=8.0cm,angle=0]{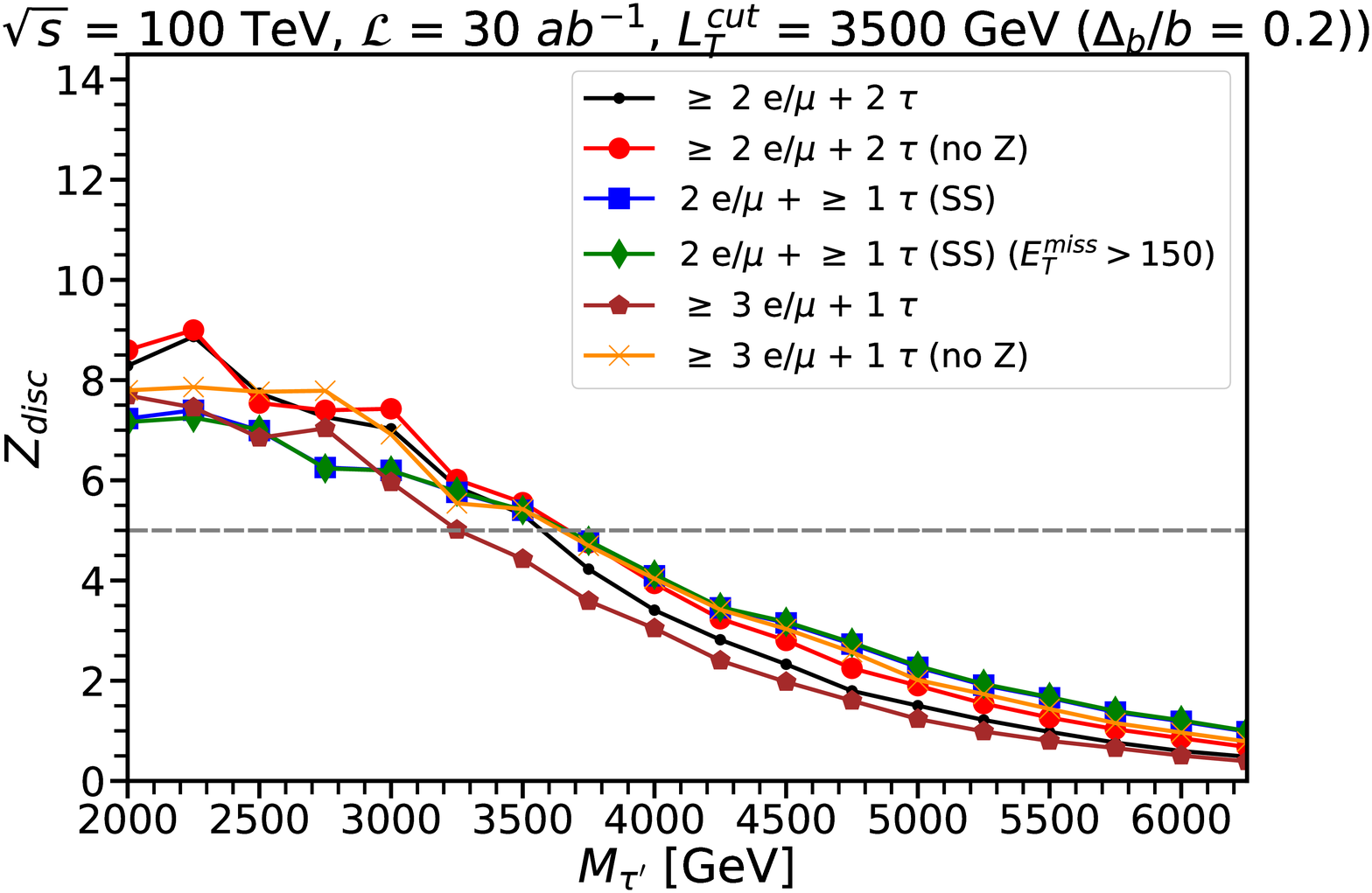}
  \end{minipage}
  \begin{minipage}[]{0.495\linewidth}
    \includegraphics[width=8.0cm,angle=0]{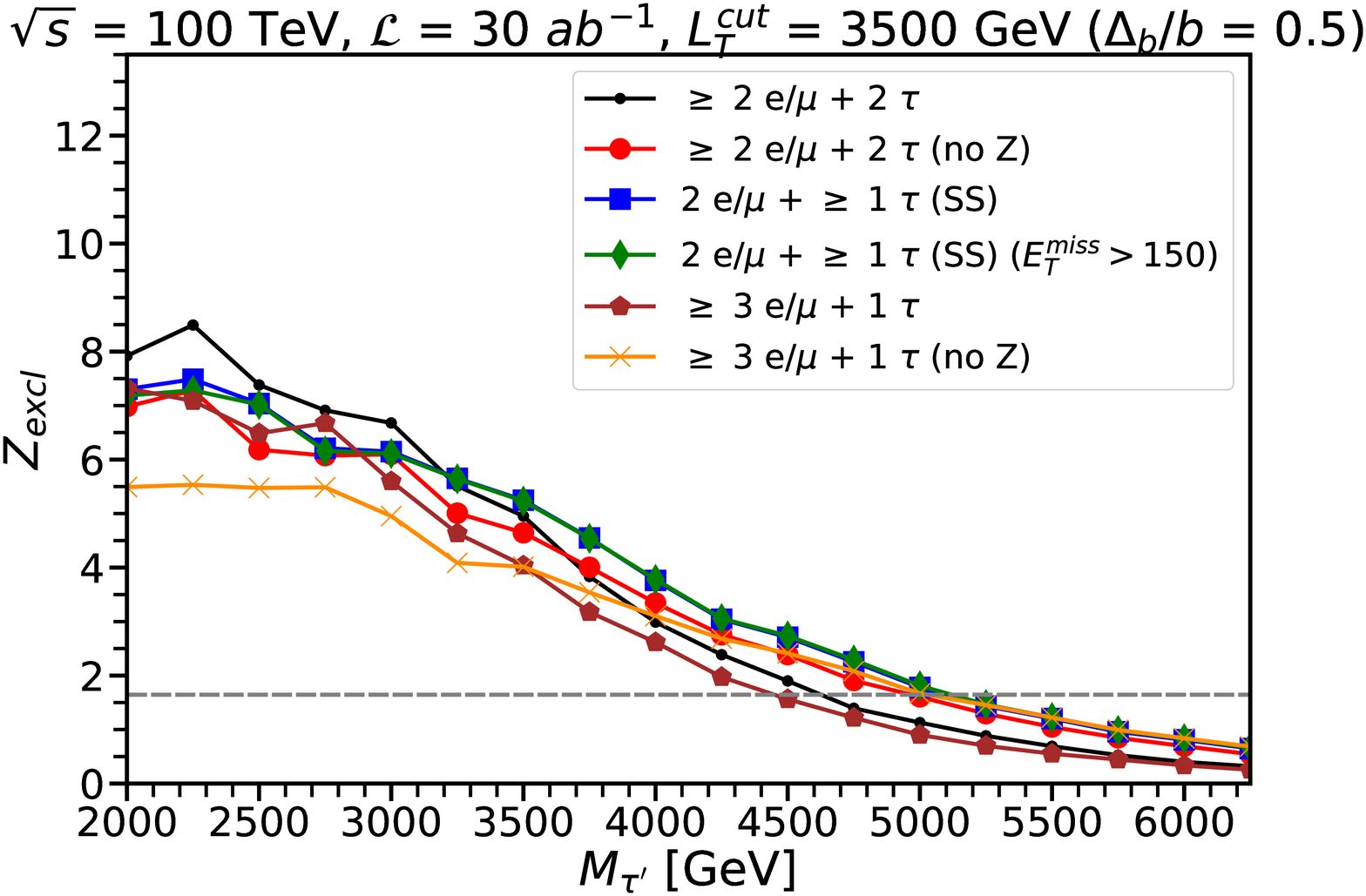}
  \end{minipage}
    \begin{minipage}[]{0.495\linewidth}
    \includegraphics[width=8.0cm,angle=0]{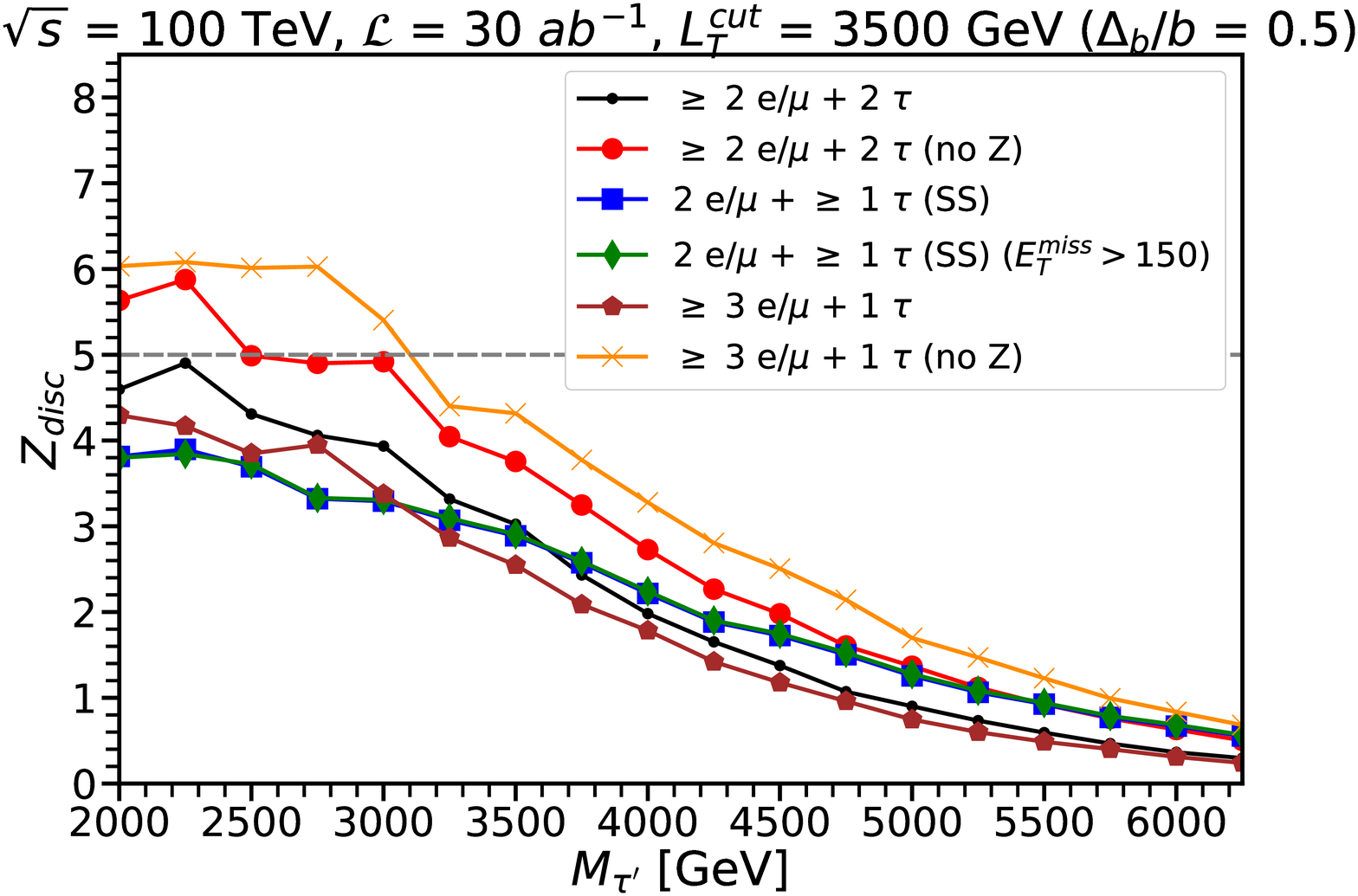}
  \end{minipage}
\begin{center}\begin{minipage}[]{0.95\linewidth}
\caption{
\label{fig:Z_100TeV} 
The median expected significances for exclusion $Z_{\rm excl}$ 
(left panels) and discovery $Z_{\rm disc}$ (right panels) as a function of
$M_{\tau^{\prime}}$ in the Doublet VLL model, for $pp$ collisions at
$\sqrt{s} = 100$ TeV with integrated luminosity $\mathcal{L} = 30$ 
ab$^{-1}$, for six different signal regions as described in the text, each including a cut $L_T>$
3500 GeV. The fractional uncertainty in the background is assumed to be
$\Delta_b/b = 0.1$ (top row), $0.2$ (middle row), and $0.5$ (bottom
row).
}
\end{minipage}\end{center}
\end{figure}

Figure \ref{fig:Z_100TeV} shows the median expected significances for exclusion $Z_{\rm excl}$ (left panels) and discovery
$Z_{\rm disc}$ (right panels) as a function of $M_{\tau'}$, for 
$\Delta_b/b = 0.1$ (top row), $0.2$ (middle row), and $0.5$ (bottom row),
with the cut $L_T > 3500$ GeV imposed. The 
signal regions which require 2 SS $e/\mu\> +\!\geq 1 \tau$ usually provide the
best exclusion and discovery reaches. An exception is that with the fractional uncertainty in the background $\Delta_b/b = 0.5$, 
the signal regions with $\geq 3e/\mu + 1 \tau$ have the farthest discovery reach. 

From Figure \ref{fig:Z_100TeV}, we conclude that a 100 TeV $pp$ collider with  30 ab$^{-1}$ could 
exclude Doublet VLLs with $M_{\tau'}$ up to about 5750 GeV or discover them if the mass is less than about 4000 GeV, assuming the fractional uncertainty in the background to be 0.1. For $\Delta_b/b = 0.5$, one can still expect to exclude Doublet VLLs if the mass is up to about 5100 GeV, or discover them if the mass is less than about 3100 GeV. Again, as a recurring theme at all the collider options considered, a larger uncertainty in the background produces a 
moderate reduction of the exclusion reach, 
but has a much greater impact on the prospects for discovery. 

\begin{figure}[!tb]
  \begin{minipage}[]{0.505\linewidth}
  \begin{flushleft}
    \includegraphics[width=8.0cm,angle=0]{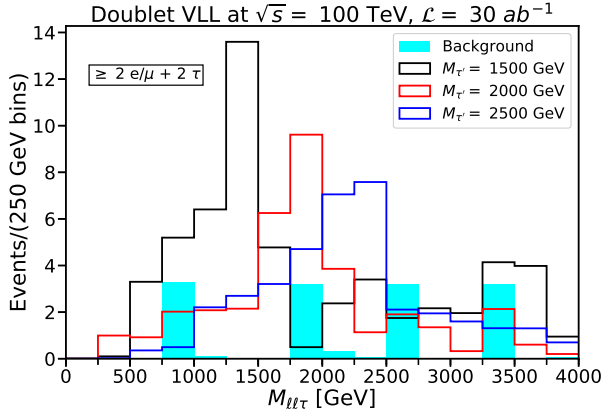}
  \end{flushleft}
  \end{minipage}
\begin{minipage}[]{0.445\linewidth}
\caption{\label{fig:inv_mass_ztau_100tev} 
The event distributions for 3-body invariant masses of 
$\tau^\pm e^+ e^-$ or $\tau^\pm \mu^+\mu^-$, for total background (shaded) and Doublet VLLs (lines), such that the $e^{+}e^{-}$ or $\mu^{+}\mu^{-}$ pair have an invariant mass within 10 GeV of $M_Z$ in a signal region with $\geq 2e/\mu + 2 \tau$, with the cut $L_T > 3500$ GeV imposed. Three different masses $M_{\tau'} = M_{\nu'} = 1500,$ 2000, and 2500 GeV are shown in the plot.
}   
  \end{minipage}
\end{figure}

We again consider the possibility of observing a mass peak for the $\tau'$ when a clear discovery can be made.
Figure \ref{fig:inv_mass_ztau_100tev} shows the event distributions for 3-body invariant mass of 
$\tau^\pm e^+ e^-$ or $\tau^\pm \mu^+\mu^-$, for the signal region with 
$\geq 2e/\mu + 2 \tau$, for three different choices of $M_{\tau'}$, and for the total of 
all backgrounds shown as the shaded histogram. We require the 2-body invariant mass of 
$e^{+}e^{-}$ or $\mu^{+}\mu^{-}$ pair to be within 10 GeV of $M_Z$. We also impose the 
cut $L_T > 3500$ GeV. From Figure \ref{fig:inv_mass_ztau_100tev}, we note that there are 
peaks in the distributions for Doublet VLLs corresponding to, and slightly below, 
their respective masses, giving a possibility to measure the masses of Doublet VLLs, 
if they are indeed discovered. 

\subsection{Singlet VLL models}

We find no possible exclusion of Singlet VLLs in the minimal and the Higgs-philic Singlet VLL models, 
for $pp$ collisions with $\sqrt{s} = 100$ TeV with integrated luminosity of 30 $ab^{-1}$. 
We find some exclusion prospects for the $Z$-philic and $W$-phobic Singlet VLL models, assuming that the fractional uncertainty in the 
background is $\Delta_b/b = 0.1$.
Again, there are no discovery prospects 
in any of the Singlet VLL models. 

\begin{figure}[!t]
  \begin{minipage}[]{0.505\linewidth}
  \begin{flushleft}
    \includegraphics[width=8.0cm,angle=0]{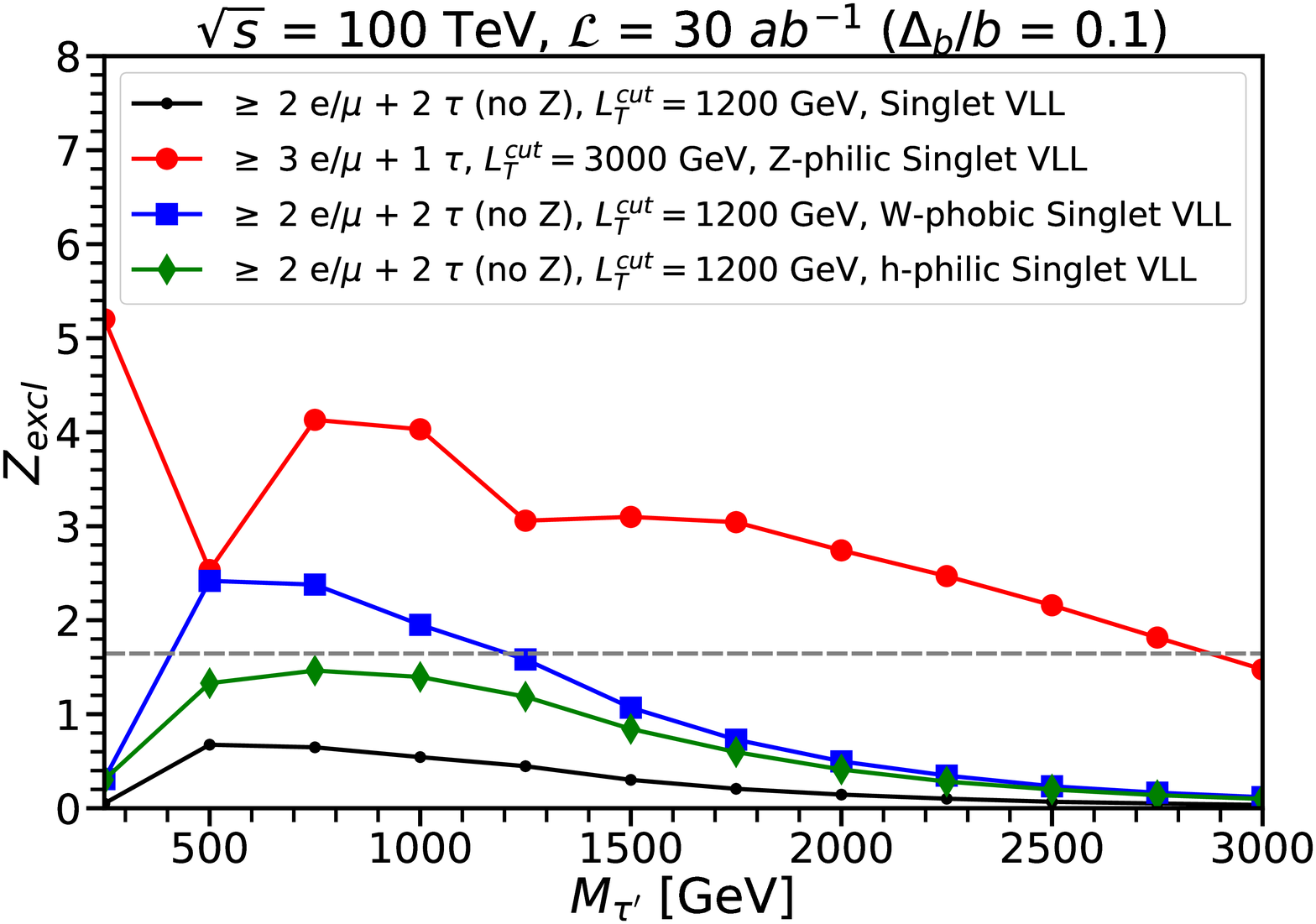}
  \end{flushleft}
  \end{minipage}
\begin{minipage}[]{0.445\linewidth}
\caption{\label{fig:Z_100TeV_Singlet} 
The median expected significances for exclusion $Z_{\rm excl}$ as a function of $M_{\tau^{\prime}}$ 
in the Singlet VLL models, for $pp$ collisions at 
$\sqrt{s} = 100$ TeV with integrated luminosity $\mathcal{L} = 30$ ab$^{-1}$, 
for the best signal region for each of the Singlet VLL models, including a cut on $L_T$ as shown in the plot.
The fractional uncertainty in the background is assumed to be $\Delta_b/b = 0.1$.
}   
  \end{minipage}
\end{figure}

We chose cuts of  $L_T > 3000$ GeV for the $Z$-philic Singlet VLL, 
and $L_T > 1200$ GeV for 
the $W$-phobic Singlet VLL, to approximately maximize 
the exclusion reach in each case. 
Figure \ref{fig:Z_100TeV_Singlet} shows 
the resulting median expected significances 
for exclusion, for $\Delta_b/b = 0.1$, for the best signal region 
for each of the Singlet VLL models,
with the cuts on $L_T$ imposed, as mentioned above. The best signal region for 
exclusion for the $Z$-philic Singlet VLL model is the 
one which requires $\geq 3 e/\mu + 1 \tau$, 
while it was $\geq 2 e/\mu + 2 \tau$, no-$Z$ 
for all other Singlet VLL models.
 
From Figure \ref{fig:Z_100TeV_Singlet},  assuming $\Delta_b/b = 0.1$, we conclude that 
a 100 TeV $pp$ collider with 30 ab$^{-1}$ should be able to exclude Singlet VLLs with 
masses up to about 2850 GeV, if they are $Z$-philic, or exclude masses up to
1200 GeV, if they are $W$-phobic. We find that there 
are no prospects for exclusion of Singlet VLLs in both minimal and the 
Higgs-philic Singlet VLL models. 

\clearpage

\section{Outlook\label{sec:Outlook}}
\setcounter{equation}{0}
\setcounter{figure}{0}
\setcounter{table}{0}
\setcounter{footnote}{1}

Vectorlike leptons are a common feature of motivated theories 
of physics beyond the Standard Model.
In this paper, we have studied the prospects for discovering 
simple VLL models at future $pp$ colliders,
assuming that the new particles decay promptly by mixing with taus,
using multi-lepton signatures at high transverse momentum.
The results of this paper should be considered as first estimates, 
which certainly should be reassessed
as the plans and parameters for future colliders and their detectors 
become clearer. We have not attempted to make use of 
$h\rightarrow b\overline b$ decays, due in part to a relative lack 
of confidence in what should be assumed about $b$-jet tagging 
capabilities compared to $e,\mu$ identification. We have also based 
our analysis entirely on a cut-and-count strategy, while 
a more sophisticated method such as a likelihood analysis 
would certainly be able to do better.

Summaries of our results are shown in Table \ref{tab:doublet_reach} for the 
Doublet VLL model, and in Table \ref{tab:singlet_reach} for the non-minimal 
Singlet VLL models.
In the weak isodoublet case, we have found excellent reach prospects for both 95\% 
exclusion (if the VLLs are absent) or $5\sigma$ discovery (if the VLLs are 
present). In the latter case, we also showed that
there is a possibility to observe an invariant mass peak, although we did not 
pursue this in detail
and the mass peak observation is often quite limited by statistics.
In the weak isosinglet case, the situation is much more difficult. We found no 
discovery or
exclusion reach at all,\footnote{The only limits on such a particle at present are 
$M_{\tau'} > 101.2$ GeV from LEP \cite{Achard:2001qw}, and our estimate 
above in a footnote in section \ref{sec:productiondecay}, inferred from the ATLAS 
long-lived chargino search of ref.~\cite{Aaboud:2019trc},
of $M_{\tau'} > 750$ GeV if it is quasi-stable on detector scales.}
at any $pp$ collider, 
for the simple and well-motivated (e.g.~\cite{Martin:2009bg}) 
case of a promptly decaying pure isosinglet $\tau'$ 
that mixes with the Standard Model tau. 
This should stand as a challenge to future work. 
\begin{table}
\caption{Summary of mass reaches for the 95\% exclusion and $5\sigma$ 
discovery of Doublet VLL at future $pp$ colliders. The fractional uncertainty 
in the background is assumed to be $\Delta_b/b = 0.1$.
\label{tab:doublet_reach}}
\centering
\begin{center}
\begin{tabular}{|c|c|c|c|}
\hline
Collider	& ~~Exclusion reach~~ & ~~Discovery reach~~ \\
\hline
14 TeV HL-LHC, 3 $ab^{-1}$ & 1250 GeV &  900 GeV\\
\hline
27 TeV HE-LHC, 15 $ab^{-1}$ & 2300 GeV &  1700 GeV\\
\hline
70 TeV $pp$ collider, 30 $ab^{-1}$ & 4700 GeV &  3400 GeV\\
\hline
~~100 TeV $pp$ collider, 30 $ab^{-1}$~~ & 5750 GeV &  4000 GeV\\
\hline
\end{tabular}
\vspace{-0.3cm}
\end{center}
\end{table}
\begin{table}
\caption{Summary of mass reaches for the exclusion of Singlet VLL in the 
non-minimal models at future $pp$ colliders. The fractional uncertainty 
in the background is assumed to be $\Delta_b/b = 0.1$.
\label{tab:singlet_reach}}
\centering
\begin{center}
\begin{tabular}{|c|c|c|c|}
\hline
Collider	& ~$Z$-philic~ & ~$W$-phobic~ & ~Higgs-philic~ \\
\hline
14 TeV HL-LHC, 3 $ab^{-1}$ & 600 GeV & 360 GeV & 300 GeV\\
\hline
27 TeV HE-LHC, 15 $ab^{-1}$ & 1200 GeV & 670 GeV & $-$\\
\hline
70 TeV $pp$ collider, 30 $ab^{-1}$ & 1700 GeV & 850 GeV & $-$\\
\hline
~~100 TeV $pp$ collider, 30 $ab^{-1}$~~ & ~~2850 GeV~~ & ~~1200 GeV~~ & $-$\\
\hline
\end{tabular}
\vspace{-0.3cm}
\end{center}
\end{table}

\section*{Appendix A: Mixed singlet/doublet VLL models\label{appendix:basisintegrals}}
\renewcommand{\theequation}{A.\arabic{equation}}
\setcounter{equation}{0}
\setcounter{figure}{0}
\setcounter{table}{0}
\setcounter{footnote}{1}

Consider a generalized model framework that contains, 
besides the Standard Model 
chiral third-family
leptons $\ell_3$ and $\overline e_3$, a 
vectorlike weak isosinglet pair $E, \overline E$ 
and a weak isodoublet pair $L, \overline L$.
(Here we use 2-component fermion notation \cite{Dreiner:2008tw,Martin:2012us}, 
so that barred fields have
electric charge $+1$ and unbarred fields have electric charge $-1$.
The bar is part of the name of the 2-component fermion field, 
and does not denote any kind of conjugation.)
After electroweak symmetry breaking, the Lagrangian mass terms for 
the charged leptons can be written in the form
\beq
-{\cal L} &=& \begin{pmatrix} \overline e_3 & \overline E & \overline L \end{pmatrix}
{\cal M} 
\begin{pmatrix}
\ell_3 
\\
E
\\
L
\end{pmatrix} 
+ {\rm c.c.},
\eeq
where the charged lepton mixing mass matrix is
\beq
{\cal M} = \begin{pmatrix}
y_\tau v & 0 & \phantom{.}\epsilon_2 v \\
\epsilon_1 v & M_1 & x_2 v \\
0 & x_1 v & M_2
\end{pmatrix}
\eeq
where $v = 174$ GeV is the Standard Model Higgs expectation value,
$y_\tau$, $\epsilon_1$, $\epsilon_2$, $x_1$, and $x_2$ are Yukawa couplings, 
and $M_1$ and $M_2$ are electroweak-singlet bare mass terms for the isosinglet and 
isodoublet vectorlike lepton pairs, respectively.
This can be diagonalized to obtain mass eigenstates according to:
\beq
R^* {\cal M} L^\dagger &=&
{\rm diag}(M_\tau, M_{\tau'}, M_{\tau''})
\eeq
where $R$ and $L$ are unitary $3\times 3$ matrices 
and, by convention, $M_\tau < M_{\tau'} < M_{\tau''}$,
where $\tau$ is the usual tau lepton with $M_\tau = 1.777$ GeV, 
and in the special cases considered here $\tau''$ will be taken to be heavy
enough to decouple from direct experimental observation. 
The neutral VLL $\nu'$ has mass $M_2$.

For the lighter new charged VLL $\tau'$, we have partial decay widths:
\beq
\Gamma (\tau' \rightarrow W \nu_\tau) &=&
\frac{M_{\tau'}}{32 \pi}
(1 - r_W)^2 (1 + 2 r_W) |g^W_{\tau' \nu_\tau^\dagger}|^2/r_W
,
\\
\Gamma (\tau' \rightarrow Z \tau) &=&
\frac{M_{\tau'}}{32 \pi}
(1 - r_Z)^2 (1 + 2 r_Z) 
\left (|g^Z_{\tau' \tau^\dagger}|^2 + |g^Z_{\overline\tau' \overline\tau^\dagger}|^2 \right )/r_Z
,
\\
\Gamma (\tau' \rightarrow h \tau) &=&
\frac{M_{\tau'}}{32 \pi}
(1 - r_h)^2
\left (|y^h_{\tau' \overline\tau}|^2 + |y^h_{\tau\overline\tau' }|^2 \right )
,
\eeq
where $r_X = M_X^2/M_{\tau'}^2$ for each of $X=h,Z,W$, and
\beq
g^W_{\tau_j \nu_\tau^\dagger} &=& \frac{g}{\sqrt{2}} L^*_{j1}
,
\\
g^Z_{\tau_j \tau^\dagger_k} &=& \frac{g}{\cos\theta_W} \left [ \frac{1}{2} L^*_{j2} L_{k2} + 
\Bigl (-\frac{1}{2} + \sin^2\theta_W \Bigr ) \delta_{jk} \right ]
,
\\
g^Z_{\overline\tau_j \overline\tau^\dagger_k} &=& \frac{g}{\cos\theta_W} \left [
\frac{1}{2} R^*_{j3} R_{k3} - \sin^2\theta_W \delta_{jk} \right ]
,
\\
y^h_{\tau_j\overline\tau_k} &=& \frac{1}{\sqrt{2}} \left [
y_\tau L^*_{j1} R^*_{k1} + \epsilon_2 L^*_{j3} R^*_{k1} + \epsilon_1 L^*_{j1} R^*_{k2}
+ x_2 L^*_{j3} R^*_{k2} + x_1 L^*_{j2} R^*_{k3} \right ]
,
\eeq
for $(\tau_1, \tau_2, \tau_3)$ corresponding to mass eigenstates 
$(\tau, \tau', \tau'')$ respectively.

In general, the decay widths to Standard Model states are quadratic in 
$\epsilon_1, \epsilon_2$ for small values of those parameters.
For example, expanding to obtain the terms proportional to 
$\epsilon_1^2$, $\epsilon_1 \epsilon_2$, and $\epsilon_2^2$, 
and then keeping only the leading order in a further expansion in
$y_\tau$, $x_1$, and $x_2$ in each of these terms, we obtain
for $M_2^2 \ll M_1^2$:
\beq
|g^W_{\tau' \nu_\tau^\dagger}|^2/r_W 
&\approx&
\left [
\frac{\epsilon_1 v (x_1 M_1 + x_2 M_2)}{M_1^2} 
- \frac{\epsilon_2 y_\tau v}{M_2}  
\right ]^2
,
\\
\left (|g^Z_{\tau' \tau^\dagger}|^2 + |g^Z_{\overline\tau' \overline\tau^\dagger}|^2 \right )/r_Z
&\approx&
\epsilon_1^2 v^2 \frac{M_2^2 (x_1 M_2 + x_2 M_1)^2}{2 M_1^6}
- \epsilon_1 \epsilon_2 y_\tau v^2 \frac{(x_1 M_1 + x_2 M_2)}{M_1^2 M_2}
+ \frac{\epsilon_2^2}{2}
,
\phantom{xxxx}
\\
|y^h_{\tau' \overline\tau}|^2 + |y^h_{\tau\overline\tau' }|^2 
&\approx& 
\epsilon_1^2 v^2 \frac{(2 x_1 M_1 + x_2 M_2)^2}{2 M_1^4}
- 3 \epsilon_1 \epsilon_2 y_\tau v^2 \frac{(x_1 M_1 + x_2 M_2)}{M_1^2 M_2}
+ \frac{\epsilon_2^2}{2}
,
\phantom{xxxx}
\eeq
and for $M_1^2 \ll M_2^2$:
\beq
|g^W_{\tau' \nu_\tau^\dagger}|^2/r_W &\approx& 
\left [
\epsilon_1 - \epsilon_2 y_\tau v^2 \frac{(x_1 M_2 + x_2 M_1)}{M_1 M_2^2}
\right ]^2
,
\\
\left (|g^Z_{\tau' \tau^\dagger}|^2 + |g^Z_{\overline\tau' \overline\tau^\dagger}|^2 \right )/r_Z
&\approx& 
\frac{\epsilon_1^2}{2}
- \epsilon_1 \epsilon_2 y_\tau v^2 \frac{(x_1 M_2 + x_2 M_1)}{M_1 M_2^2}
+ \epsilon_2^2 v^2 \frac{M_1^2 (x_1 M_1 + x_2 M_2)^2}{2 M_2^6}
,
\phantom{xxxx}
\\
|y^h_{\tau' \overline\tau}|^2 + |y^h_{\tau\overline\tau' }|^2 
&\approx& 
\frac{\epsilon_1^2}{2} - 3 \epsilon_1 \epsilon_2 y_\tau v^2
\frac{(x_1 M_2 + x_2 M_1)}{M_1 M_2^2}
+ \epsilon_2^2 v^2 \frac{(2 x_1 M_2 + x_2 M_1)^2}{2 M_2^4}
.
\eeq
Some special limits of interest follow:
\begin{itemize}
\item If $M_2^2 \ll M_1^2$ and $\epsilon_1 = 0$, then 
the isosinglet heavier fermion mass eigenstate
$\tau''$ decouples from experiment, and the lighter states $\tau'$ and $\nu'$ 
form the minimal Doublet VLL model 
as discussed 
above in section \ref{sec:productiondecay}
with $\epsilon = \epsilon_2$ and $M_{\tau'} = M_{\nu'} =M_2$,
and branching ratios that asymptotically approach 
${\rm BR}(\tau'\rightarrow h\tau) = {\rm BR}(\tau'\rightarrow Z\tau) = 0.5$ 
and ${\rm BR}(\nu' \rightarrow W \tau) = 1$.
\item If $M_1^2 \ll M_2^2$ and $\epsilon_2 =  0$, then 
the heavy isodoublet fermions 
$\tau''$ and $\nu'$ decouple,
and the result is the minimal Singlet VLL model as discussed above 
in section \ref{sec:productiondecay}
with $\epsilon = \epsilon_1$ and $M_{\tau'} = M_1$, and branching ratios that
asymptotically approach
${\rm BR}(\tau'\rightarrow h\tau) = {\rm BR}(\tau'\rightarrow Z\tau) = 0.25$ 
and ${\rm BR}(\tau' \rightarrow W \nu_\tau) = 0.5$.
\item If $M_1^2 \ll M_2^2$ and $\epsilon_1 = 0$ and $\epsilon_2 \not=0$, 
then the lightest new fermion will again be a mostly isosinglet $\tau'$
vectorlike lepton with mass approximately $M_{\tau'} = M_1$.
Since the $\tau'$ is mostly isosinglet, its production cross-section
is nearly the same as in the minimal Singlet VLL model.
However, due to its mixing with the heavier 
isodoublets rather than direct mixing
with the Standard Model tau lepton, the decay 
$\tau' \rightarrow W \nu_\tau$ is highly suppressed. 
The possibilities include the sub-cases:
\begin{itemize}
\item[$\star$] $W$-phobic Singlet VLL:
if $x_1=0$, 
then BR$(\tau' \rightarrow h\tau) \approx {\rm BR}(\tau' \rightarrow Z \tau) \approx 0.5$.
\item[$\star$] Higgs-philic Singlet VLL:
if $x_2 = 0$, then 
BR$(\tau' \rightarrow h\tau) \approx 1$.
\item[$\star$] $Z$-philic Singlet VLL: 
if $x_1 \approx -x_2 M_1/2 M_2$, then 
${\rm BR}(\tau' \rightarrow Z \tau) \approx 1$.
\end{itemize}
\end{itemize}
These of course do not exhaust the possibilities, and in a more general search 
it would be sensible to simply take $\Gamma(\tau' \rightarrow Z \tau)$ and
$\Gamma(\tau' \rightarrow h \tau)$ and
$\Gamma(\tau' \rightarrow W \nu_\tau)$ to be free parameters.

\vspace{0.3cm}

{\it Acknowledgments:} P.N.B. thanks Olivier Mattelaer for all the helpful suggestions regarding the usage of Madgraph. P.N.B. also thanks Sergey A.~Uzunyan for his kind assistance in using the NICADD compute cluster at Northern Illinois University, and 
Ramanpreet Singh and Elliot Parrish
for technical advice. 
P.N.B. would also like to express special thanks to Nicholas T.~Karonis, John Winans and the entire support team of Gaea cluster at Northern Illinois University for their help and support in using their high performance computing facility. This work was supported in part by the National
Science Foundation grant number PHY-1719273.


\end{document}